\documentclass[11pt,a4paper]{article}
\usepackage{jheppub}
\pdfoutput=1

\usepackage{amsmath,amssymb,amsfonts,graphics,graphicx,amscd,amsfonts,epsf,epsfig,color}
\usepackage{caption}
\usepackage{subcaption}

\allowdisplaybreaks[4]

\usepackage[normalem]{ulem}

\newcommand{\vev}[1]{\left\langle #1 \right\rangle}

\subheader{LLNL-JRNL-782380, RIKEN-iTHEMS-Report-19}

\title{Thermal phase transition in Yang-Mills matrix model}

\author[a]{Georg Bergner,}
\author[b]{Norbert Bodendorfer,}
\author[c]{Masanori Hanada,}
\author[d,e]{Enrico Rinaldi,}
\author[b]{Andreas Sch\"{a}fer,}
\author[f,g]{and Pavlos Vranas}

\affiliation[a]{
University of Jena, Institute for Theoretical Physics,\\
Max-Wien-Platz 1, D-07743 Jena, Germany}
\affiliation[b]{
University of Regensburg, Institute of Theoretical Physics,\\
Universit\"{a}tsstrasse 31, D-93053, Germany}
\affiliation[c]{
School of Physics and Astronomy and STAG Research Centre,\\
University of Southampton, Southampton, SO17 1BJ, UK}
\affiliation[d]{
Arithmer Inc., R\&D Headquarters, Minato, Tokyo 106-6040, Japan}
\affiliation[e]{
RIKEN iTHEMS Program, Wako, Saitama 351-0198, Japan}
\affiliation[f]{
Nuclear and Chemical Sciences Division, Lawrence Livermore National Laboratory,\\
Livermore CA 94550, USA}
\affiliation[g]{
Nuclear Science Division, Lawrence Berkeley National Laboratory,\\
Berkeley, CA 94720, USA}

\abstract{
We study the bosonic matrix model obtained as the high-temperature limit of two-dimensional maximally supersymmetric SU($N$) Yang-Mills theory.
So far, no consensus about the order of the deconfinement transition in this theory has been reached and this hinders progress in understanding the nature of the black hole/black string topology change from the gauge/gravity duality perspective.
On the one hand, previous works considered the deconfinement transition consistent with two transitions which are of second and third order.
On the other hand, evidence for a first order transition was put forward more recently.
We perform high-statistics lattice Monte Carlo simulations at large $N$ and small lattice spacing to establish that the transition is really of first order.
Our findings flag a warning that the required large-$N$ and continuum limit might not have been reached in earlier publications, and that was the source of the discrepancy.
Moreover, our detailed results confirm the existence of a new partially deconfined phase which describes non-uniform black strings via the gauge/gravity duality.
This phase exhibits universal features already predicted in quantum field theory.
}
\arxivnumber{1909.04592}
\begin{document}
\maketitle

\section{Introduction}\label{sec:introduction}

Bosonic matrix models in one dimension have been studied in various contexts.
Despite their simple structure, they display a rich non-trivial phase diagram that can be accessed by analytical methods only in certain limiting cases.
An important motivation to study these theories arises from their connections to supersymmetric Yang-Mills theories (SYM) in one and two dimensions, which have various applications to quantum gravity via the gauge/gravity duality~\cite{Maldacena:1997re}.

The Euclidean action of the gauged bosonic U($N$) matrix model is\footnote{
The U(1) part of the gauge group is decoupled from the SU($N$) part.
Therefore, our results in this paper are valid also for the SU($N$) theory.
The only technical difference is that the center symmetry becomes $\mathbb{Z}_N$ instead of U(1).
Note also that the U(1) part of the scalars are decoupled.
In order to remove the trivial flat direction associated with the U(1) part, we impose
$\int_0^\beta dt X_I(t)=0$ for each $I$.
}
\begin{eqnarray}
S = \frac{N}{2\lambda}
\int_0^\beta dt\ {\rm Tr}\left\{
(D_t X_I)^2
-
\frac{1}{2}[X_I,X_J]^2 \label{eq:MainAction}
\right\},
\end{eqnarray}
where $\lambda=g^2_{\textrm{YM}}N$ is the 't~Hooft coupling, $\beta$ is the inverse temperature, $I, J =1,2,\cdots, d$ with $d=D-1$,
and $X_I$ are $N\times N$ hermitian matrices.
The covariant derivative $D_t$ is defined by $D_tX_I=\partial_tX_I-i[A_t,X_I]$, where $A_t$ is the gauge field.
We will mainly consider the case $D=10$, since it is the bosonic version of the BFSS model~\cite{Banks:1996vh}.
In addition to gauge symmetry, this theory has the U(1) center symmetry.
An order parameter associated with the center symmetry is the Polyakov loop,
\begin{equation}\label{eq:ploop}
P = \frac{1}{N} {\rm Tr}  {\cal P} e^{i\int_0^\beta dtA_t}
\quad ,
\end{equation}
where ${\cal P}$ denotes the path ordering.
The Polyakov loop transforms as $P\to e^{i\theta}P$ under the U(1) transformation.
Another important symmetry is SO($d$) which rotates the $X_I$ scalars as $d$-dimensional vectors.
In this paper, we will focus on the breaking of the center symmetry.

While the full supersymmetric BFSS model in the large-$N$ limit has a dual gravity description~\cite{Itzhaki:1998dd}, no weakly-curved gravity dual is known for its bosonic part.
Still, this model can provide insights into gravity, as we will see shortly.
For comparison and to study the large-$D$ limit, we will also investigate the case of $D=26$.

This theory is deconfined at high temperature, and confined at low temperature for any $D\ge 3$.
Based on large-$D$ and large-$N$ analytical techniques~\cite{Mandal:2009vz} and, partially, on numerical Monte Carlo results at finite $N$~\cite{Kawahara:2007fn}, it was initially believed that the deconfinement transition is not a first order one but rather there are two phase transitions, one of second order and one of third order, in close proximity.
More recently however, evidence was presented that there may be only one transition of first order~\cite{Azuma:2014cfa}.
We study numerically the phase diagram of this model in the large-$N$ limit in order to confront the numerical results with analytical predictions, in particular taking into account the continuum limit which was not previously investigated.
In this paper, we will provide robust numerical evidence that for $D=10$ there is only one deconfinement transition and it is of first order in the large-$N$ limit.

There are several reasons to be interested in the order of the phase transitions for this model.
Firstly, let us point out the connection to the topology change between a black hole and a black string~\cite{Gregory:1993vy,Kol:2002xz}.
In order to understand it, note that the bosonic matrix model in Eq.~\eqref{eq:MainAction} is the high-temperature limit of two-dimensional maximally supersymmetric Yang--Mills (SYM) theory compactified on the spatial circle S$^1$.
The bosonic matrix model is obtained by shrinking the temporal circle of the two-dimensional SYM theory to a point.
Consequently, what is called the ``temporal circle'' in the bosonic matrix model actually corresponds to the ``spatial circle'' in the compactified two-dimensional SYM theory.
The phase transition we study in this paper is the remnant of center symmetry breaking/restoration along the spatial S$^1$ in this higher-dimensional theory, which can be regarded as the black hole/black string topology change in the ${\mathbb R}^{1,8}\times S^1$ spacetime of the dual gravitational description~\cite{Aharony:2004ig}.

In the studies of the black hole/black string transition based on general relativity, a reliable analysis of the topology change is out of reach at the moment due to the unavoidable curvature singularity.
This problem can be avoided by using the dual gauge theory, namely the 2d maximal SYM theory mentioned above.
The 2d maximal SYM theory contains information about the stringy corrections and can teach us how the singularity can be resolved.
At low temperatures and large $N$, where the stringy corrections are small, the dual gravity description predicts a first order transition when the radius of S$^1$ is small~\cite{Aharony:2004ig}.
However, the details of how the topology change takes place is out of reach in the gravitational approach.
At intermediate and high temperatures, the only practical tool to gain insights into the dynamics of the theory is a numerical approach.

\begin{figure}[htbp]
\begin{center}
\rotatebox{0}{
\scalebox{0.4}{
\includegraphics{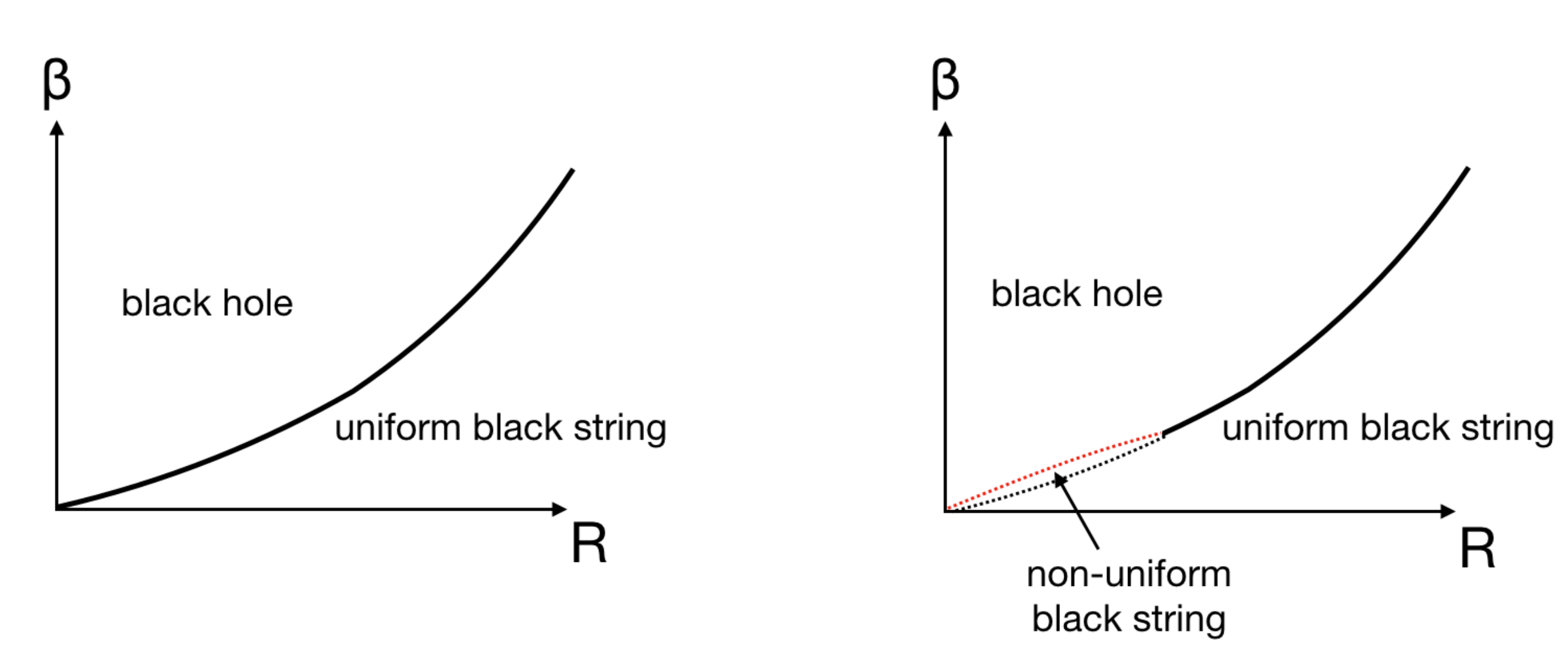}}}
\end{center}
\caption{
Possible phase diagrams of 2d maximal SYM on a circle in the canonical ensemble~\cite{Aharony:2004ig,Aharony:2005ew}.
The vertical axis is the inverse temperature, $\beta=T^{-1}$ and the horizontal axis is the radius $R$ of the spatial circle.
The phase transition is related to the breaking of center symmetry along the spatial circle, analogous to confinement and deconfinement.
Black hole, uniform black string and non-uniform black string phases on the gravity side correspond to the ``completely'' deconfined, confined and ``partially'' deconfined phases, respectively.
(See Sec.~\ref{sec:partial_deconfinement} for the meaning of ``complete'' and ``partial'' deconfinement.)
Our numerical simulations tell us that the left figure is more likely to be true.
}\label{fig:2dSYM_phase_diagram}
\end{figure}

Lattice approaches (for example see e.g. Refs.~\cite{Joseph:2015xwa,Bergner:2016sbv,Hanada:2016jok,Schaich:2018mmv} for recent reviews) are applicable to 2d maximal SYM~\cite{Catterall:2010fx,Catterall:2017lub,Giguere:2015cga,Kadoh:2017mcj}.
Numerical approaches are computationally expensive and it is hard to take $N$ sufficiently large at this stage.
A theory that is more computationally tractable is the one-dimensional bosonic matrix model Eq.~\eqref{eq:MainAction}.
Depending on the nature of the transition in the bosonic matrix model, the dual gravity prediction of a first order transition may survive at high temperature (left of figure~\ref{fig:2dSYM_phase_diagram}), or it may fail and the transition will split into two, a third order and a second order one as depicted in the right panel of figure~\ref{fig:2dSYM_phase_diagram}.
In the former case, we might be able to obtain valuable intuition into the low-temperature region, which is not easy to access with currently available computational resources, from the high-temperature region, which is relatively easier to study.
In the latter case, the ``non-uniform string'' becomes stable due to stringy effects.

Similar problems are of interest in the context of the application of holography to QCD.
In fact, this strategy was also used by Witten~\cite{Witten:1998zw} to study 4d Yang-Mills (YM) as the high-temperature limit of 5d maximal SYM, for which the dual gravity analysis is tractable.
Depending on whether the qualitative features of the phase transition change or not, the scope of the holographic \emph{approximation} can change. (See e.g. Ref.~\cite{Mandal:2011ws} for detailed discussions regarding this point.)
Another example is SYM on ${\mathbb R}^{3}\times$S$^1$ deformed by the gaugino mass~\cite{Poppitz:2012nz}.
The deconfinement transition on a small three-sphere was studied in the same manner~\cite{Sundborg:1999ue,Aharony:2003sx}.

The nature of the deconfinement transitions in various gauge theories fits to the framework of partial deconfinement~\cite{Hanada:2018zxn} that we will explain in Sec.~\ref{sec:theory}.
The matrix model provides us with the simplest setup to study it in a non-perturbative way.
A good understanding of these transitions might shed light on the deconfinement transition in gauge theories or the microscopic nature of the QCD crossover from the hadronic phase to the quark-gluon-plasma phase.

A related motivation is provided by the black hole information problem.
When the phase transition is of first order, there is an unstable phase~\cite{Hanada:2018zxn} which is analogous to the Schwarzschild black hole with negative specific heat~\cite{Hawking:1974sw} in the standard setup of the AdS/CFT duality~\cite{Witten:1998zw}.
Therefore, the identification of simple models exhibiting such behavior is the first step towards investigating this issue with a detailed numerical study.

The theory we consider in this paper (Eq.~\eqref{eq:MainAction}) admits an analytic treatment in terms of a large-$D$ expansion~\cite{Mandal:2009vz}, which predicts two transitions, one of second and one of third order.
Numerical Monte Carlo studies at $D=10$~\cite{Kawahara:2007fn}, up to $N=32$, looked consistent with this large-$D$ analysis, while more recent studies~\cite{Azuma:2014cfa} for the same value of $N$ seemed to contradict with that analysis.

In the following we investigate some reasons that motivate our investigation.
Firstly, finite-$N$ corrections become more relevant near the critical point, in analogy to finite-volume effects of a statistical system near criticality.
The numerical data of Ref.~\cite{Kawahara:2007fn} did not include a dedicated study of finite-$N$ corrections, showing only two values of $N$.
In particular, the transition was studied at $N=32$ and we will show that this value is not sufficiently large to reveal the order of the transition by looking at $\vev{|P|}(T)$.
Other papers~\cite{Azeyanagi:2009zf,Filev:2015hia} which numerically backed up the observation in Ref.~\cite{Kawahara:2007fn} were not able to study the large-$N$ limit either.
Evidence for a first order transition was presented in Ref.~\cite{Azuma:2014cfa}, along with a discussion of why the large-$D$ analysis may fail to correctly predict the order of the transition for small $D$.
The first order signal shown in that study was however not completely clear.
Moreover, the study did not consider discretisation effects which may affect the nature of the transition by shuffling the order of temperatures if there are multiple transitions as expected from the large-$D$ analysis (see figure \ref{fig:Pol-vs-T-possibilities}).
This was due to the limited computational resources available to deal with the considerable increase in numerical cost for larger $N$ and smaller lattice spacing.

Secondly, $D=10$ may not be sufficiently large to make contact to the large-$D$ expansion and therefore to trust the analytical expectation.
In order to get a rough intuition, let us consider an analogous gravity problem, the black hole/black string transition in general relativity on $\mathbb{R}^{D-1}\times S^1$.
In this case, the transition is first order at $D\le 13$~\cite{Sorkin:2004qq} and \emph{large enough $D$} means $D>13$.
This suggests that it might be dangerous to trust the large-$D$ analysis at $D=10$.
Note that the order of the transition is particularly sensitive to the value of $D$.
The large-$D$ analysis seems to be more reliable all the way down to small $D$ for other properties of the theory like the approximate location of the transition~\cite{Hanada:2016qbz}, see also the discussion in \cite{Azuma:2014cfa}.
It is therefore important to study in detail the large-$N$ limit at fixed $D=10$.

In this paper, we present numerical evidence that the transition in the $D=10$ model is of first order.
Therefore, in figure~\ref{fig:2dSYM_phase_diagram}, the left diagrams are more likely to be true.
(Although, strictly speaking, our findings also allow for the possibility that the transition is of higher order at intermediate values of $R$.)
We study $D=26$ as well.
Somewhat surprisingly, we have observed signals consistent with a first order transition, which suggests the large-$D$ approximation is not precise even at $D=26$.

The organization of this paper is as follows.
We start with explaining theoretical expectations in Sec.~\ref{sec:theory}, in order to define the strategy of our numerical simulations.
Then, we will show the results of the simulations and their implications in Sec.~\ref{sec:numerical-results}.
Sec.~\ref{sec:discussions} is devoted to discussions and conclusions.

\section{Theoretical expectations}\label{sec:theory}

In this Section, we summarize the theoretically expected features of the theory, which will be compared to numerical data and used to determine the order of the phase transition.
We introduce three different patterns characterizing the deconfinement transition~\cite{Hanada:2018zxn} and explain how they can be distinguished using numerical simulations.
These patterns can be explained by introducing the concept of \emph{partial deconfinement}~\cite{Hanada:2016pwv,Hanada:2018zxn,Berenstein:2018lrm}.
Since our main goal is to identify the order of the transition, we only include relevant details about partial deconfinement and we refer the interested reader to the papers cited above for a comprehensive review.

\subsection{Partial deconfinement and possible phase structures}\label{sec:partial_deconfinement}

\begin{figure}[htbp]
\begin{center}
\scalebox{0.4}{
\includegraphics{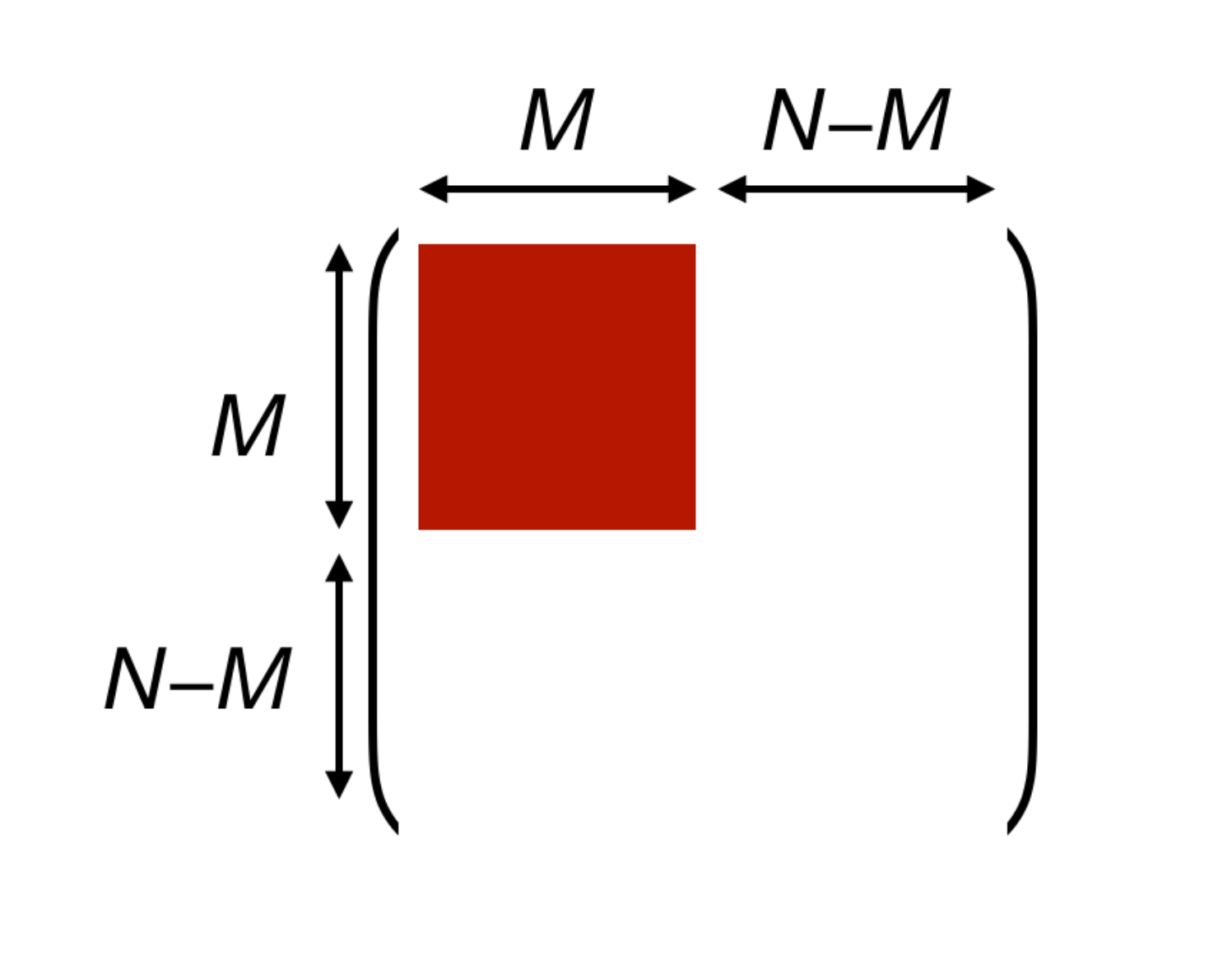}}
\end{center}
\caption{
Matrix representation of the partial deconfinement proposal~\cite{Hanada:2016pwv,Hanada:2018zxn,Berenstein:2018lrm}.
The matrix represents a U($M$) subgroup of the U($N$) gauge group which is `deconfined'.
Hence the full gauge group is only `partially deconfined'.
}\label{fig:partial_deconfinement}
\end{figure}

The word `partial deconfinement'~\cite{Hanada:2016pwv} is used to characterize a \emph{phase} where a subset of the U($N$) group, which we denote as U($M$) in the following, is deconfined.
A \emph{matrix representation} of this concept is illustrated in figure~\ref{fig:partial_deconfinement}.
The `completely deconfined' and `confined' phases correspond to, as expected, $M=N$ and $M=0$, respectively.
Intuitively, the reason why partial deconfinement takes place is simpler to understand when working in the large-$N$ limit: complete deconfinement requires an energy of order $N^2$, and hence, if the energy is much smaller (say $E\sim \epsilon N^2$), only a part of the color degrees of freedom can be excited, U($M$) with $M\sim\sqrt{\epsilon}N$.
The initial motivation of partial deconfinement was to understand the gauge theory description of the Schwarzschild black hole with negative specific heat via the gauge/gravity duality, closely following a very similar mechanism based on partial Higgs-ing~\cite{Berkowitz:2016znt,Berkowitz:2016muc} in gauge theories.
There have been various consistency checks for several theories, both at weak coupling and strong coupling \cite{Hanada:2016pwv,Hanada:2018zxn}, and explicit demonstrations based on state counting are available for several theories~\cite{Hanada:2019czd}.

\begin{figure}[htbp]
\begin{center}
\scalebox{0.2}{
\includegraphics{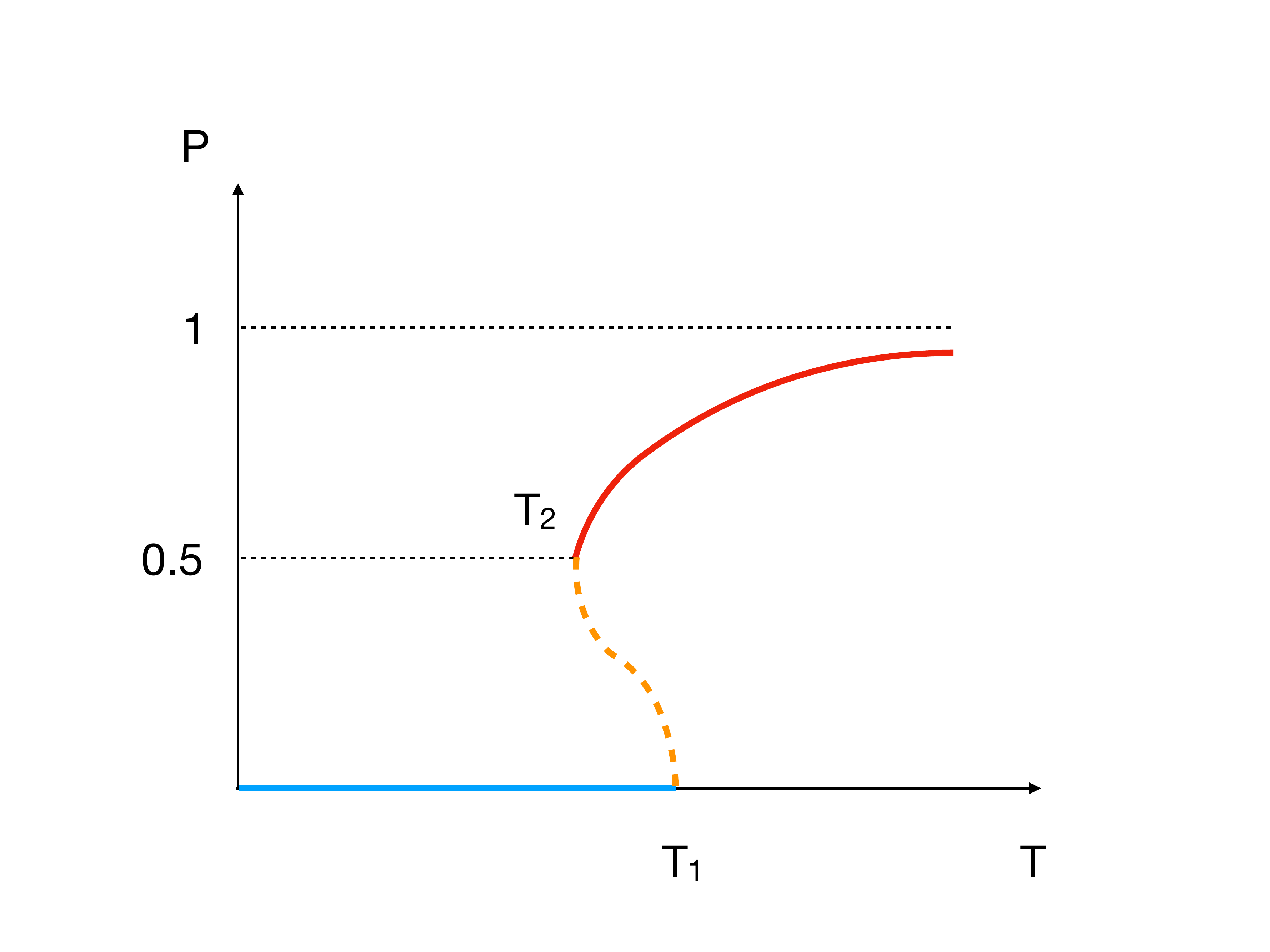}}
\scalebox{0.2}{
\includegraphics{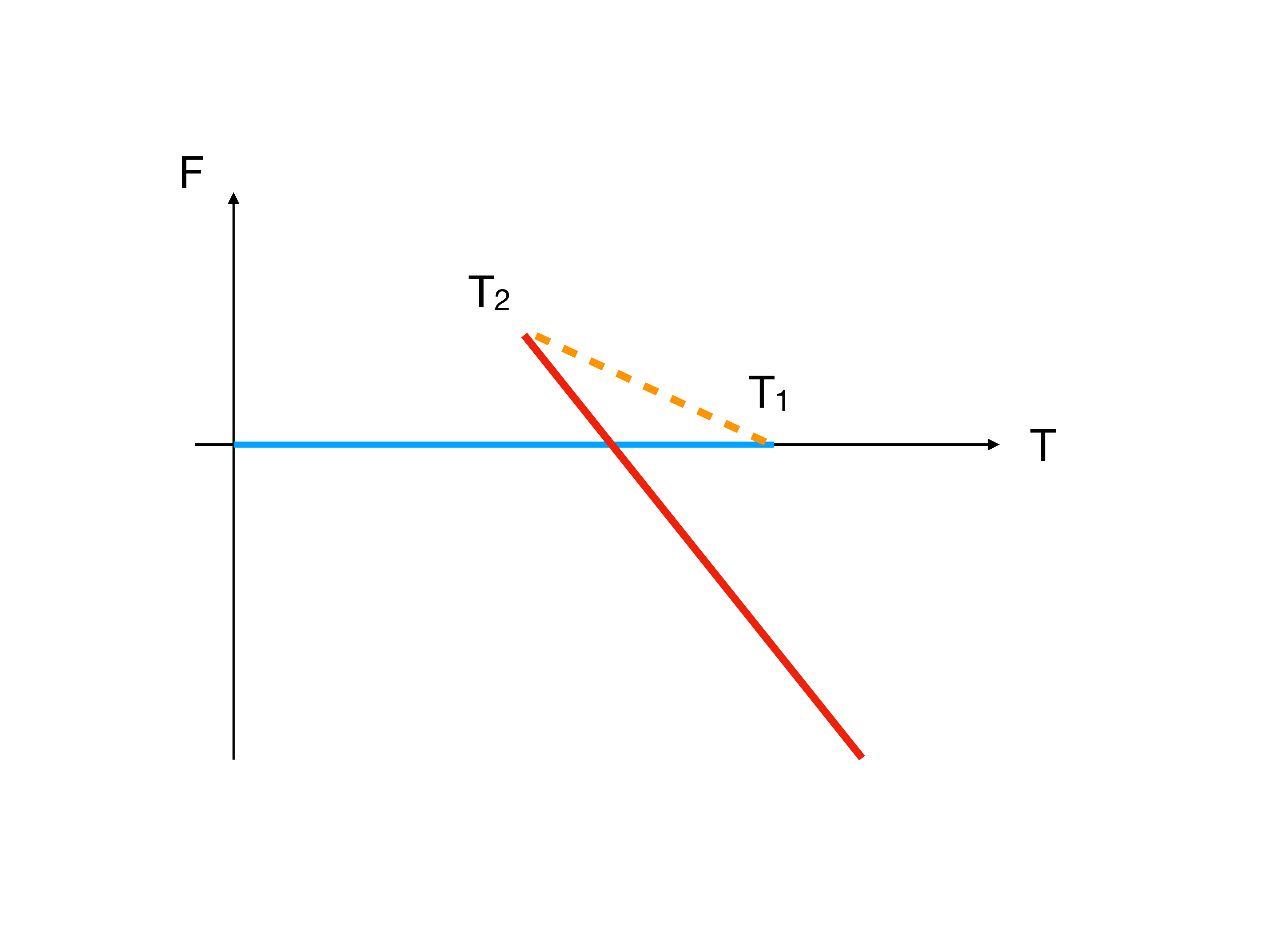}}
\scalebox{0.2}{
\includegraphics{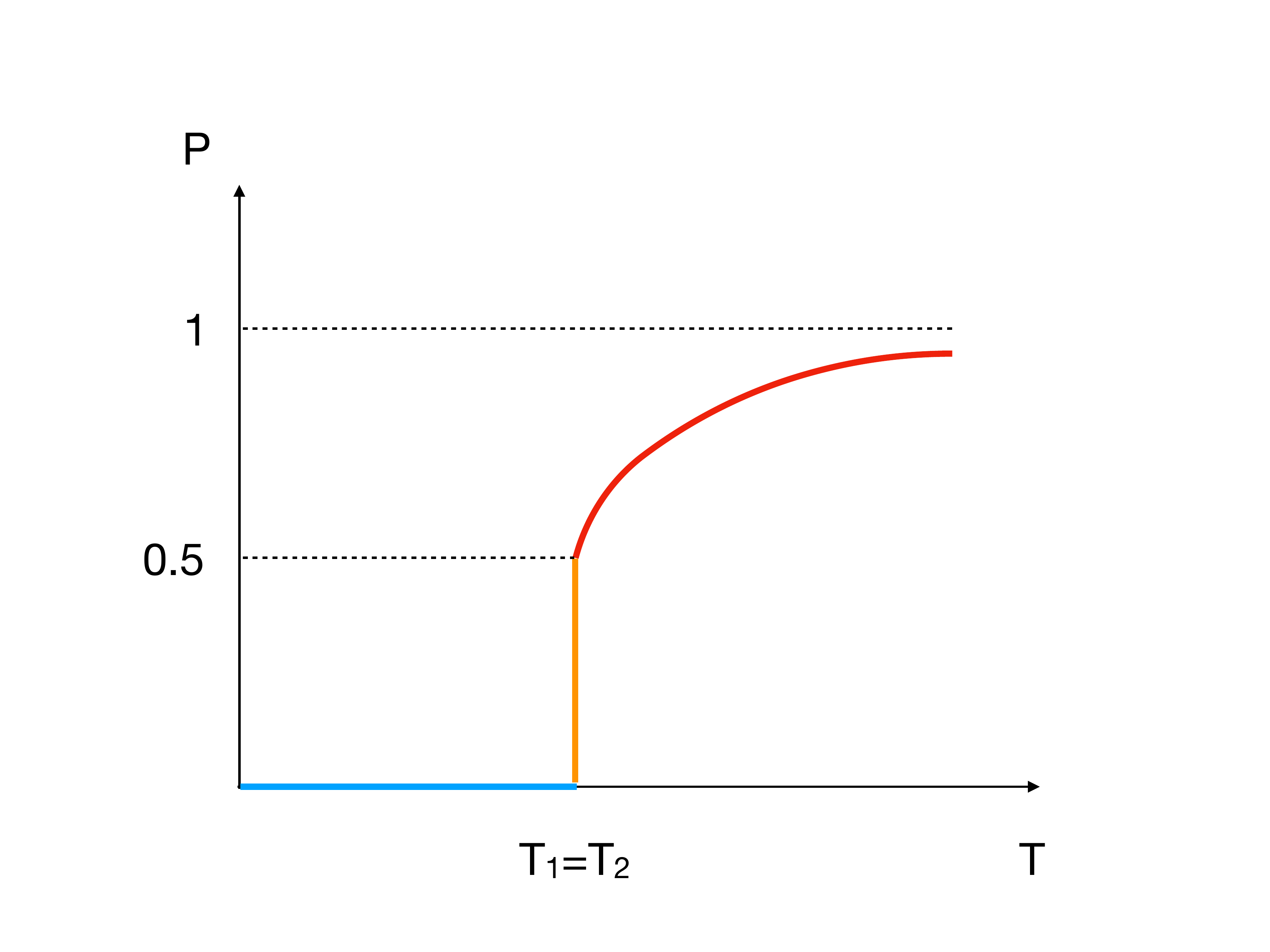}}
\scalebox{0.2}{
\includegraphics{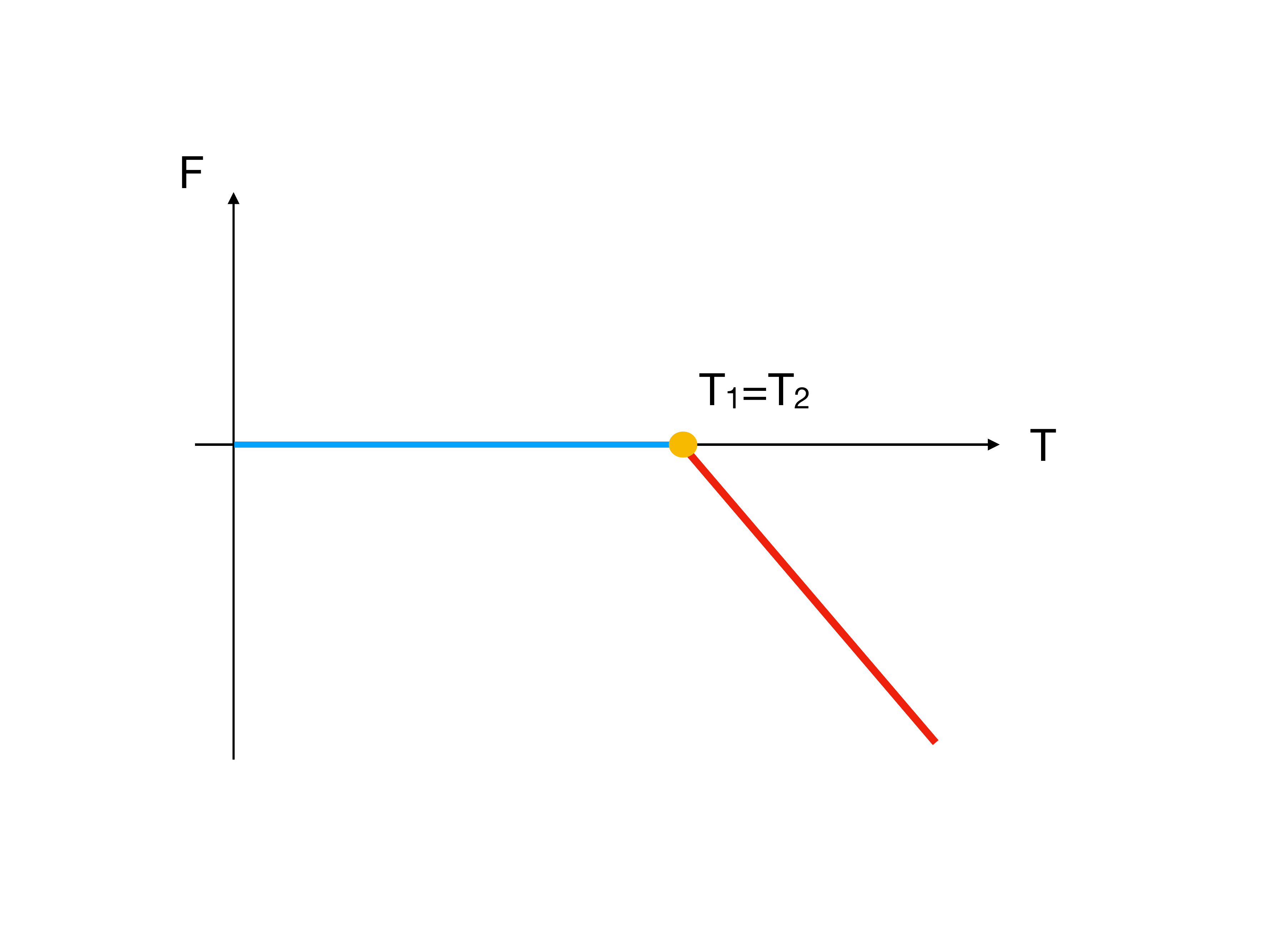}}
\scalebox{0.2}{
\includegraphics{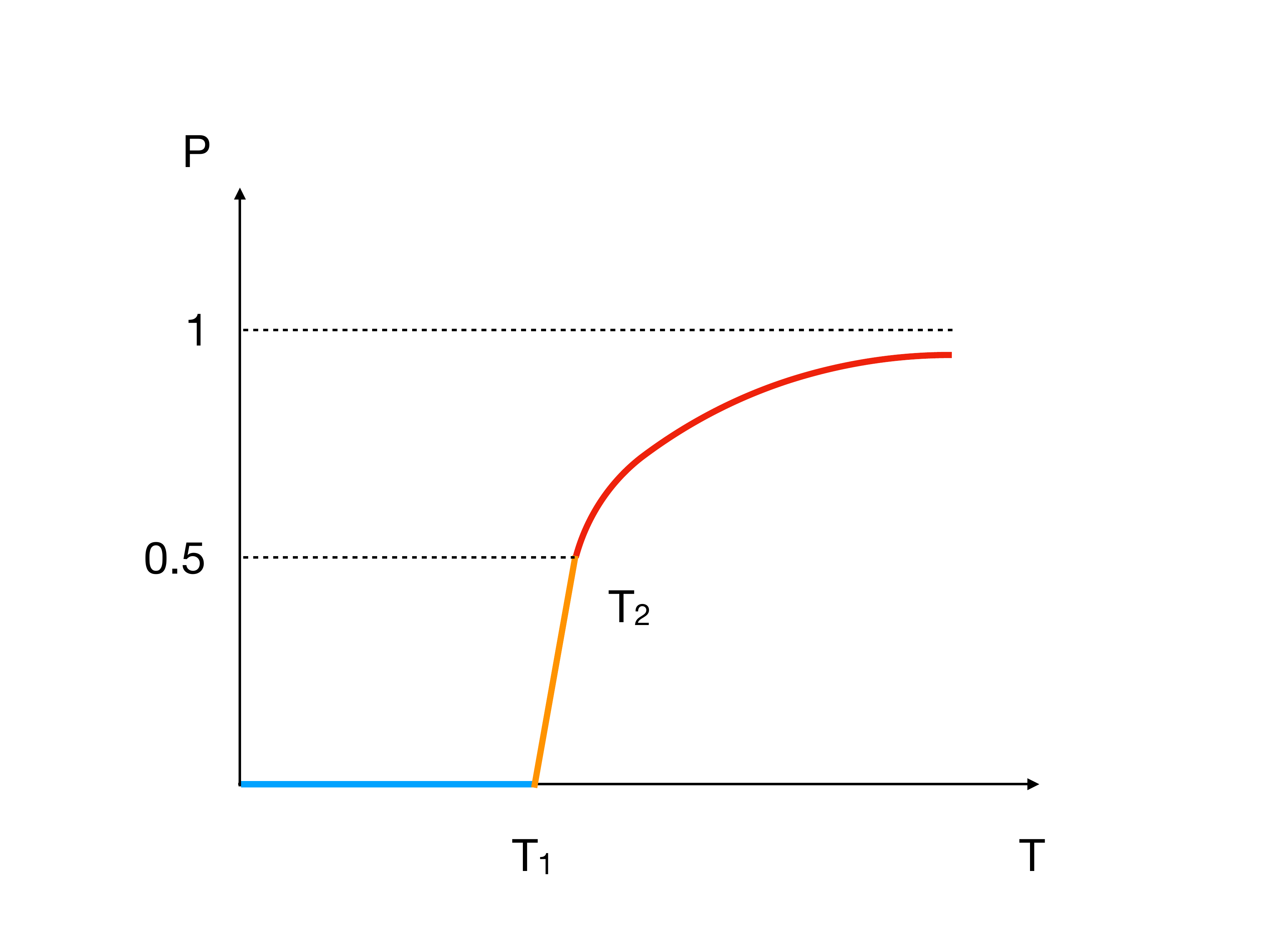}}
\scalebox{0.2}{
\includegraphics{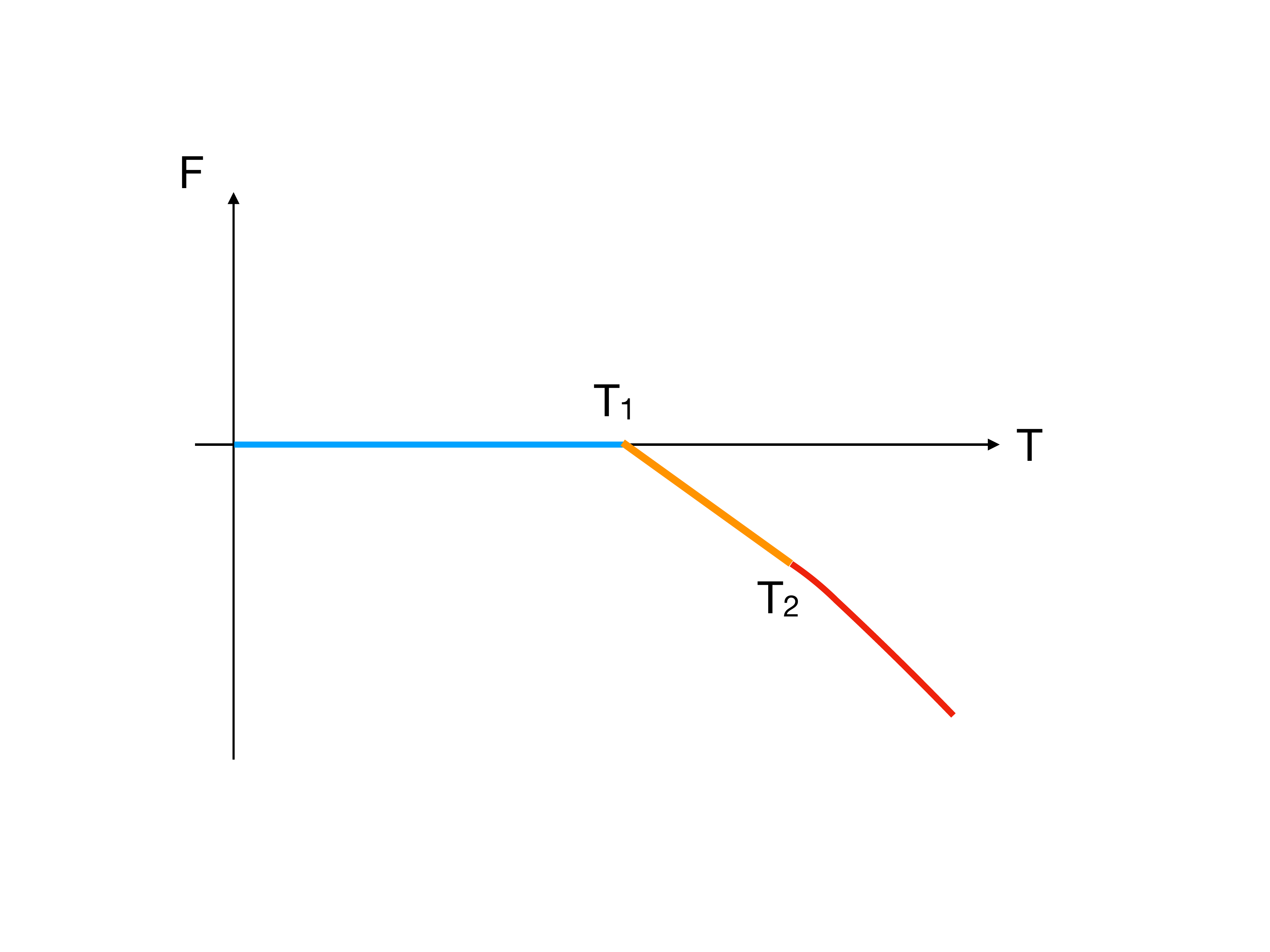}}
\end{center}
\caption{Left: Cartoon pictures of the possible phase diagrams of the bosonic BFSS matrix model in the canonical ensemble.
In order, the three rows correspond to three deconfinement transitions: first order with hysteresis ($T_1>T_2$), first order without hysteresis ($T_1=T_2$), and two transitions of second and third orders ($T_1<T_2$).
Dashed lines represent unstable phases, while solid lines represent stable or metastable phases.
The blue, orange and red lines correspond to the confined, partially deconfined and completely deconfined phases.
Right: Corresponding free energies for the three types of scenarios. See text for details. Here we use $|P|=P$.
}\label{fig:Pol-vs-T-possibilities}
\end{figure}

In the large-$N$ limit, quite generally, the deconfinement transition in gauge theories was classified in three types~\cite{Hanada:2018zxn} according to the picture of canonical ensemble and the concept of partial deconfinement introduced above.
Given that there are three phases, we will introduce two temperatures $T_1$ and $T_2$, the first separating the `completely confined` and the `partially deconfined` regions, and the second separating the partially deconfined and the completely deconfined regions.
The phase diagram is represented by its absolute value $|P|$ as a function of the temperature.
In figure~\ref{fig:Pol-vs-T-possibilities}, the three rows represent the following three distinct possibilities for the order of the phase transition:
\begin{itemize}
\item
{\bf First order with hysteresis.}
There is a local maximum of the free energy corresponding to the partially deconfined phase separating two minima, the completely deconfined phase and the confined phase.
A hysteresis sets in at $T_2\le T\le T_1$.
In the microcanonical ensemble, when the volume is sufficiently large, the confined and completely deconfined phases can occupy most of the space, and the partially deconfined phase appears at the interface.
In a matrix model, the partially deconfined phase is stable in the microcanonical ensemble because there are no spatial dimensions and a separation in volume cannot take place.

\item
{\bf First order without hysteresis.}
At the transition temperature $T = T_1 = T_2$, there is a Hagedorn string with degenerate free energy~\cite{Sundborg:1999ue,Aharony:2003sx}.
It corresponds to the partially deconfined phase denoted by the orange line.
When the volume is sufficiently large, different vacua can appear at different locations, but this is only realized in systems with spatial dimensions.

\item
{\bf Two transitions of second and third orders.}
There are three stable phases with $T_1 \le T_2$: even the partially deconfined phase is stable, both in canonical and microcanonical ensemble.
\end{itemize}

A convenient way to distinguish the three different phases (confined, partially deconfined and completely deconfined) is to look at the distribution of the phases of the Polyakov loop $\rho(\theta)$ \cite{Hanada:2018zxn}.
Note that this is just one of the characterizations of the phases.
In fact, partial deconfinement does not necessarily require center symmetry, and the Polyakov loop is not necessarily an order parameter for some theories with partial deconfinement~\cite{Hanada:2019czd}.
By definition, $\theta$ is distributed between $+\pi$ and $-\pi$.
In the confining phase, the distribution is uniform at large $N$:
\begin{eqnarray}
\rho_{\rm c}(\theta)=\frac{1}{2\pi}.
\end{eqnarray}
Here, the subscript c stands for `confined'. We will use p and d for `partially deconfined' and `deconfined', respectively.
The transition between the partially and completely deconfined phases is the Gross-Witten-Wadia (GWW) transition~\cite{Gross:1980he,Wadia:2012fr}, as found in~\cite{Hanada:2018zxn}.
Namely, in the partially deconfined phase, the distribution is not uniform, but also not gapped, i.e. $\rho_{\rm p}(\theta)>0$ everywhere in $-\pi\le\theta\le\pi$, while in the completely deconfined phase $\rho_{\rm d}$ is gapped.
It is natural to expect \cite{Hanada:2018zxn}
\begin{eqnarray}
\rho_{\rm p}(\theta)
=
\left(
1-\frac{M}{N}
\right)
\cdot
\rho_{\rm c}(\theta)
+
\frac{M}{N}
\cdot
\rho_{\rm GWW}(\theta),
\end{eqnarray}
where $\rho_{\rm GWW}(\theta)$ is the distribution at the GWW transition point.

In many cases, the distribution takes a simple form:
\begin{eqnarray}
\rho_{\rm p}(\theta)=\frac{1}{2\pi}\left(1+A\cos\theta\right),
\label{rho_p}
\label{qe:GWW-partially-deconfined}
\end{eqnarray}
and
\begin{eqnarray}
\rho_{\rm d}(\theta)=
\left\{
\begin{array}{cc}
\frac{A}{\pi}\cos\frac{\theta}{2}\sqrt{\frac{1}{A}-\sin^2\frac{\theta}{2}}
&
\left(|\theta|<2\arcsin\sqrt{\frac{1}{A}}\right)\\
0 & \left(|\theta|\ge 2\arcsin\sqrt{\frac{1}{A}}\right)
\end{array}
\right.
\label{rho_d}
\end{eqnarray}
Here, we have fixed the U$(1)$ phase factor using $P=|P|$.
In the partially deconfined phase, $A=0$ at $T=T_1$ and $A=1$ at $T=T_2$ (the GWW transition point), and $0<A<1$ otherwise.
In the transition region, we can interpret $A=\frac{M}{N}$.
In the completely deconfined phase, $A\ge 1$, and $M=N$ regardless of the value of $A$.

For the bosonic matrix model, the form of $\rho_{\rm d}(\theta)$ in Eq.~\eqref{rho_d} has been confirmed by previous studies in a wide region of parameter space~\cite{Kawahara:2007fn,Hanada:2018zxn}.
In Ref.~\cite{Kawahara:2007fn}, the form of $\rho_{\rm p}(\theta)$ in Eq.~\eqref{rho_p} has also been observed and used as the evidence for the absence of a first order transition.
However, as we will see, this is an artifact of the finite-$N$ correction described below in Sec.~\ref{sec:finiteN}.

For our purposes it is important to note, for example from figure~\ref{fig:Pol-vs-T-possibilities}, that in the case of the first order transition with hysteresis the deconfined phase becomes unstable below $P=\frac{1}{2}$.
In other words, if the value of $P$ is stable at $0< P < \frac{1}{2}$, the transition cannot be of first order.

\subsection{Large-$D$, large-$N$ analysis}\label{sec:LargeD}

An analytic approach to understanding the thermal phase transition in Yang-Mills matrix models with action Eq.~\eqref{eq:MainAction} has been developed in Ref.~\cite{Mandal:2009vz}.
The key technical tool employed was an expansion of the functional integral around a non-trivial saddle point in the limit of a large number of matrices $d$ (where $d=D-1$).
Around the saddle point and in this limit, fluctuations are suppressed by powers of $1/d$.
A priori, it is not obvious what value of $d$ is large enough to justify this expansion, although hints may be obtained from gravity computations, as noted in Sec.~\ref{sec:introduction}.
Our numerical results suggest that $d=9$ is definitely too small, and even at $d=25$ our simulations show a qualitatively different behavior.
The analytic agreement with numerical studies~\cite{Kawahara:2007fn} mentioned in Ref.~\cite{Mandal:2009vz} might be attributed to fact that $N\leq32$ is too small to reveal the nature of the large-$N$ transition.
On the other hand, simulations at $N=32$ up to $d=15$ were interpreted as consistent with a first order transition~\cite{Azuma:2014cfa}, in disagreement with the analytical approach.

The results of Ref.~\cite{Mandal:2009vz} can be summarized as follows.
Throughout the analysis, a gauge is adopted where $A_t$ is time-independent and diagonal.
At low temperatures, the eigenvalue distribution $\rho(\theta)$ of the Polyakov loop $P$ becomes constant as $N\rightarrow \infty$.
As a consequence, the Polyakov loop vanishes.
This behavior persists up to a temperature $T_{1}$, where a second order phase transition happens.
The large $N$ eigenvalue distribution is now given by\footnote{Note that we consider $|P|$ to be normalized to scale as $N^0$.}
\begin{equation}
	\rho_{T_{1}\leq T \leq T_{2}}(\theta) =\frac{1}{2\pi} \left( 1+ 2 |P| \cos\theta\right)
\end{equation}
and $|P|$ continuously increases form $0$ to $1/2$ as $T$ is increased to a second critical temperature $T=T_{2}$.
With the identification $|P|=\frac{A}{2}=\frac{M}{2N}$, this is the same distribution as $\rho_{\rm p}$ given by Eq.~\eqref{qe:GWW-partially-deconfined}.
At $T=T_2$, the third order Gross-Witten-Wadia type~\cite{Gross:1980he, Wadia:2012fr} phase transition occurs after which $|P|$ increases further but with a smaller slope.
The eigenvalue distribution becomes gapped at $T=T_{2}$ and eventually approaches a single delta function at very high temperatures.

The predictions for the critical temperatures including the first $1/d$ corrections at large $N$ are given by~\cite{Mandal:2009vz}
\begin{equation}
T_{1}=\frac{d^{1/3}}{\log d}\left[1+\frac{1}{d}\left(\frac{203}{160}-\frac{\sqrt{5}}{3}\right)\right]^{-1},
\label{eq:large-d-1}
\end{equation}
and
\begin{equation}
\frac{1}{T_{2}}-\frac{1}{T_{1}}=\frac{\log d}{d^{4/3}}
\left[
-\frac{1}{6}+\frac{1}{d}\left(
\left(
-\frac{499073}{460800}
+
\frac{203\sqrt{5}}{480}
\right)\log d
-
\frac{1127\sqrt{5}}{1800}
+
\frac{85051}{76800}
\right)
\right].
\label{eq:large-d-2}
\end{equation}
We note that for $d \rightarrow \infty$, $\Delta T_c := T_{2}-T_{1} \rightarrow 0$, i.e. the two transitions occur in a very narrow temperature regime, making quantitative numerical checks difficult.
For the cases considered in this paper, one obtains the values in Tab.~\ref{tab:largeD}.
\begin{table}[ht]
\centering
\begin{tabular}{c|c|c}
 & $D=10$ & $D=26$   \\ \hline
$T_{1}$ & 0.895 & 0.890   \\ \hline
$T_{2}$ & 0.917 &  0.897  \\ \hline
$\Delta T_{c}$ & 0.022 & 0.007
\end{tabular}
\caption{Values of $T_1$, $T_2$ and their difference $\Delta T_{c}$ from Eqs.~\eqref{eq:large-d-1}-\eqref{eq:large-d-2} with $D=10$ and $D=26$.}\label{tab:largeD}
\end{table}

\subsection{Finite-$N$ effects}\label{sec:finiteN}

By definition, there is no spatial extent in a matrix model.
Hence, the thermodynamic limit has to be realized as the large-$N$ limit.
The finite-$N$ corrections can obscure the phase transitions, and it is important to know what kind of corrections are expected, in order to determine the nature of the phase transition numerically.

The situation is easier to understand when the large-$N$ transition is not of first order.
Because the large-$N$ limit is the thermodynamic limit, the transition becomes sharper gradually as $N$ increases.

Some caution is required when the large-$N$ transition is of first order.
The free energy is of order $N^2$ also in this case, and the fluctuations about the minima are $1/N$-suppressed.
Therefore, at sufficiently large $N$, the distributions of the observables such as $|P|$ and $E/N^2$ should have a two-peak structure since tunneling between them is suppressed as $e^{-{\rm const.}\times N^2}$.
However, when $N$ is not sufficiently large, the tunneling probability might be so large that the two peaks merge and become one single wide peak.
In this way, the first order nature of the transition gets completely hidden.
Even worse, because the confining and deconfining phases give $\rho_{\rm c}(\theta)=\frac{1}{2\pi}$ and $\rho_{\rm d}(\theta)=\frac{1}{2\pi}(1+\cos\theta)$, the mixture of two phases -- say of the confining phase with probability $1-A$ and of the deconfining phase with probability $A$ -- gives $\rho(\theta)=\frac{1}{2\pi}(1+A\cos\theta)$, which is exactly the same as $\rho_p$ of the partially deconfined phase.
Therefore, observing a distribution compatible with $\rho(\theta)=\frac{1}{2\pi}(1+A\cos\theta)$ is not enough to claim that the transition is not of first order.

\section{Numerical results}\label{sec:numerical-results}

\subsection{The order of the phase transition and the large-$D$ limit}\label{sec:NumericsOrder}

In this Section we investigate numerically the phase transition of the bosonic matrix model.
We are in particular interested in the $D=10$ case, where the large-$D$ approximation might no longer be applicable.
The smooth behavior of the order parameter observed at small $N$ turns into a signal for a transition only in the large-$N$ limit.
In this limit, two possible scenarios can be discriminated:
\begin{enumerate}
\item {\bf Two distinct transitions become visible:}
The lower one will be of second order and the higher one will be indicated by an expectation value of the Polyakov loop $\langle | P | \rangle=\frac{1}{2}$.
Due to the continuous behavior of the Polyakov loop and the small difference of the transition temperatures, the two transitions can only be distinguished at large enough $N$.
\item {\bf One first order transition appears:}
In case of the first order transition, the signal will be quite similar to the one of a second order transition.
Starting from the low temperature confined phase, there will be a broadening of the minimum of the constraint effective potential and an increase of the susceptibility.
However, before the actual second order transition (Hagedorn transition) occurs, a second minimum of the effective potential induces a first order transition.
\end{enumerate}
Because of the nature of this phenomenon, it is difficult to decide about the order of the transition based on the susceptibility, but a two peak signal in the histogram of the order parameter or a hysteresis effect allows a clear distinction of the two scenarios.
Therefore, we concentrate on the appearance of these signals at large $N$ in the following investigations.

Our lattice action includes the bosonic part and the gauge fixing part of the BFSS lattice action defined in Ref.~\cite{Berkowitz:2016jlq}, which was also used in the numerical study of Ref.~\cite{Berkowitz:2018qhn}.
We consider scans in $T$ at a fixed number of lattice points $L$ and matrix size $N$.
The action is parameterized in such a way that the lattice spacing scales as $1/L$ and quantities like the temperature are all provided in units of the 't Hooft coupling $\lambda$, which is set to unity in the numerical simulations.

\subsubsection{$D=10$ model}\label{sec:NumericsD10}

\begin{figure}[htbp]
	\centering
	\begin{subfigure}{.48\textwidth}
		\rotatebox{0}{
			\scalebox{0.4}{	\includegraphics{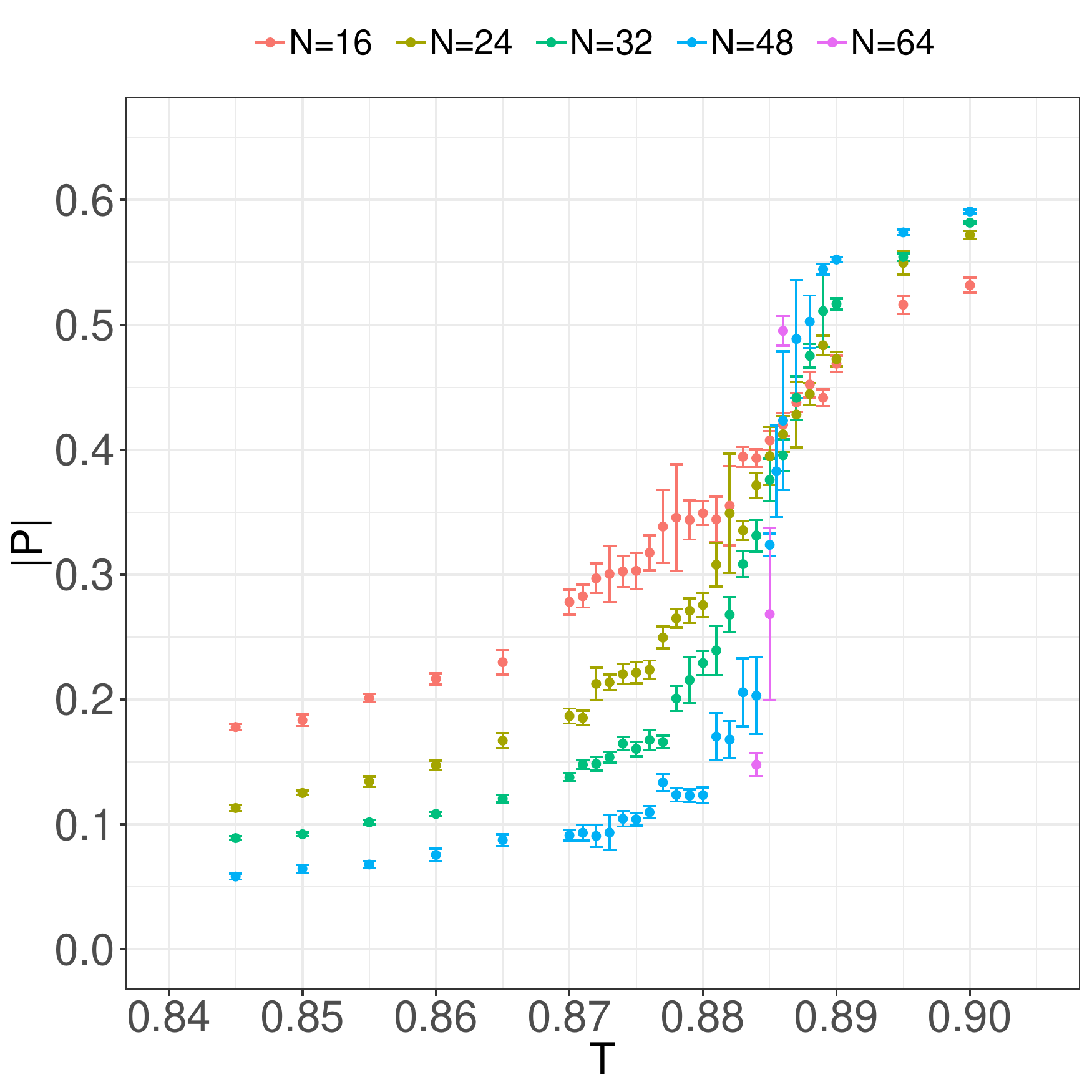}}}
		\caption{\mbox{}}
	\end{subfigure}
	\begin{subfigure}{.48\textwidth}
		\rotatebox{0}{
			\scalebox{0.4}{
				\includegraphics{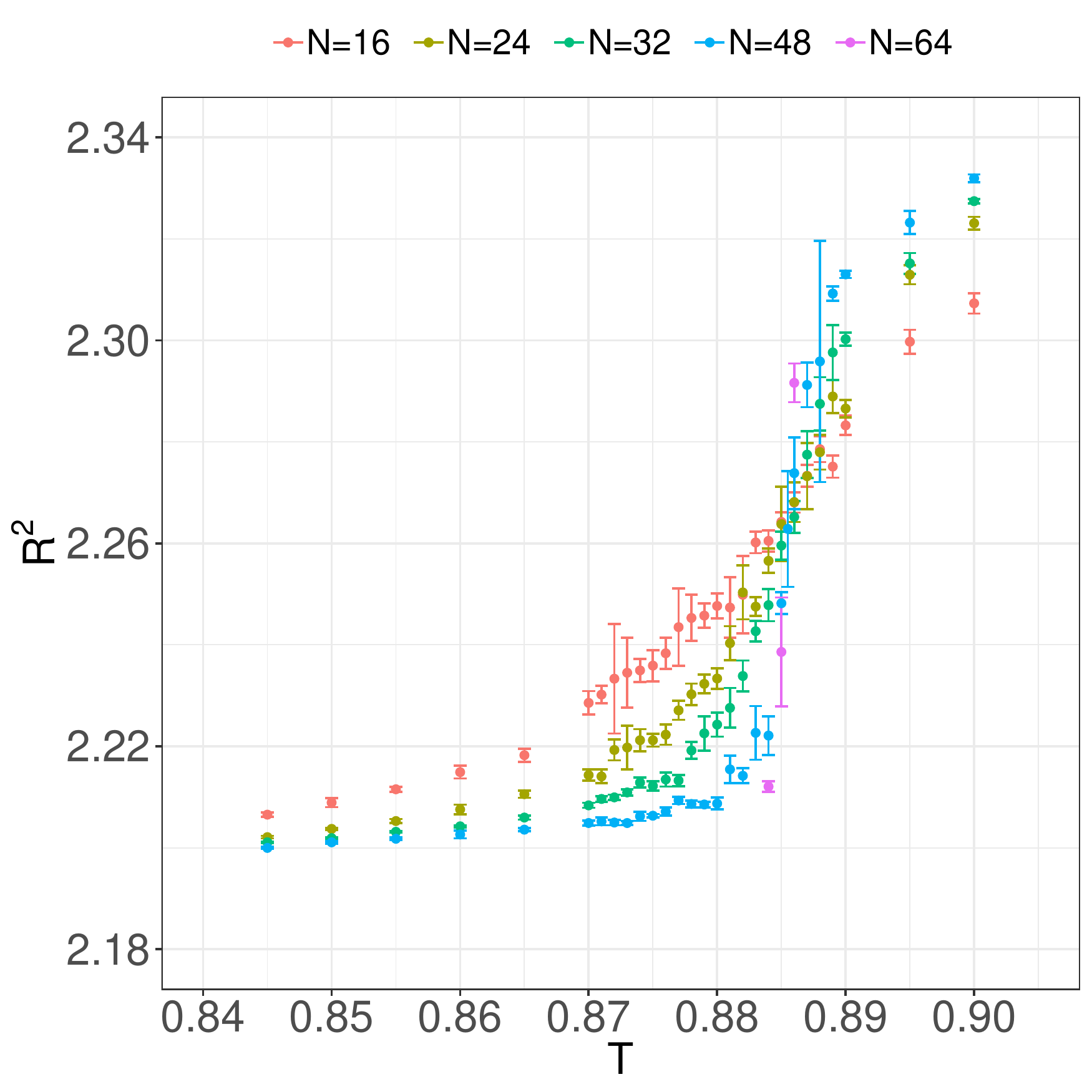}}}
		\caption{\mbox{} }
	\end{subfigure}
	\caption{a) $|P|$ vs $T$ and b) $R^2$ vs $T$, both for $L=24$, $D=10$ and various $N$.}\label{fig:P-vs-T}
\end{figure}

The first step of our numerical investigation is the measurement of the temperature dependence of the expectation value of the Polyakov loop as one approaches the large-$N$ limit.
This order parameter will show a smooth behavior at finite $N$ and signal one or two phase transitions in the large-$N$ limit.
As shown in figure~\ref{fig:P-vs-T}, there are good indications for a transition in the range $0.884<T<0.890$.
This means a rough agreement of $T_c$ with the predictions of the large-$D$ expansion, but the deviations from large-$D$ predictions are already comparable to $\Delta T_c = |T_2 - T_1|$.
This indicates that we can not completely rely on the large-$D$ prediction concerning the realization of one or the other scenario at $D=10$.

If we assume the scenario of two separate transitions, the increase of the Polyakov loop would have a finite width $\Delta T_c$, with a rise starting at $T_1$ and stopping at $T_2$.
The slope of $|P|$ at $T_1$ would increase with $N$, but the point with $|P|=\frac{1}{2}$ at $T_2$ would remain at a finite distance $\Delta T_c$ from $T_1$.
Based on this assumptions, we can deduce a rough estimate of $\Delta T_c$ and its large-$N$ extrapolation from the width of the transition.
At $N=32$, which has been the maximal $N$ in previous numerical studies, the obtained width is still compatible with the large-$D$ prediction $\Delta T_c=0.022$.
However, the extrapolation towards the large-$N$ limit does not support a finite $\Delta T_c$ required by the scenario of two separate transitions.
In order to substantiate these findings, a more detailed analysis is necessary.

As pointed out above, the best way to discriminate the two scenarios is the two-state signal or a hysteresis of the order parameter.
The two-state signal can be deduced from a two-peak structure in the histograms of the order parameter that persists in the large-$N$ limit.
In addition, we consider possible effects of the finite lattice spacing by comparing histograms from simulations with a different number of lattice points $L$.
In case of a first order transition, the separation of the two peaks becomes more pronounced at large $N$ since the tunneling rate between the two states is exponentially suppressed with $N^2$ in this limit.
The tunneling is also visible in Monte-Carlo time, but this effect is not unambiguous since it has an algorithm dependence.

\begin{figure}
	\centering
	\begin{subfigure}[b]{0.35\textwidth}
		\includegraphics[width=\textwidth]{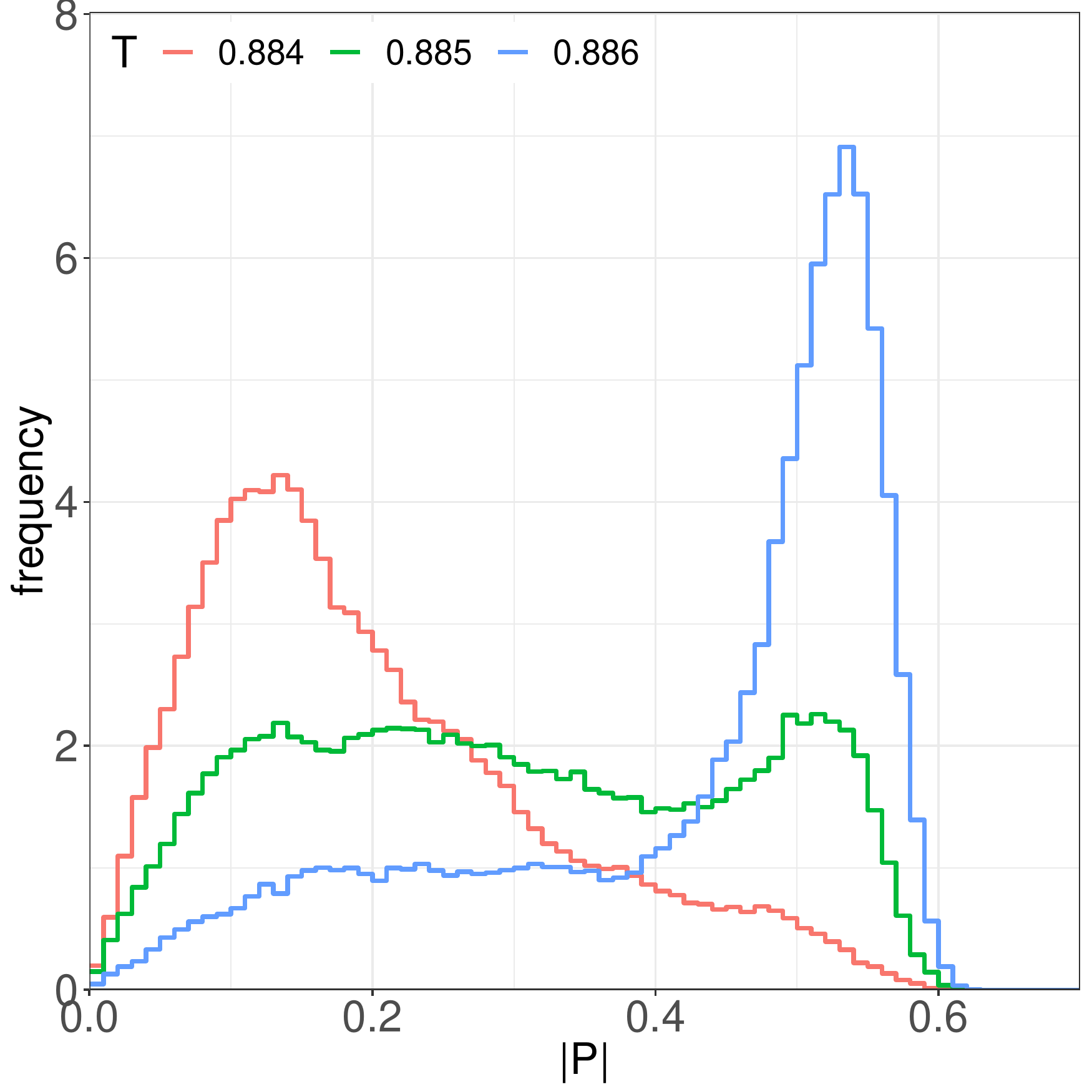}
		\caption{$N=48$, $L=24$}
	\end{subfigure}
	\begin{subfigure}[b]{0.35\textwidth}
		\includegraphics[width=\textwidth]{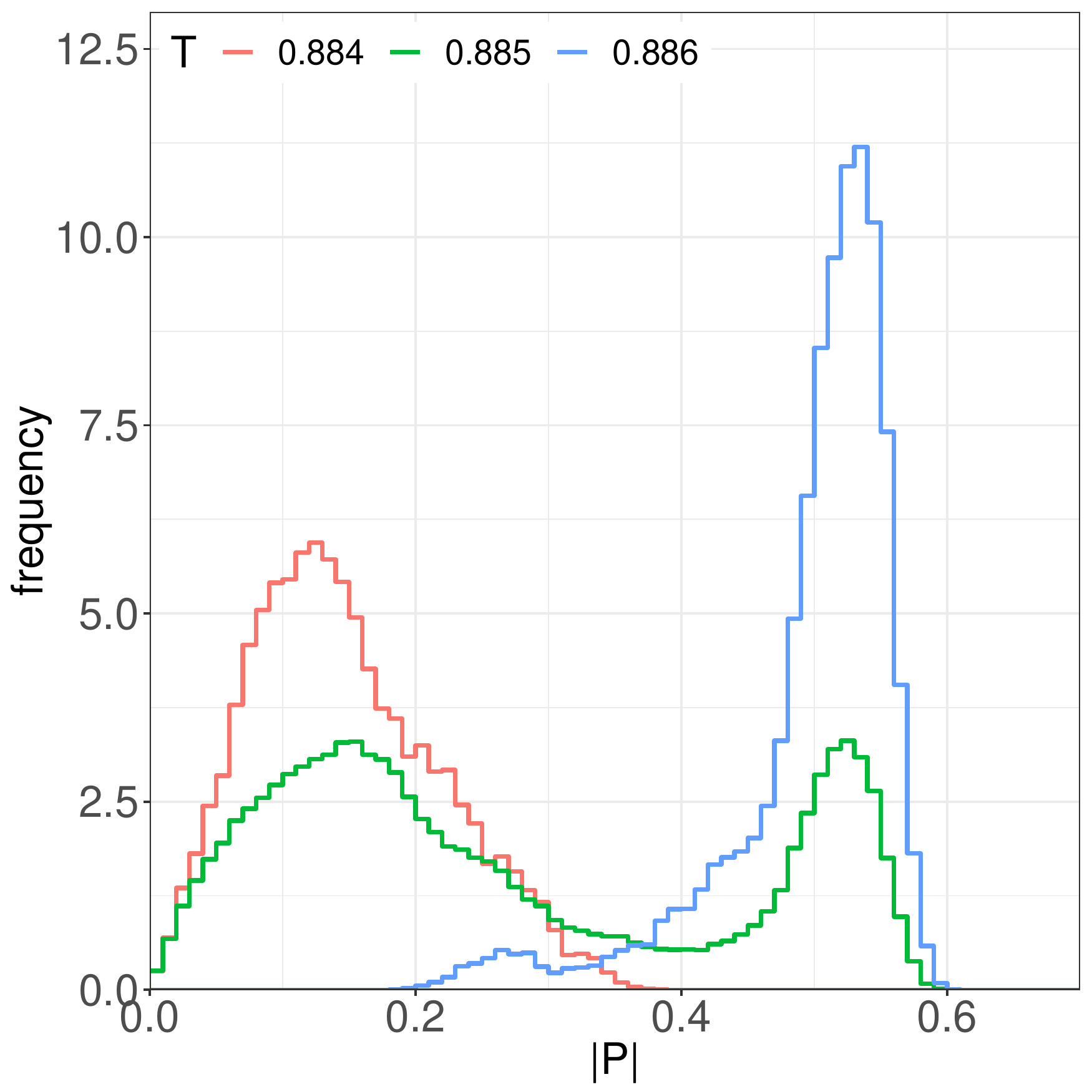}
		\caption{$N=64$, $L=24$}
	\end{subfigure}

	\begin{subfigure}[b]{0.35\textwidth}
		\includegraphics[width=\textwidth]{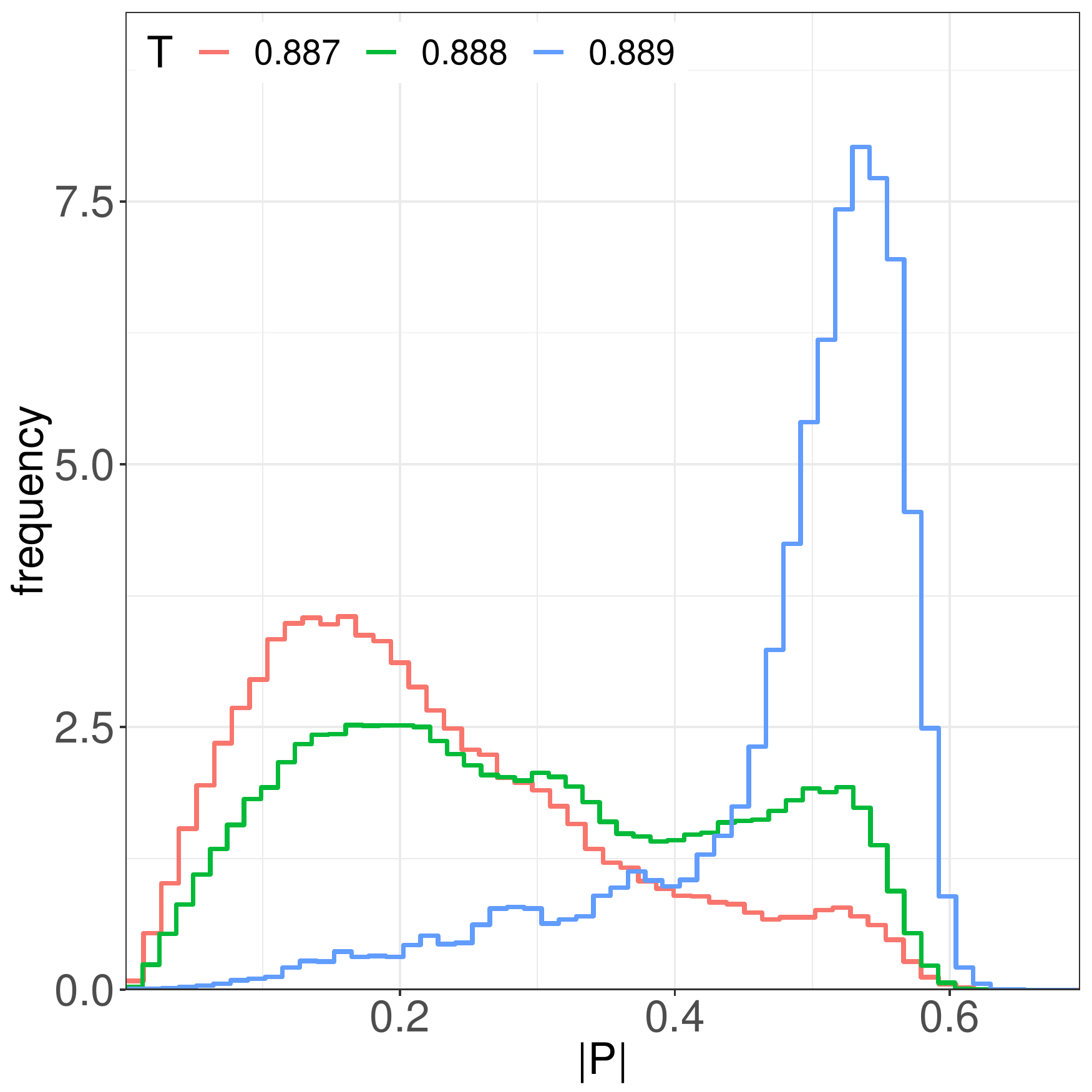}
		\caption{$N=48$, $L=32$}
	\end{subfigure}
	\begin{subfigure}[b]{0.35\textwidth}
		\includegraphics[width=\textwidth]{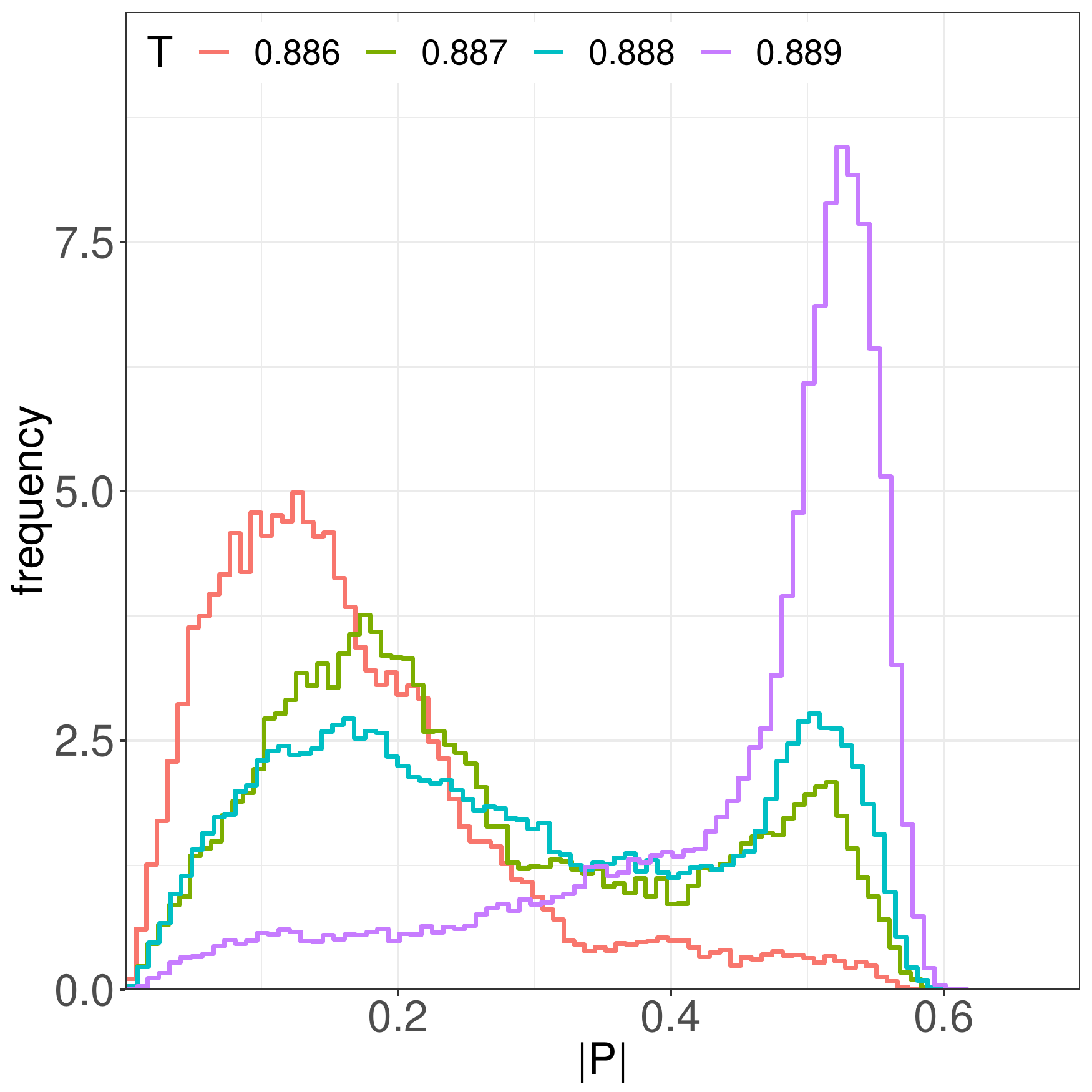}
		\caption{$N=64$, $L=32$}
	\end{subfigure}

	\begin{subfigure}[b]{0.35\textwidth}
		\includegraphics[width=\textwidth]{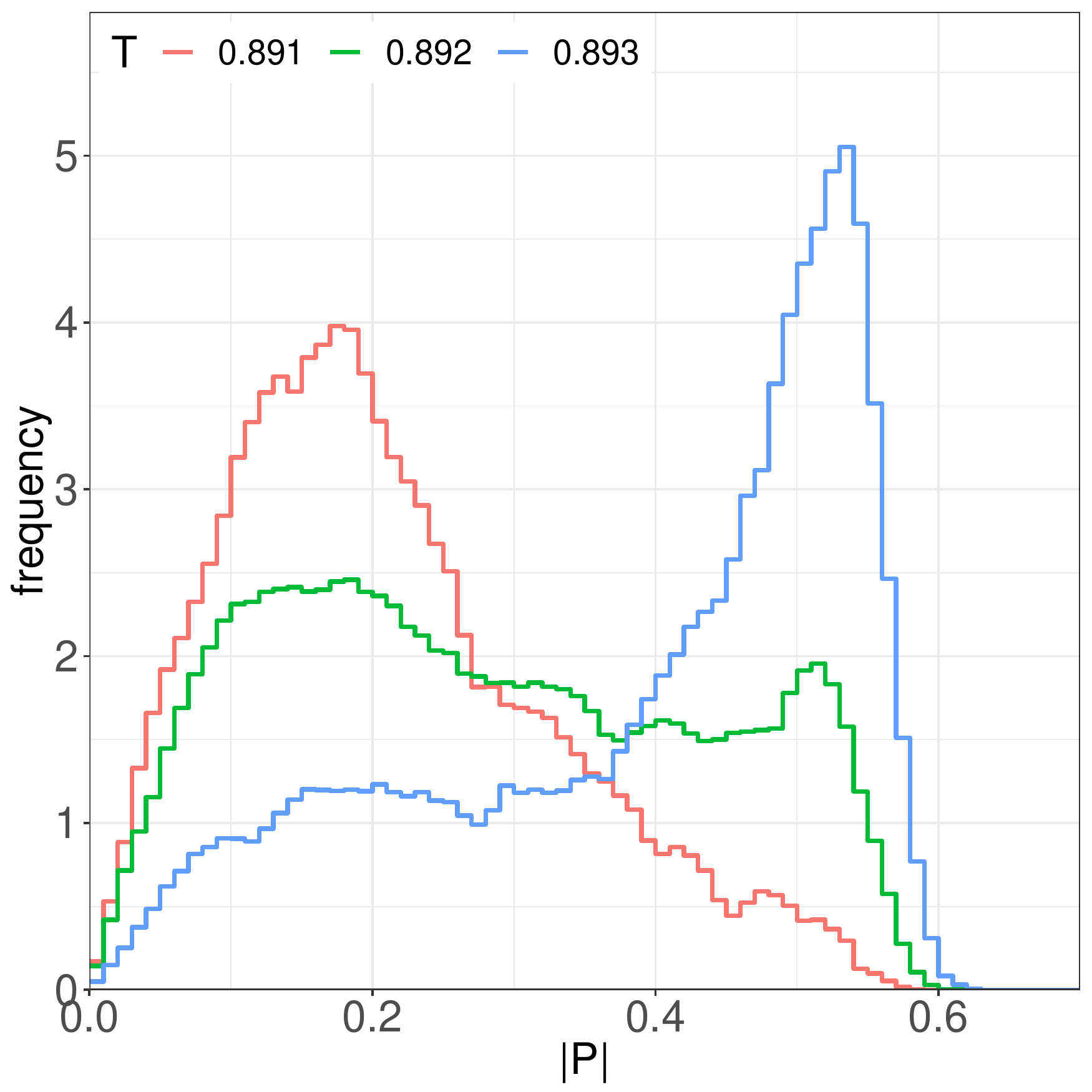}
		\caption{$N=48$, $L=48$}
	\end{subfigure}
	\begin{subfigure}[b]{0.35\textwidth}
		\includegraphics[width=\textwidth]{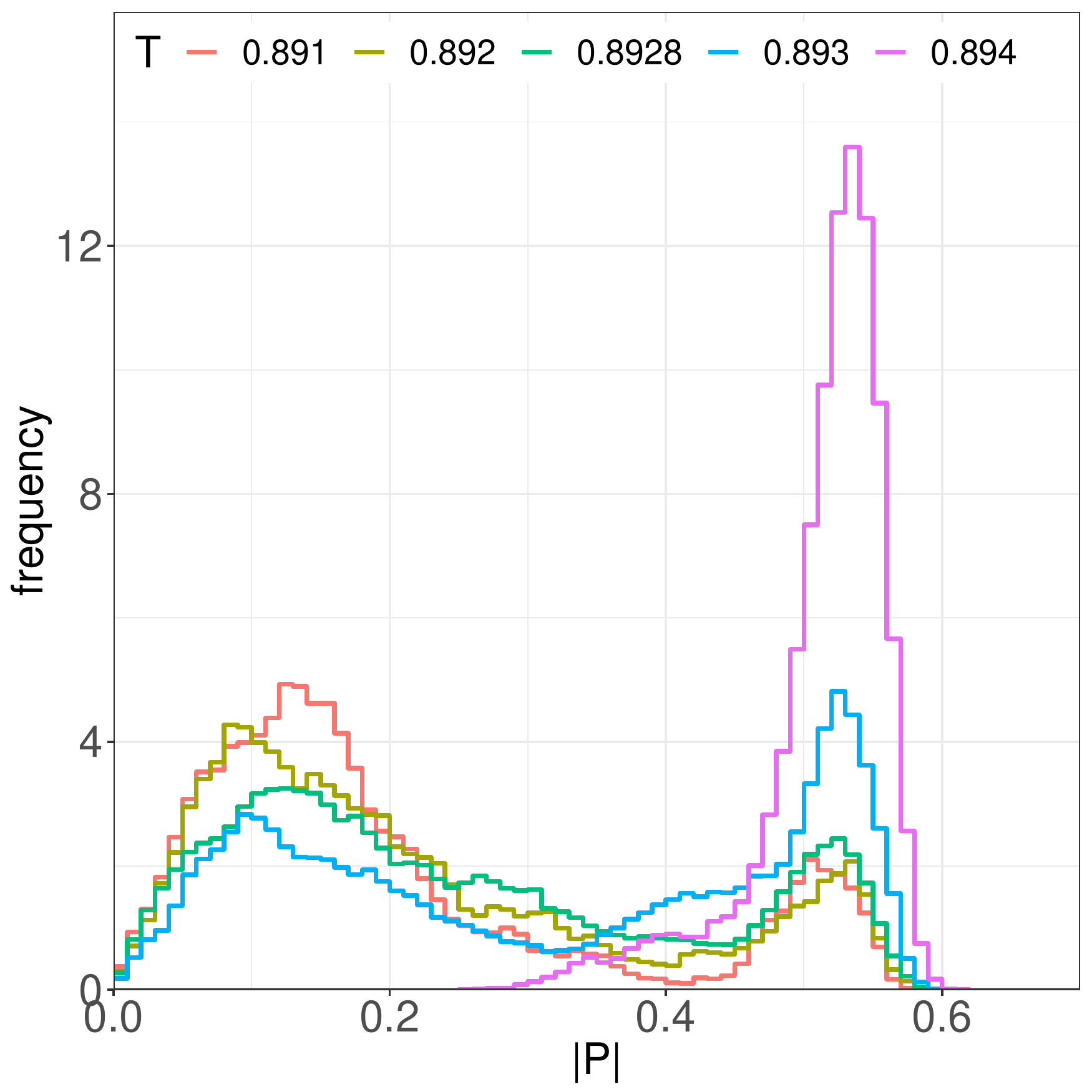}
		\caption{$N=64$, $L=48$}
	\end{subfigure}
	\caption{Histogram of the order parameter $|P|$ close to the transition temperature for the $D=10$ theory.}\label{fig:TwoState}
\end{figure}

\begin{figure}[htbp]
	\centering
		\rotatebox{0}{
			\scalebox{0.85}{
				\includegraphics{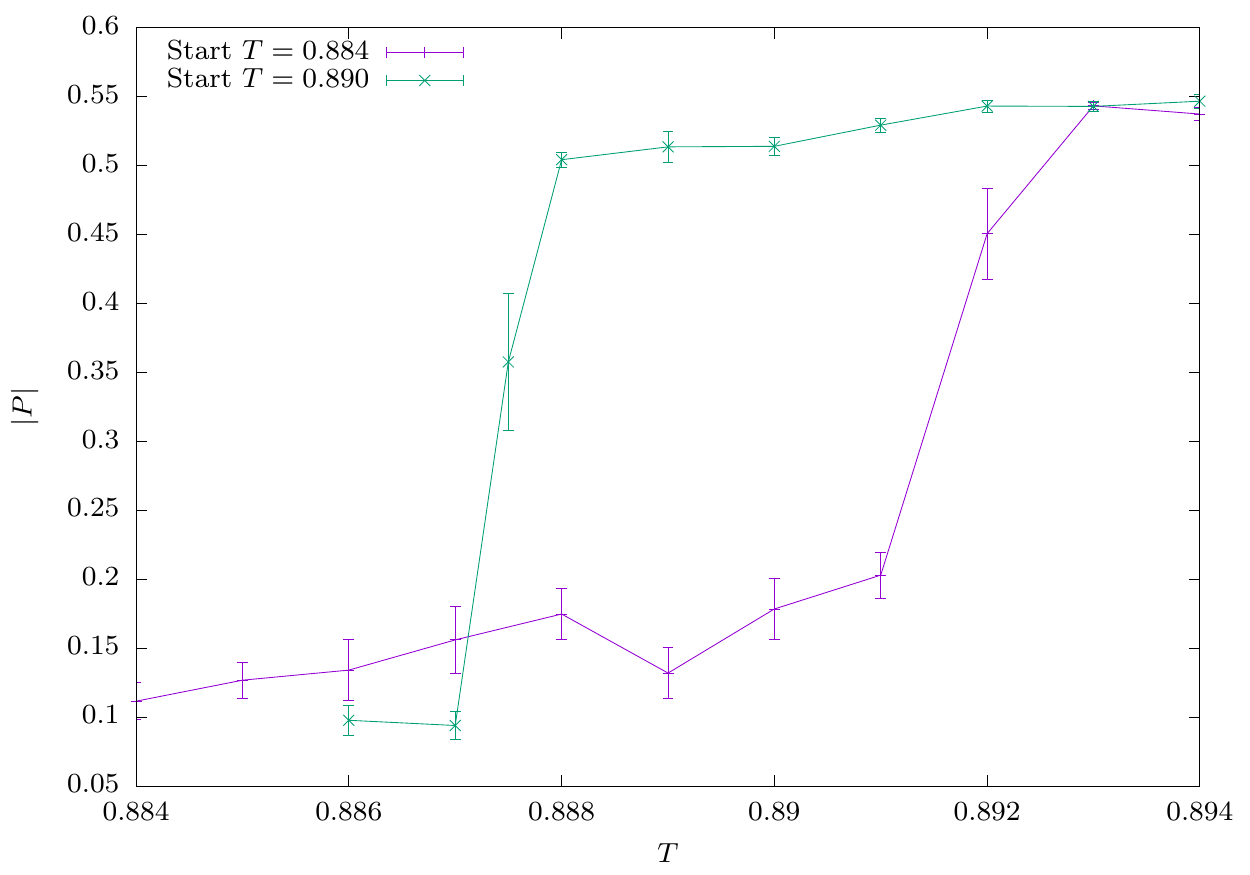}}}
	\caption{Hysteresis cycle for the order parameter $|P|$ with $N=64$, $L=32$ in the $D=10$ theory.}\label{fig:N64d9BinningGeorg}
\end{figure}

For $N\leq 32$ we can not observe a clear two-peak signal in the histogram (although hints of a two-peak structure are visible, see the appendix), but at $N=48$ a two-state signal can be observed that becomes more pronounced at $N=64$, see figure~\ref{fig:TwoState}.
There is evidence for the existence of two phases, one with small $|P|$ and the other with $|P|\approx 0.5$.
Consequently, a hysteresis of the order parameter is found, see figure~\ref{fig:N64d9BinningGeorg}.
We have investigated three different lattice spacings at $N=64$ in order to show that the effect persists in the continuum limit.
The complete Monte Carlo history for $N=64$, $L=24$ is shown in figure~\ref{fig:HistoryD10} in the appendix for the transition temperature $T=0.885$.
It displays repeated tunneling events between the two phases.

\begin{figure}[htbp]
	\centering
		\rotatebox{0}{
			\scalebox{0.4}{
				\includegraphics{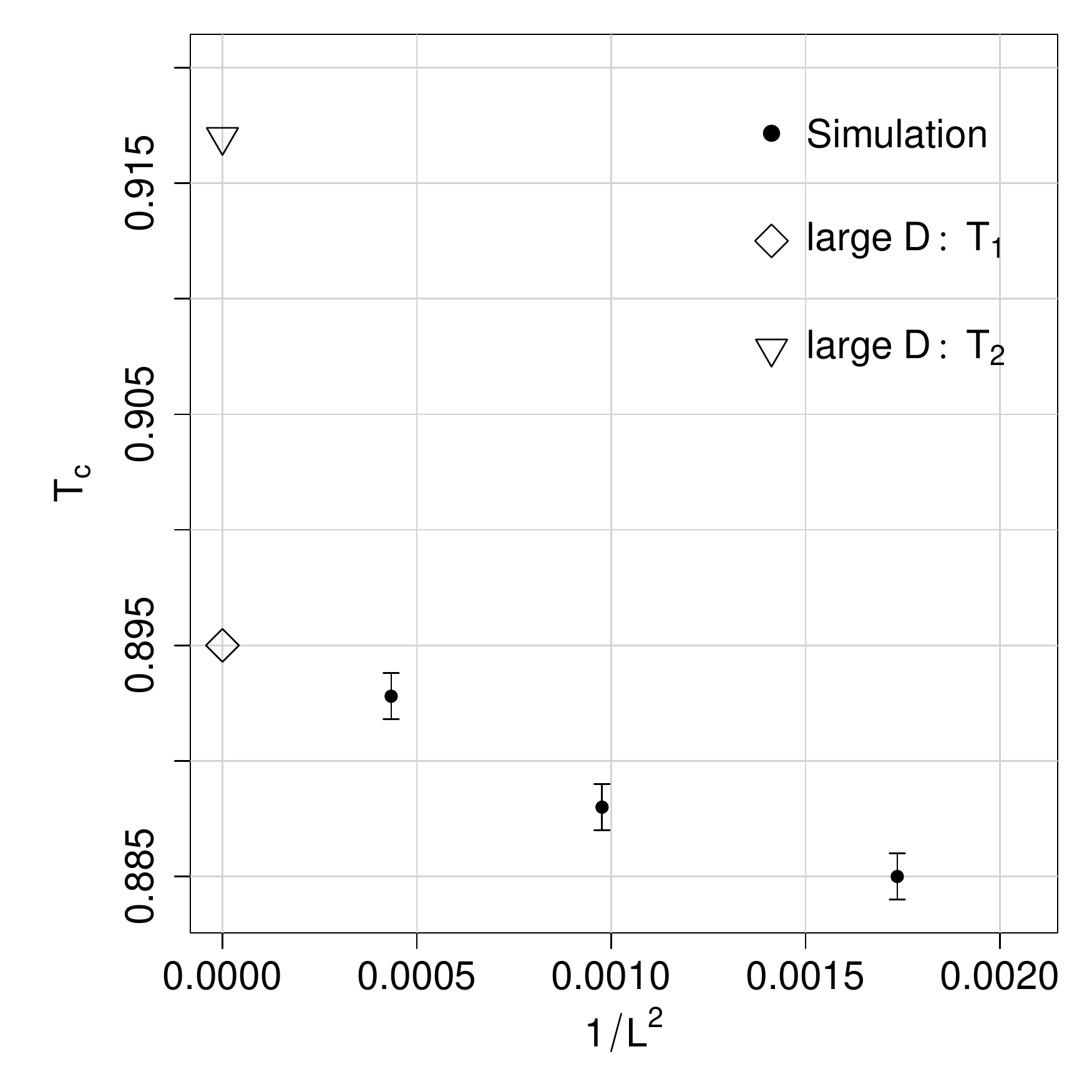}}}
	\caption{Values for $T_c$ extracted from the two-peak histograms extrapolated to the large-$N$ limit for various values of $L$.
	On the left, we show the continuum results from the next-to-leading large-$D$ expansion at $D=10$.}\label{fig:tc_continuum}
\end{figure}

Overall, we conclude from these data that there is a first order transition at $D=10$ and thus a considerable qualitative deviation from the large-$D$ expansion.
However, as figure~\ref{fig:tc_continuum} shows, a continuum extrapolation of $T_c$ (which is largely insensitive to $N\geq 48$) would lie within $T_1$ and $T_2$ obtained from the next to leading order expansion in large $D$.
This indicates that at least the location of the transition is correctly estimated by the analytic formulae in Sec.~\ref{sec:LargeD}.

Consistent with earlier numerical studies~\cite{Kawahara:2007fn}, we have seen that the signal at smaller $N$ might indicate a different scenario.
We have also found that the peak of the low temperature phase at small $|P|$ gets significantly broader around the transition temperature.
This is consistent with the expected Hagedorn instability at $T=T_1$.
Due to this phenomenon, we observe an increase of the susceptibility towards a peak when approaching the critical temperature from below.
Consequently, the first order transition occurs just before a second order transition manifests itself, as anticipated in Sec.~\ref{sec:partial_deconfinement}.
One would expect that the transition changes from first to second order at larger $D$ and the picture becomes more consistent with the large-$D$ analysis.

\subsubsection{$D=26$ model}\label{sec:NumericsD26}

The analysis of the previous Section revealed that the large-$D$ expansion fails to describe the order of the phase transition for $D=10$.
In this Section, we repeat the above investigation for $D=26$ which might be large enough for the expansion to be qualitatively applicable, but also small enough to sufficiently limit the computational costs.
It turns out that the simulation results are consistent with a first order transition.

\begin{figure}[htbp]
\begin{center}
\begin{subfigure}{.48\textwidth}
		\rotatebox{0}{
			\scalebox{0.4}{	\includegraphics{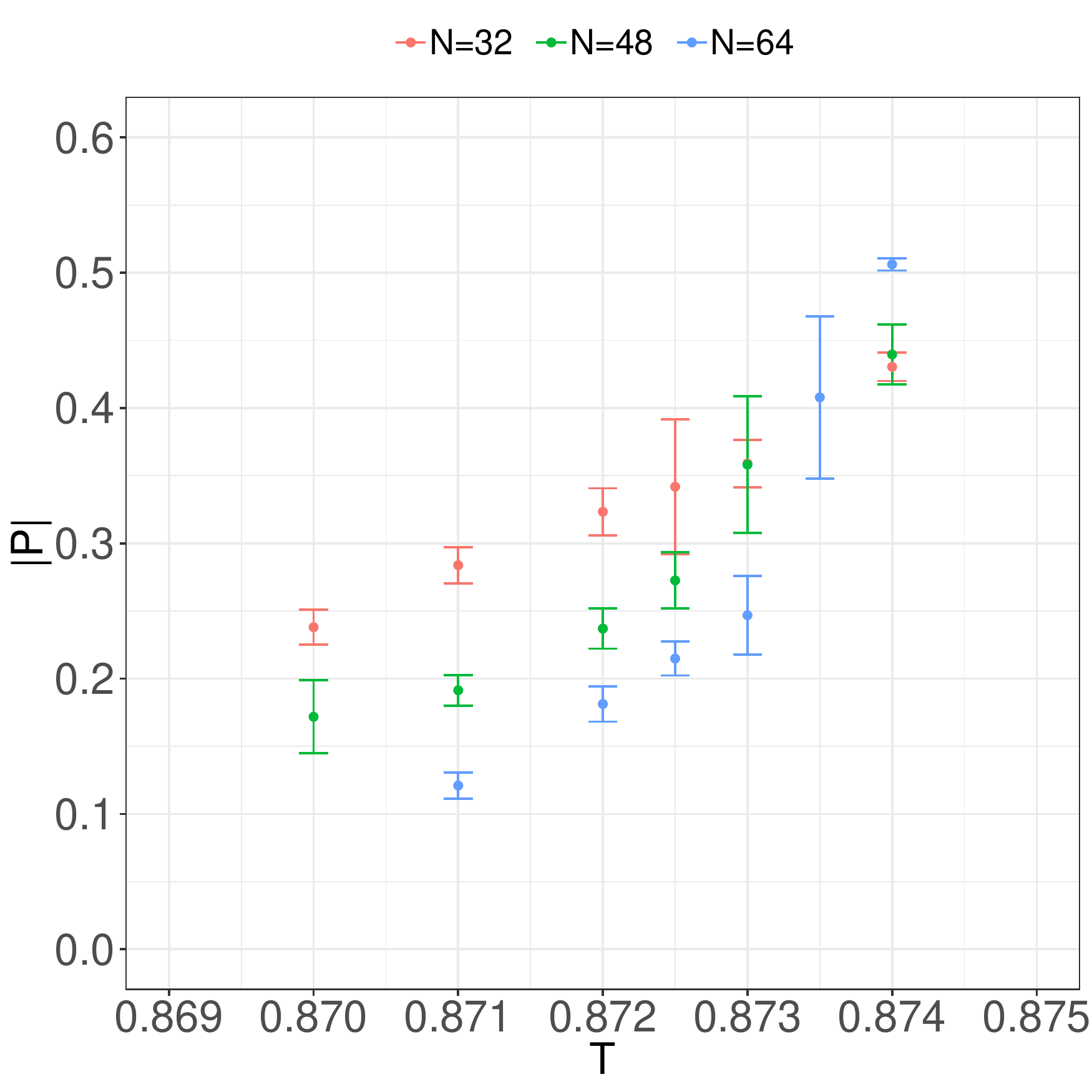}}}
		\caption{\mbox{}}
	\end{subfigure}
	\begin{subfigure}{.48\textwidth}
		\rotatebox{0}{
			\scalebox{0.4}{
				\includegraphics{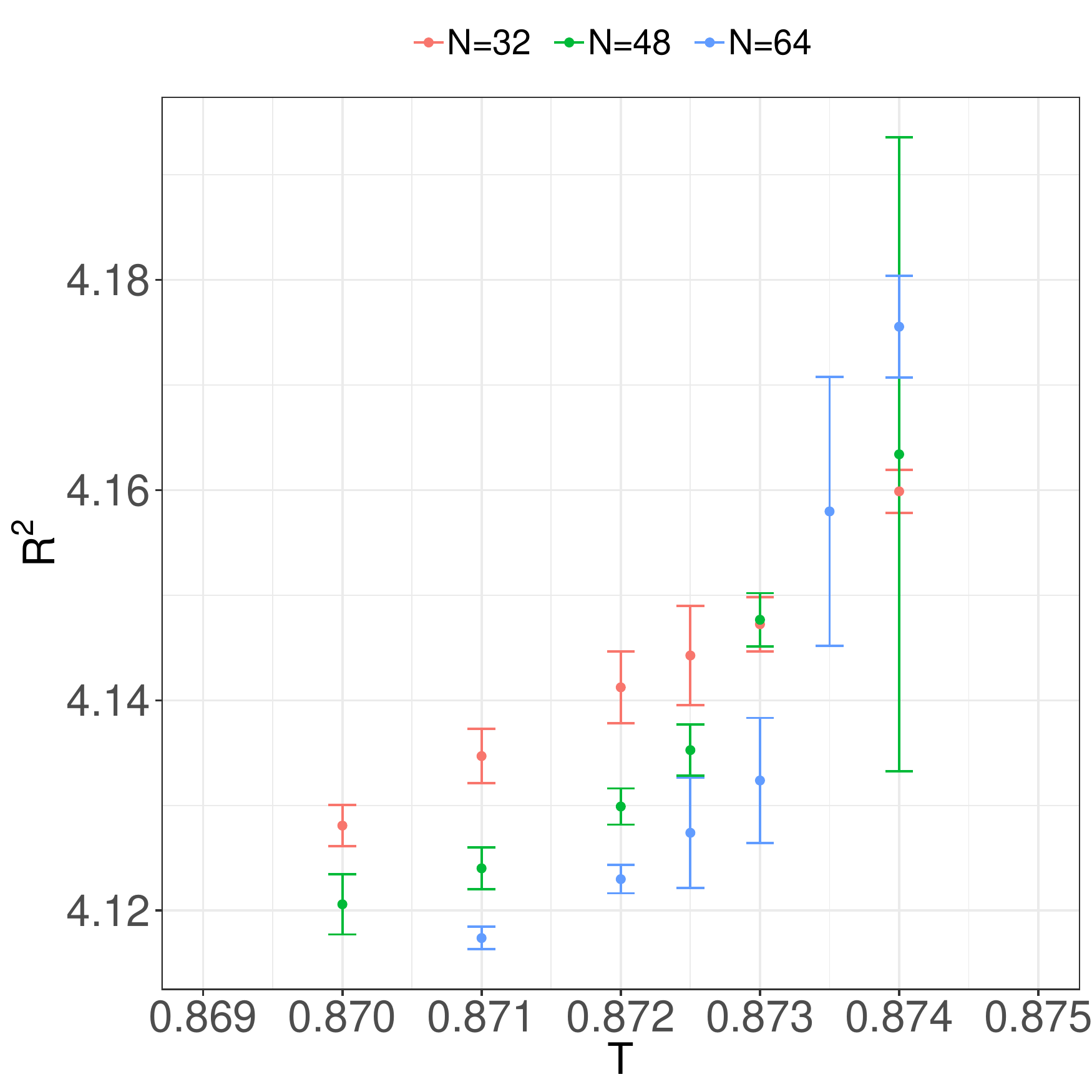}}}
		\caption{\mbox{} }
	\end{subfigure}
	\caption{a) $|P|$ vs $T$ and b) $R^2$ vs $T$, both for $L=24$, $D=26$ and various $N$. A naive large-$N$ extrapolation of $|P|$ is consistent with zero within errors for $T\leq 0.873$ (using data for $N=48$ and $N=64$ only near the transition).}\label{fig:D=26-pol}
\end{center}
\end{figure}

\begin{figure}[htbp]
	\centering
		\rotatebox{0}{
			\scalebox{0.4}{
				\includegraphics{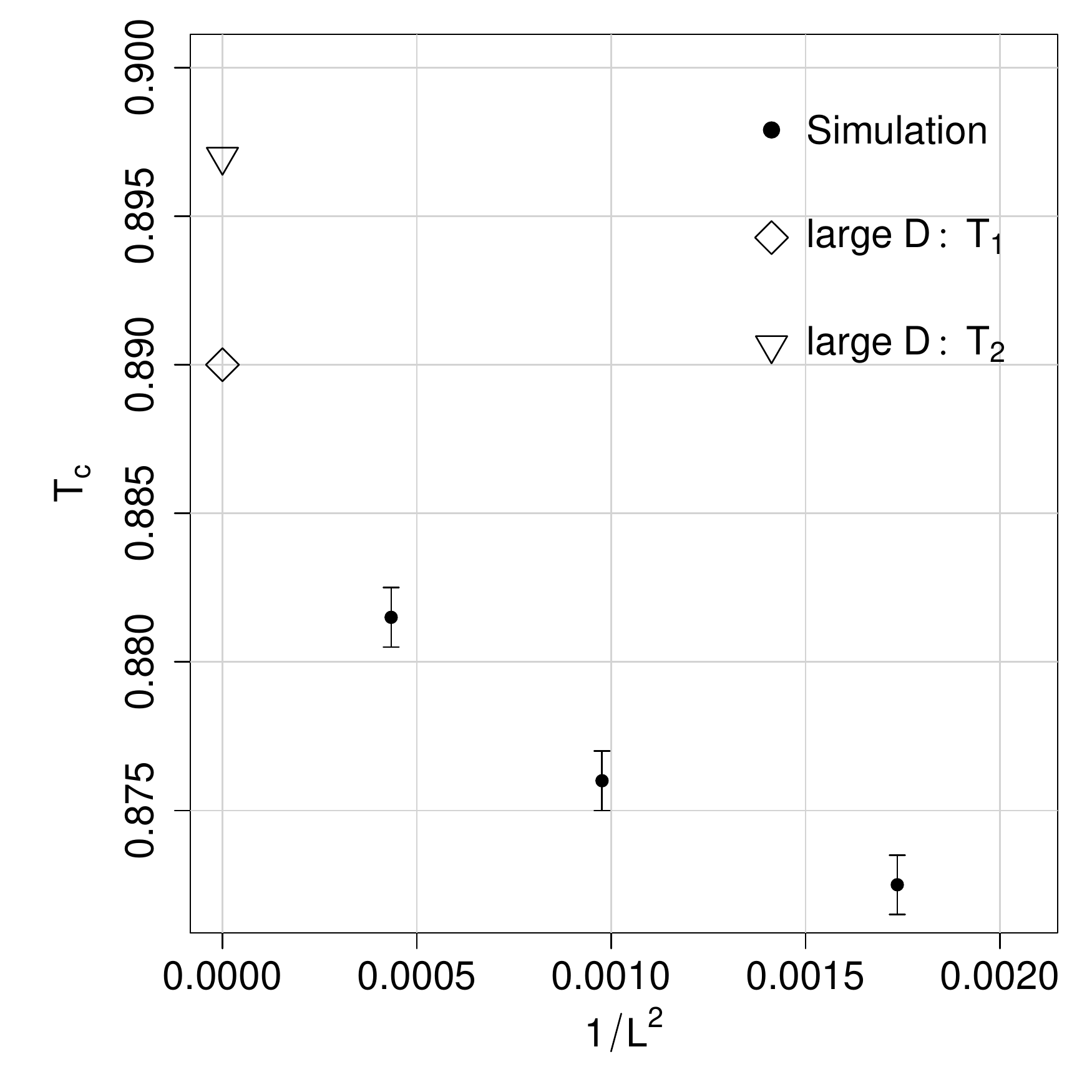}}}
	\caption{Values for $T_c$ extracted from the two-peak histograms extrapolated to the large-$N$ limit for various values of $L$.
	On the left, we show the continuum results from the next-to-leading large-$D$ expansion at $D=26$.}\label{fig:tc_continuum_D26}
\end{figure}

Figure~\ref{fig:D=26-pol} shows the dependence of the order parameter $|P|(T)$ near the transition.
A naive large-$N$ extrapolation with fixed lattice size $L=24$ locates a possible transition window to be between $T=0.873$ and $T=0.874$.
This separation is much smaller than the predicted width $\Delta T_c = 0.007$ predicted by the large-$D$ expansion (see Tab.~\ref{tab:largeD}).
There is no indication of a finite slope for large values of $N$.
Extrapolating to the continuum using $L=32$ and $L=48$, indicates that the transition is located close to $T_c=0.89$ in the continuum limit, see figure \ref{fig:tc_continuum_D26}.

\begin{figure}[htbp]
\centering
\begin{subfigure}{.43\textwidth}
\rotatebox{0}{
\scalebox{0.4}{
\includegraphics{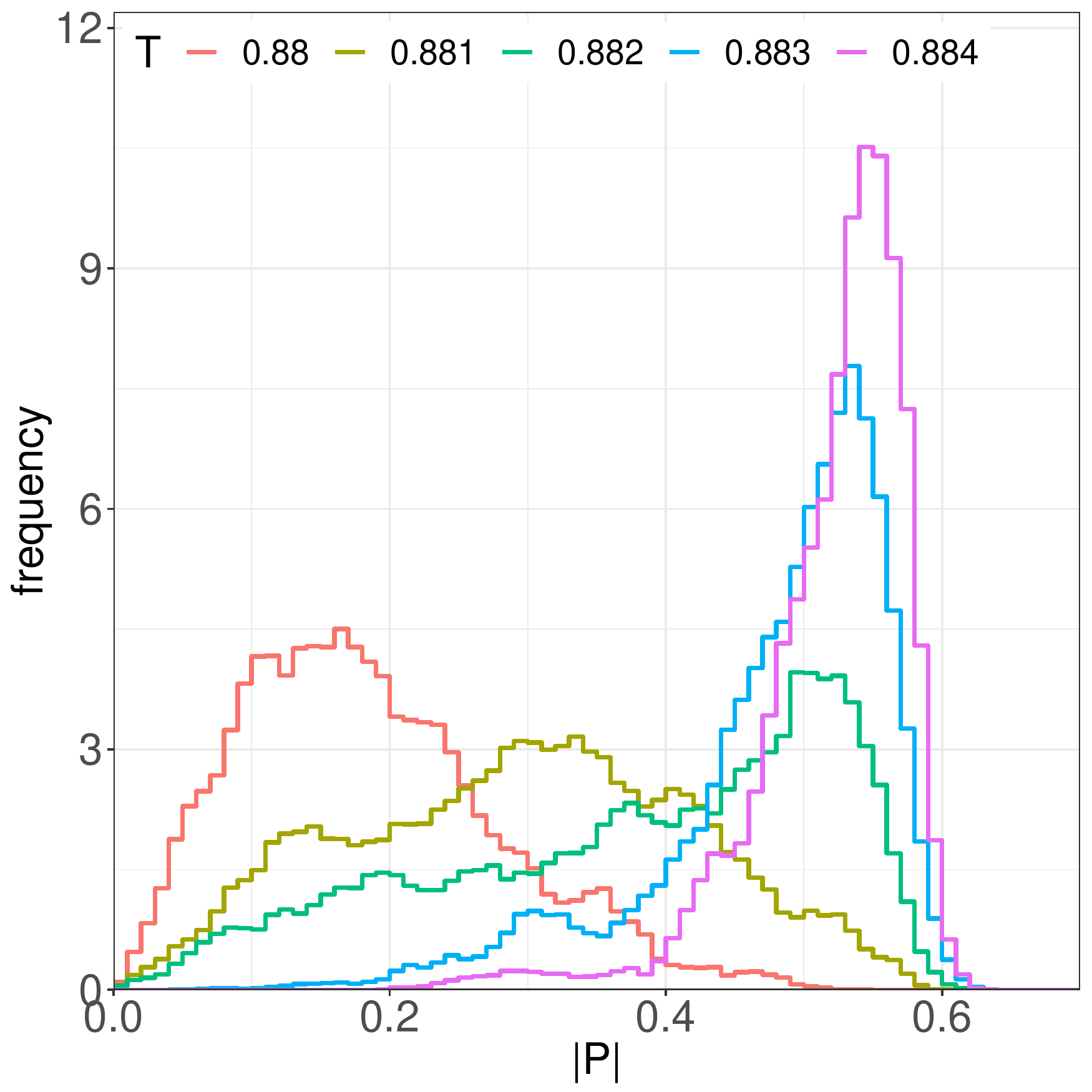}}}
\caption{\mbox{}}
\end{subfigure}
\begin{subfigure}{.43\textwidth}
\rotatebox{0}{
\scalebox{0.4}{
\includegraphics{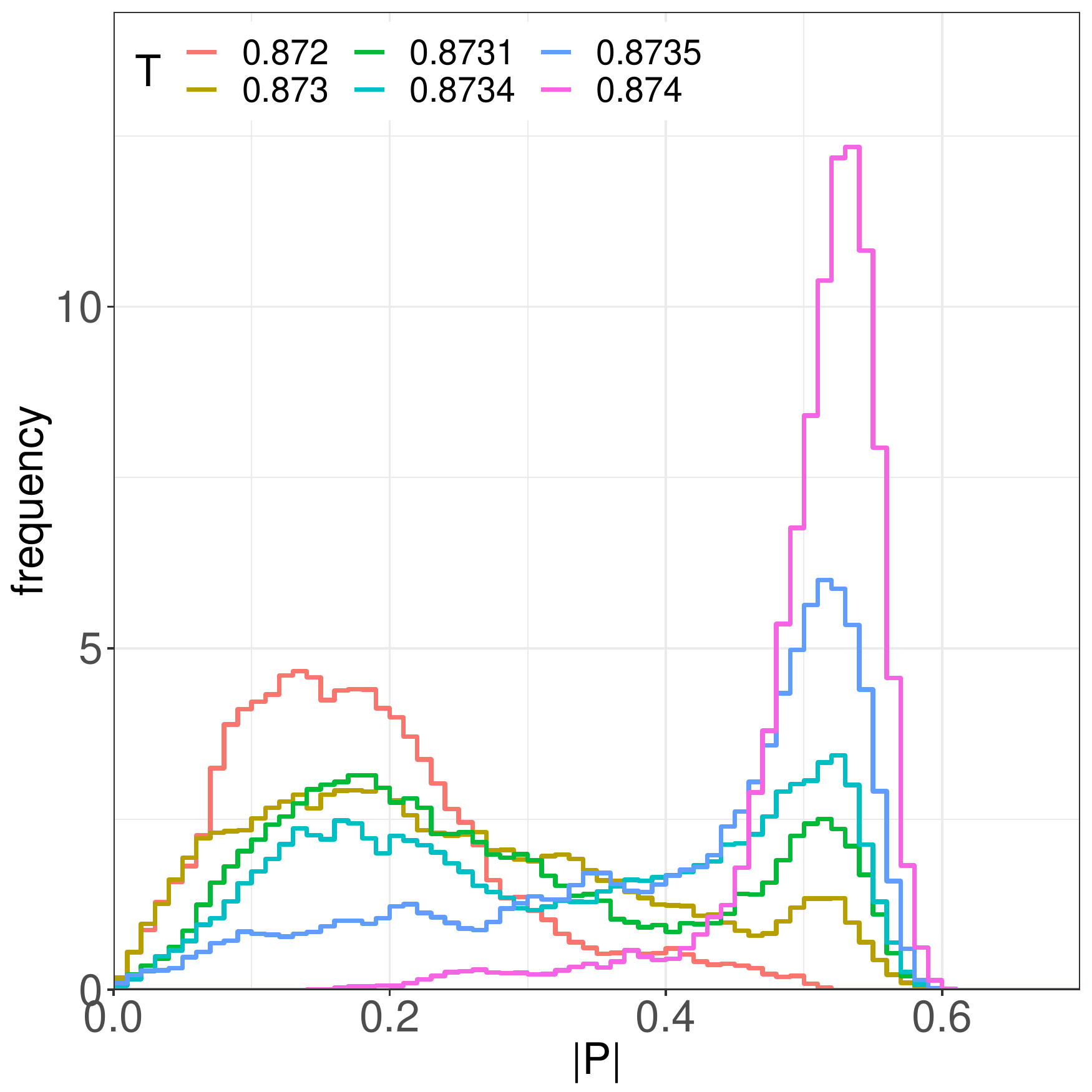}}}
\caption{\mbox{}}
\end{subfigure}

\begin{subfigure}{.43\textwidth}
\rotatebox{0}{
\scalebox{0.4}{
\includegraphics{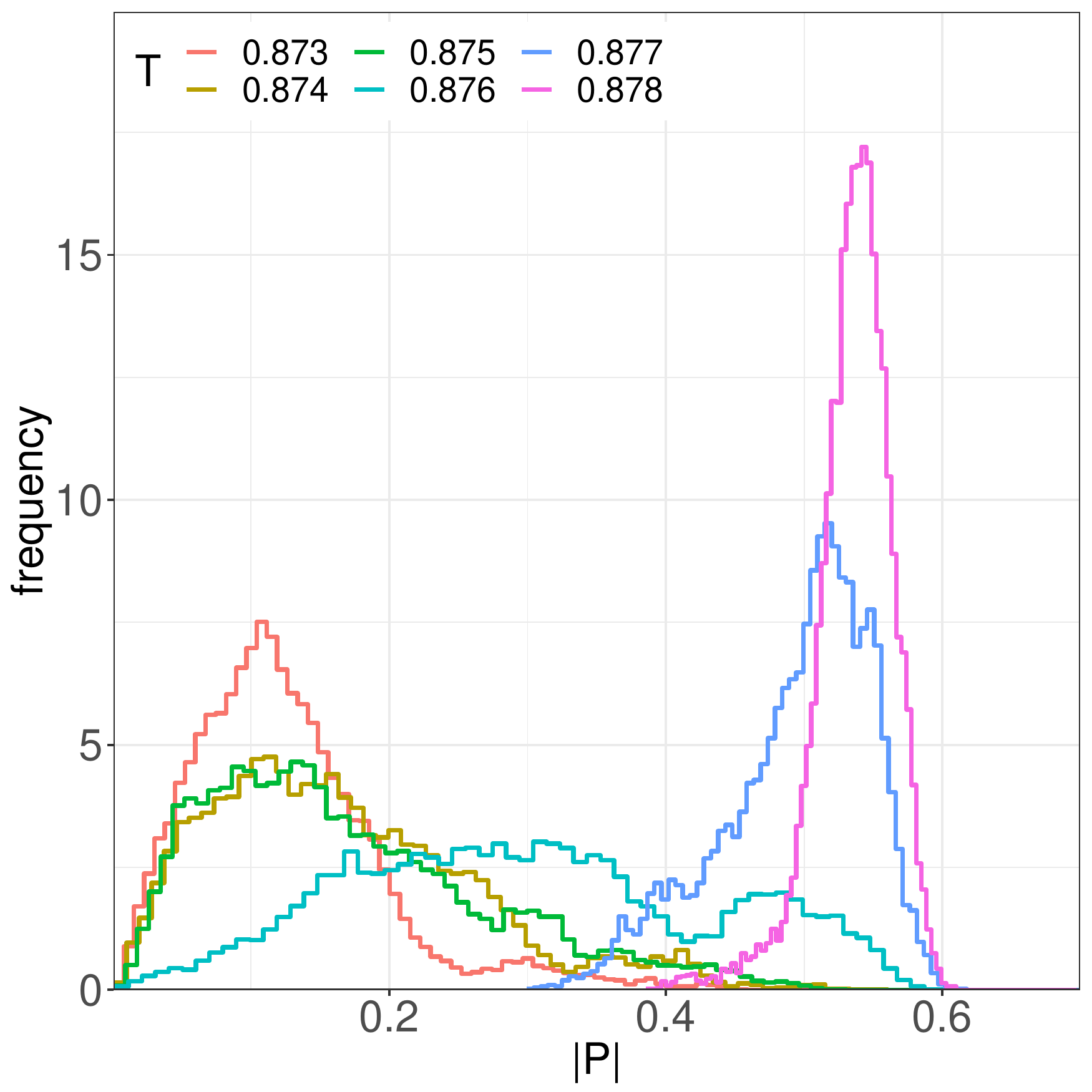}}}
\caption{\mbox{}}
\end{subfigure}
\begin{subfigure}{.43\textwidth}
\rotatebox{0}{
\scalebox{0.4}{
\includegraphics{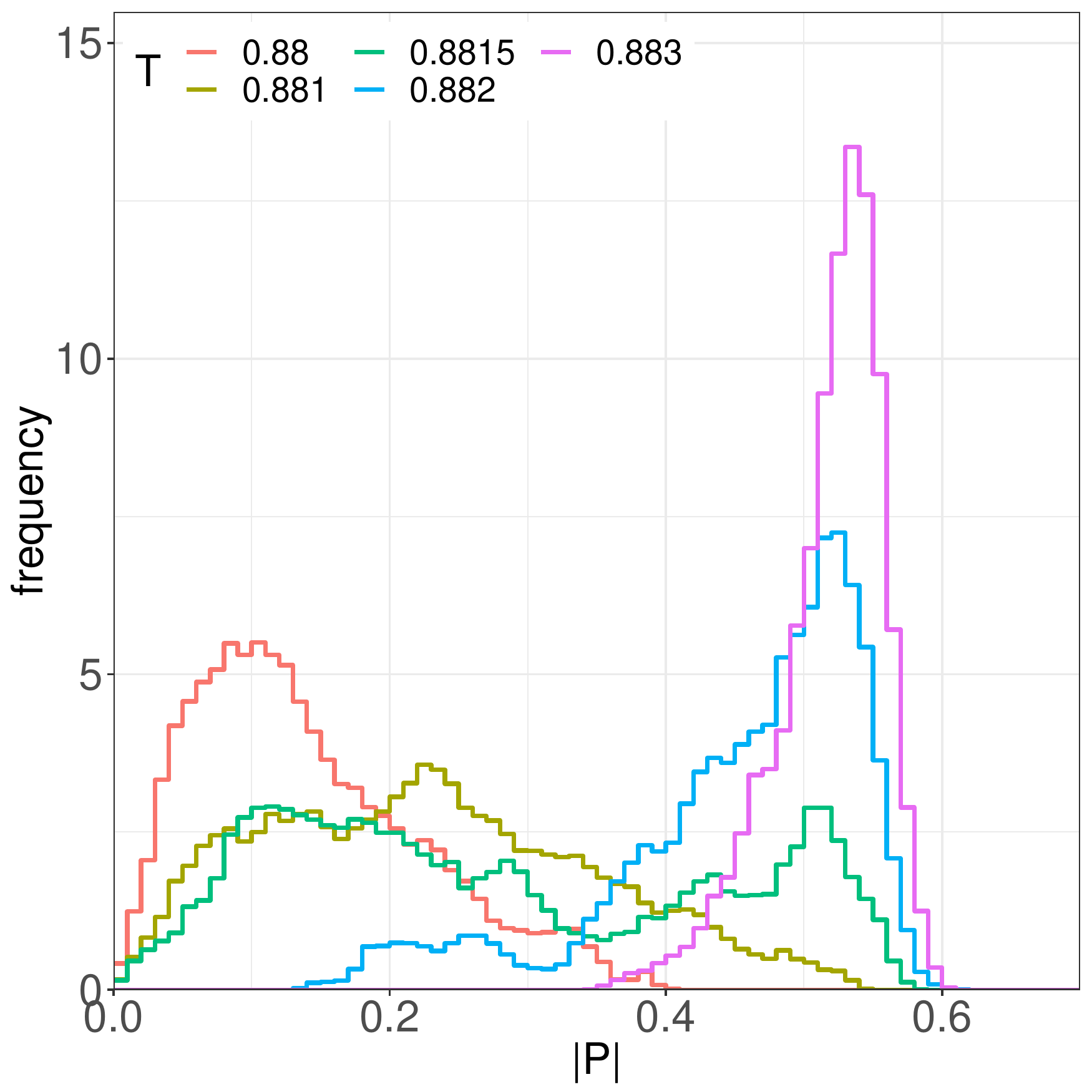}}}
\caption{\mbox{}}
\end{subfigure}
\caption{Histograms of $|P|$ for $D=26$: a) $N=48$, $L=48$, b-d) $N=64$, $L=24,32,48$.}\label{fig:N48d25Binning-Pol}
\end{figure}

Figure~\ref{fig:N48d25Binning-Pol} summarizes the histograms of $|P|$ for $N=64$, $L=24,32,48$. Compared to $D=10$, the peaks of the distributions are broader, but a two-state signal is still clearly visible. This is also supported by the Monte-Carlo history, figure~\ref{fig:HistoryD26} in the appendix.

\subsection{First order signals from other observables}\label{sec:NumericsTwoPeaks}

It is useful to study other quantities in order to provide further evidence for the existence or absence of a first order transition.
Let us consider $\frac{E}{N^2}=-\frac{3}{4N}{\rm Tr}[X_I,X_J]^2$ and $R^2\equiv\frac{1}{N}\sum_{I=1}^d{\rm Tr}X_I^2$.
If the transition is of first order, the two-peak signal should be visible for these observables as well.

\begin{figure}[htbp]
	\centering
	\begin{subfigure}{.45\textwidth}
		\rotatebox{0}{
			\scalebox{0.4}{
				\includegraphics{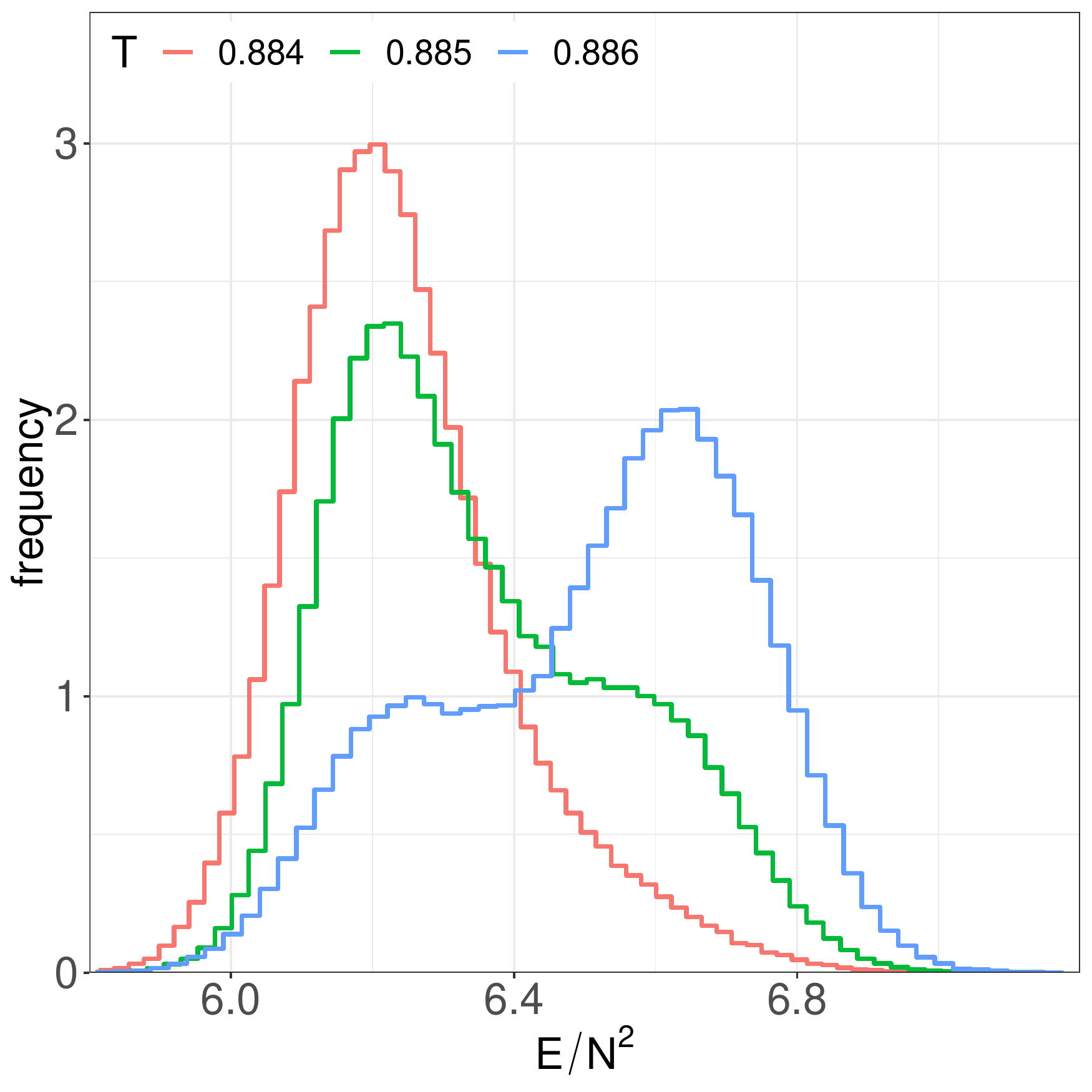}}}
		\caption{\mbox{} }
	\end{subfigure}
	\begin{subfigure}{.45\textwidth}
		\rotatebox{0}{
			\scalebox{0.4}{
				\includegraphics{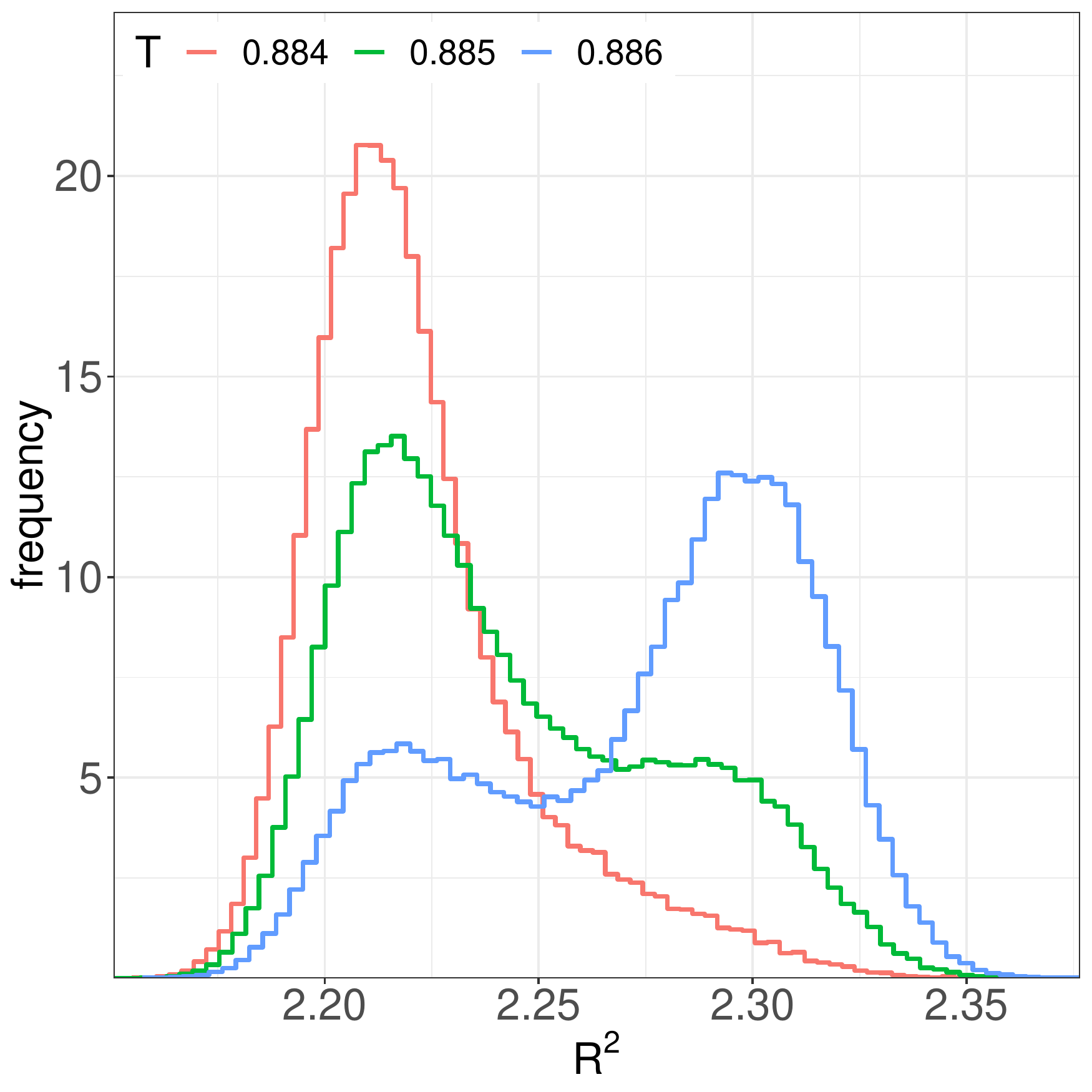}}}
		\caption{\mbox{} }
	\end{subfigure}
	\caption{$N=48$, $L=24$, $D=10$: a) binned $E$ for various $T$. b) binned $R^2$ for various $T$.  }\label{fig:N48S24d9Binning-E-and-R}
\end{figure}

\begin{figure}[htbp]
	\centering
	\begin{subfigure}{.45\textwidth}
		\rotatebox{0}{
			\scalebox{0.4}{
				\includegraphics{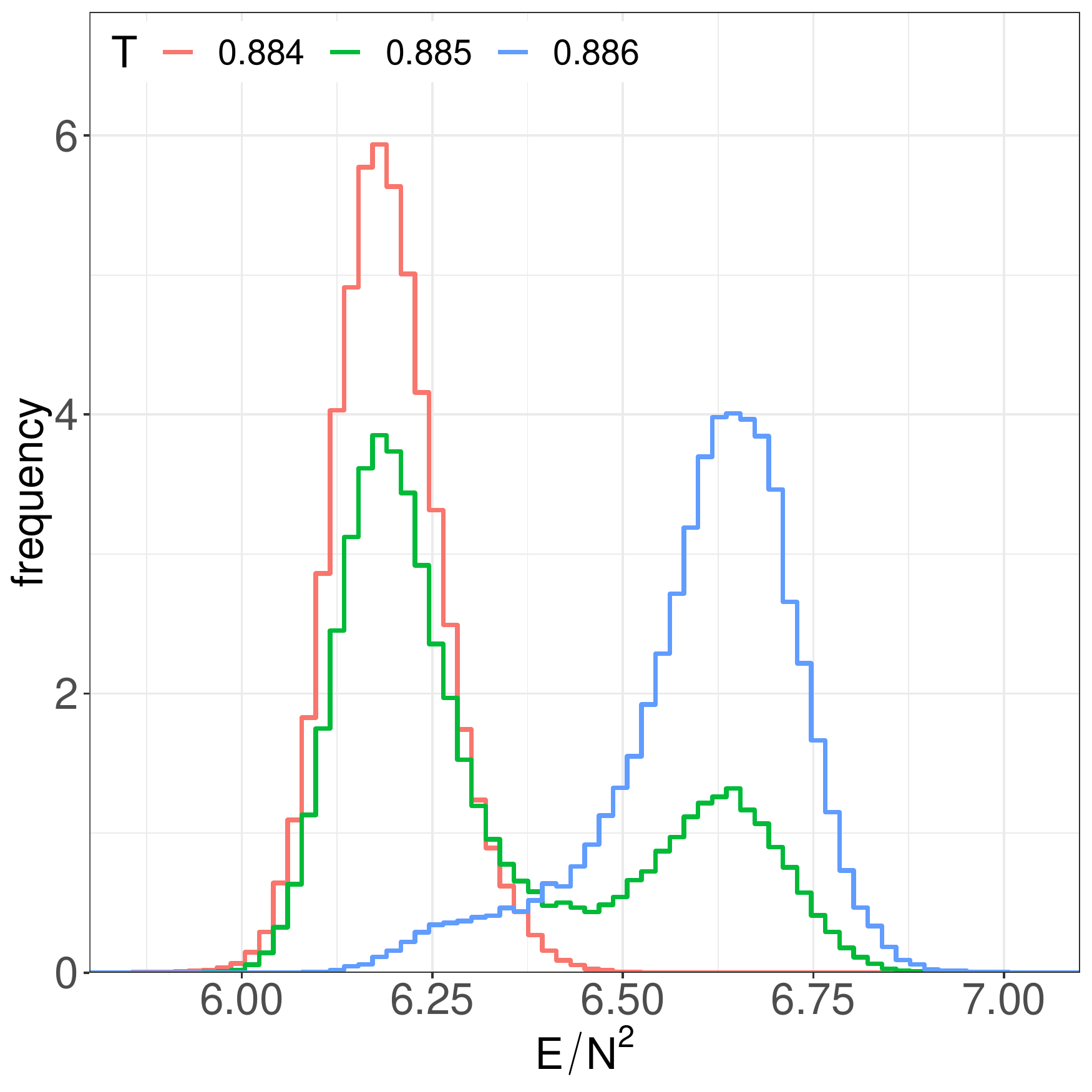}}}
		\caption{\mbox{} }
	\end{subfigure}
	\begin{subfigure}{.45\textwidth}
		\rotatebox{0}{
			\scalebox{0.4}{
				\includegraphics{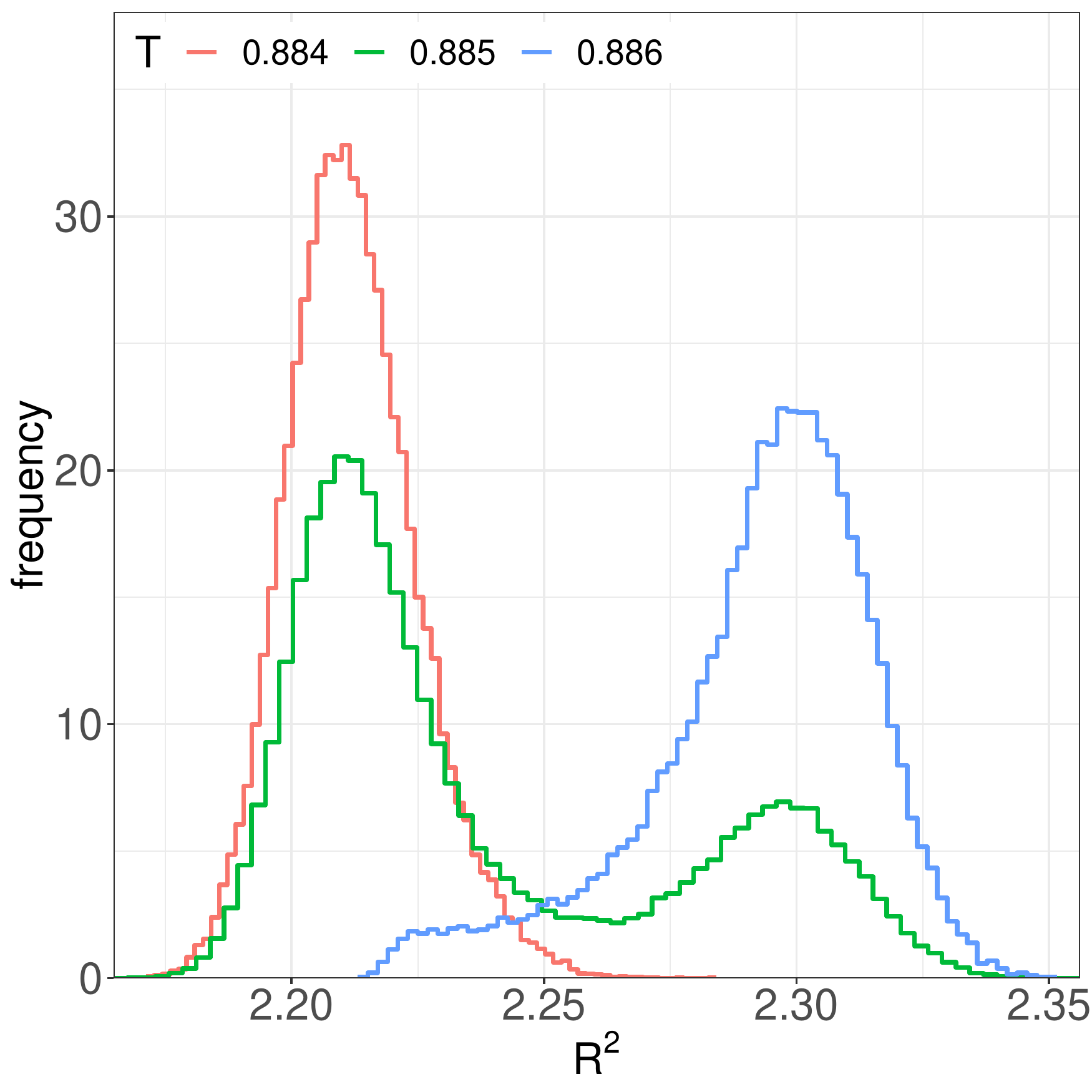}}}
		\caption{\mbox{} }
	\end{subfigure}
	\caption{$N=64$, $L=24$, $D=10$: a) binned $E$ for various $T$. b) binned $R^2$ for various $T$.  }\label{fig:N64S24d9Binning-E-and-R}
\end{figure}

In figure~\ref{fig:N48S24d9Binning-E-and-R} and \ref{fig:N64S24d9Binning-E-and-R} we plot the distribution of $\frac{E}{N^2}$ and $R^2$ for $D=10$.
Like for the order parameter, we observe a first order signal from a clear two-peak structure.
Note that the locations of the two peaks do not move significantly as a function of $N$.
This is consistent with the theoretical expectation that the maximum of the free energy (partially deconfined phase), rather than the minima (completely deconfined or confined phases), moves.

\begin{figure}[htbp]
	\centering
	\begin{subfigure}{.45\textwidth}
		\rotatebox{0}{
			\scalebox{0.4}{
				\includegraphics{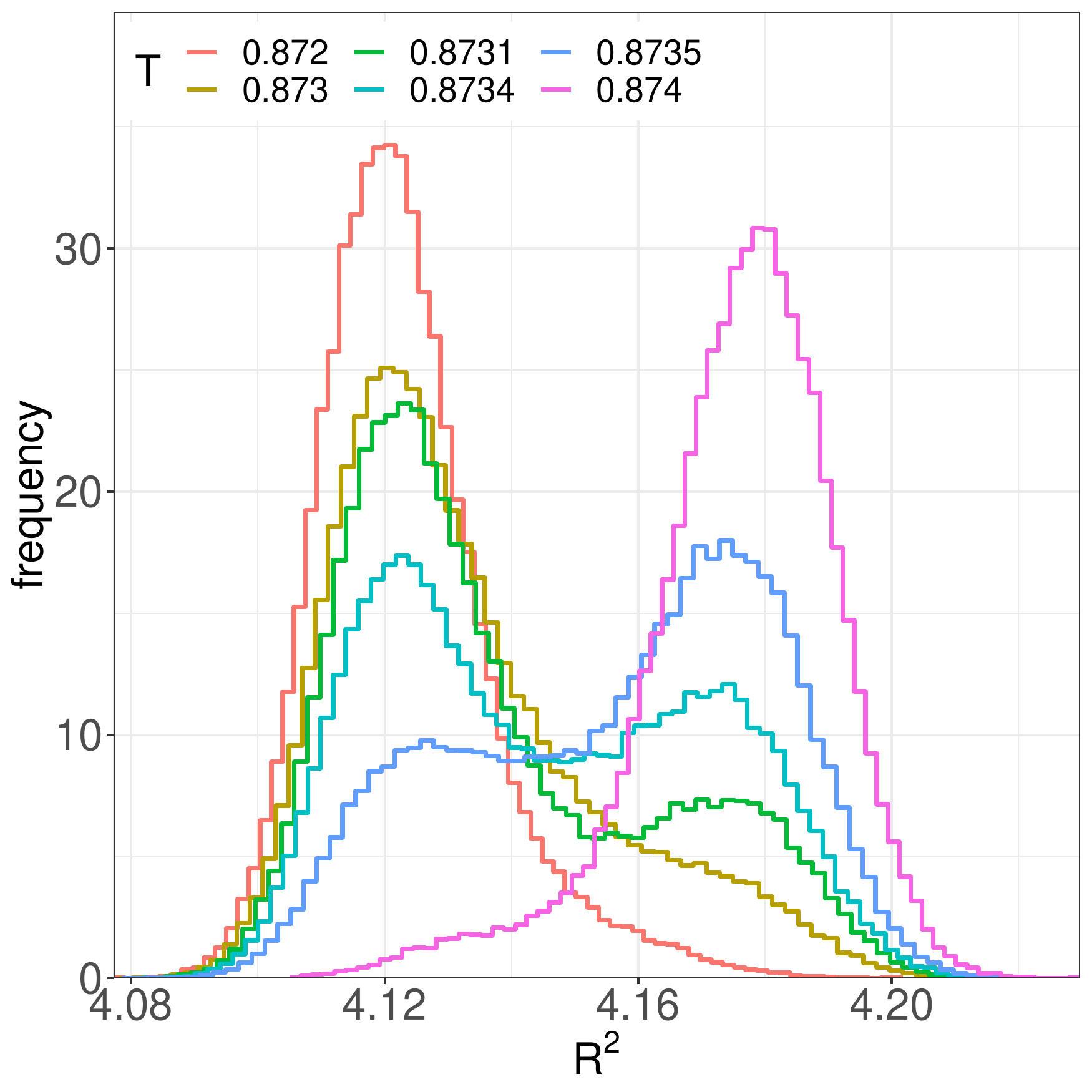}}}
		\caption{\mbox{} }
	\end{subfigure}
	\begin{subfigure}{.45\textwidth}
		\rotatebox{0}{
			\scalebox{0.4}{
				\includegraphics{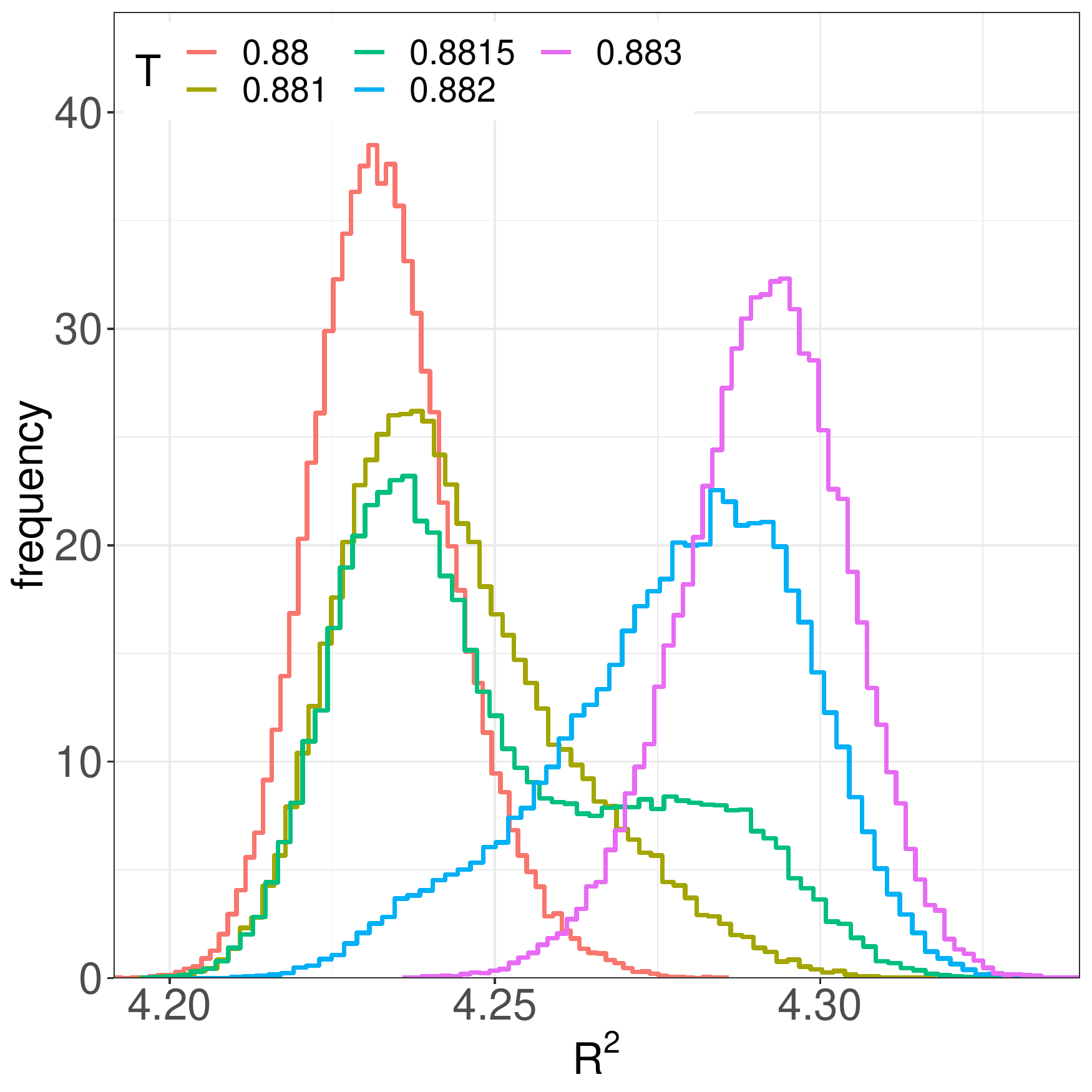}}}
		\caption{\mbox{} }
	\end{subfigure}
	\caption{Histograms of $R^2$ for $N=64$, $D=26$ for various $T$: a) L=24, b) L=48.  }\label{fig:N64d25Binning-E-and-R}
\end{figure}

The first order signal from $E/N^2$ and $R^2$ is less pronounced at $D=26$, as shown in figure~\ref{fig:N64d25Binning-E-and-R} for $N=64$.
We have only presented $R^2$ since the histograms of $E/N^2$ are similar.
For $N=48$, no clear signal can be observed, but for $N=64$, a two-peak structure is visible.
It is the clearest for $L=24$, where we collected higher statistics.
The results for $L=48$ suggest that the two-peak signal persists in the continuum limit.
This is also supported by the Monte Carlo history shown in figure~\ref{fig:HistoryD26} of the Appendix.
The separation of the phases at $D=26$ is less pronounced compared to $D=10$, which might be consistent with the transition developing towards a combination of higher order transitions at some critical $D>26$.

\section{Partial deconfinement}\label{sec:correlations}

In the previous Section, we have determined the order of the deconfinement phase transition.
Let us go one step further and study details of the phase transition.
In particular, we use our numerical data to test the partial deconfinement~\cite{Hanada:2016pwv,Hanada:2018zxn,Berenstein:2018lrm}, reviewed in Sec.~\ref{sec:theory}.

Partial deconfinement is the proposal that the deconfinement transition happens gradually so that at an intermediate stage only $M < N$ color degrees of freedom are deconfined (figure~\ref{fig:partial_deconfinement}).
We will now derive some consequences of this assumption for correlations between different observables and the eigenvalue distribution of the Polyakov loop, and test them against our numerical data.

\begin{figure}[htbp]
\begin{center}
\scalebox{0.3}{
\includegraphics{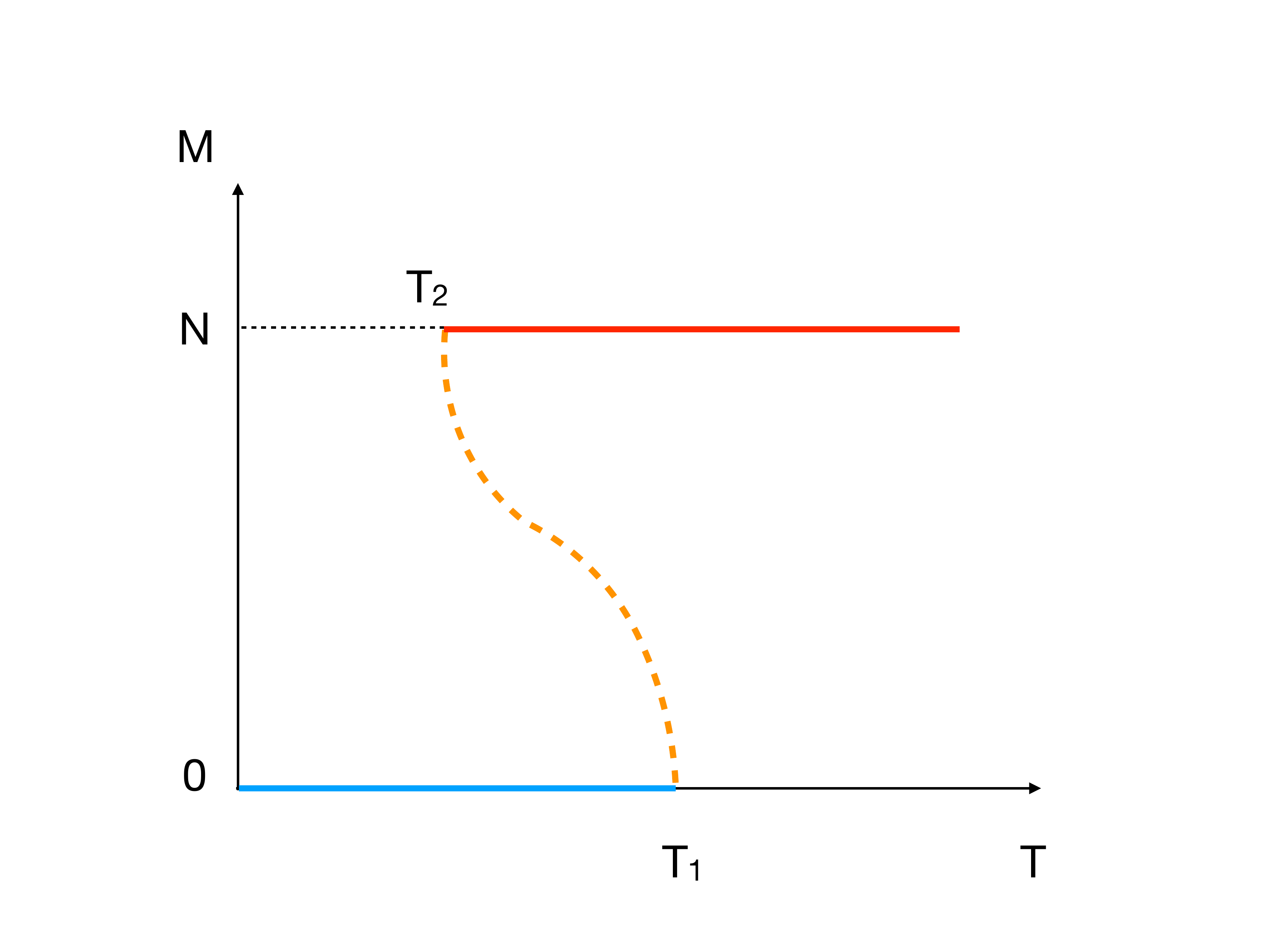}}
\end{center}
\caption{
In the theories with a first order phase transition, there is an unstable saddle (maximum of the free energy) with negative specific heat in the canonical ensemble.
In this phase, the SU($M$) subgroup of the SU($N$) gauge group is deconfined.
This phase connects two stable phases, the confining phase (blue line) and the completely deconfining phase (red line).
}\label{fig:partial_deconfinement_1}
\end{figure}

As we have seen, the phase transition is of first order.
Then the size of the deconfined sector $M$ should change with temperature as visualized in figure~\ref{fig:partial_deconfinement_1}.
At finite $N$, there are non-negligible fluctuations around the saddles, and hence, the size of the deconfined sector $M$ fluctuates configuration-by-configuration during the Monte Carlo simulation.
Below, we will relate the value of $M$ to observables (Polyakov loop $P$, energy $E$ and the extent of space $R^2$).
This leads to nontrivial relations between observables, which can be used for the consistency check of the partial deconfinement proposal.
Then we will confirm those relations numerically.

\begin{figure}[htbp]
	\begin{center}
		\rotatebox{0}{
			\scalebox{0.6}{
				\includegraphics[trim=0mm 30mm 0mm 30mm, clip, scale=1.0]{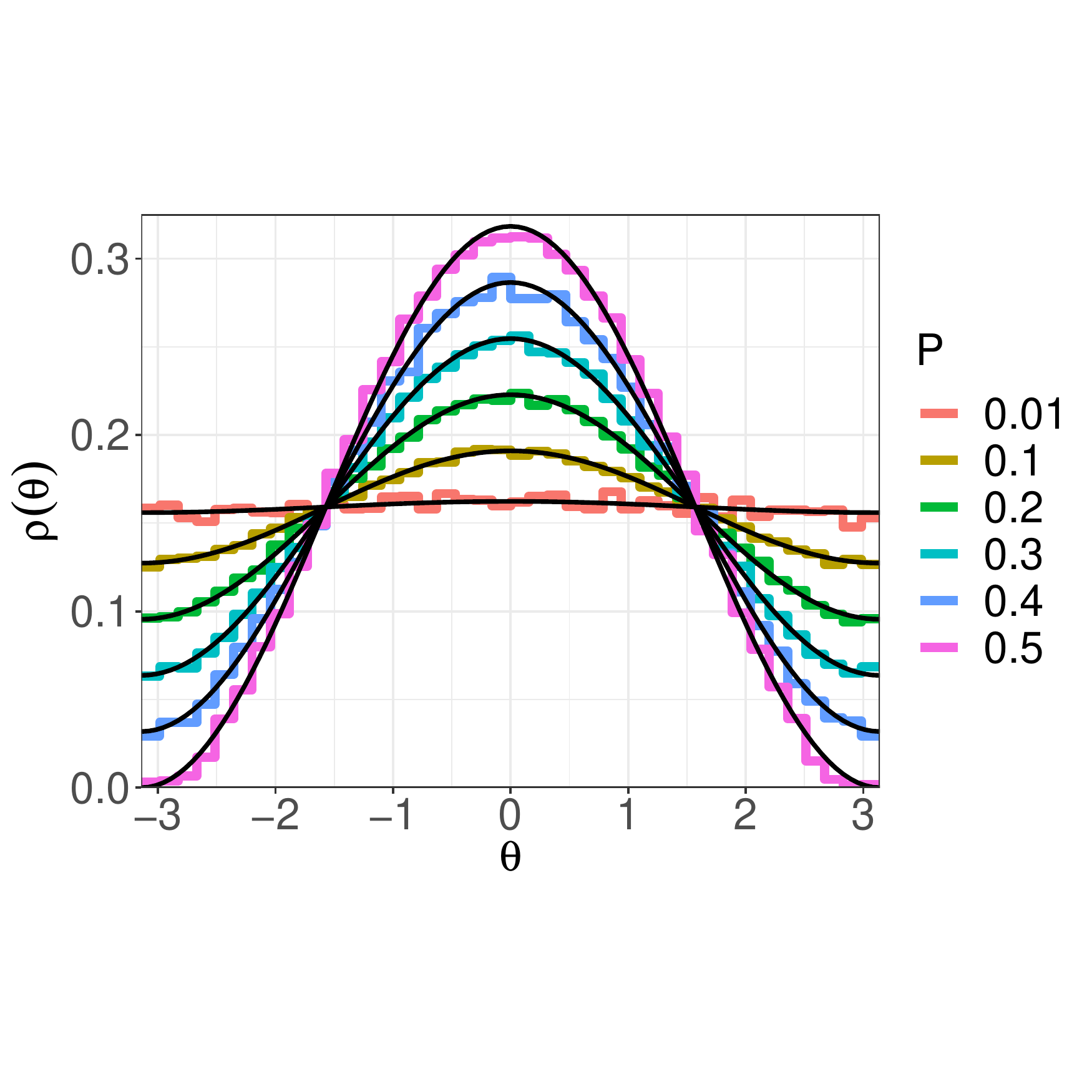}}}
	\end{center}
	\caption{Distribution of Polyakov loop eigenvalues compared to the partial deconfinement prediction (black lines) for different $P$ values, where $\rho_P(\theta)$ is defined in Eq.~\eqref{eq:PartialDecPhaseDistr}.
	The Polyakov loop eigenvalues are binned around the indicated value of $P$ with width $0.01$.}\label{fig:Pol-phase-dist-vs-P}
\end{figure}

Let us assume that Eq.~\eqref{qe:GWW-partially-deconfined} holds, at least approximately, at finite $N$.
By using $\frac{M}{N}=2|P|$, we rewrite Eq.~\eqref{qe:GWW-partially-deconfined} as
\begin{eqnarray}
	\rho_P(\theta) = \frac{1}{2}\left(1+2|P|\cos\theta\right). \label{eq:PartialDecPhaseDistr}
\end{eqnarray}
This relation can easily be tested with our numerical data of the Polyakov phase distribution once the configurations are separated according to their value of $|P|=P$.
We have plotted the distribution $\rho_P(\theta)$ for each bin of $|P|$ in figure~\ref{fig:Pol-phase-dist-vs-P} obtained from the ensemble at $N=64$, $L=24$, $T=0.885$.
The parameters have been chosen to be at the point where the two-state signal indicates a first order transition.
Thus our numerical data provides reasonable evidence that Eq.~\eqref{eq:PartialDecPhaseDistr} holds.
This shows two things: first, the partial deconfinement prediction Eq.~\eqref{eq:PartialDecPhaseDistr} of superposed eigenvalue distributions holds.
Second, the perturbative form of $\rho_{\rm d}(\theta)$~\cite{Sundborg:1999ue,Aharony:2003sx} seems to hold (within errors) also in the strongly coupled regime.

Since Eq.~\eqref{eq:PartialDecPhaseDistr} holds, it is reasonable to assume partial deconfinement with $\frac{M}{N}=2|P|$.
Consequently the partial deconfinement proposal provides further nontrivial predictions:
\begin{enumerate}
\item At fixed temperature, the energy per excited degree of freedom is fixed, and hence the energy above the ground state is proportional to $M^2$.
Therefore, as a function of $|P|$ and $T$, we expect
\begin{eqnarray}
\frac{E(|P|,T)}{N^2}
=
\varepsilon_0
+
f(T)\cdot |P|^2,
\label{eq:E-vs-P}
\end{eqnarray}
where $\varepsilon_0$ represents the zero-point energy.
We expect that this relation is precise at large $N$, where the fluctuation about the planar limit is suppressed.
\item The same counting holds for $R^2\equiv\frac{1}{N}\sum_{I=1}^d{\rm Tr}X_I^2$. Intuitively,
$\frac{1}{g_{\rm YM}^2}|X_I^{ij}|^2=N|X_I^{ij}|^2$ corresponds to the number of open strings excited between $i$-th and $j$-th D-branes, which is a function of $T$.
Therefore, $N\sum_{I=1}^d{\rm Tr}X_I^2$ should be of order $M^2$, plus the zero-mode contribution which is proportional to $N^2$.
Hence we expect
\begin{eqnarray}
R^2(|P|,T)
=
R_0^2
+
g(T)\cdot |P|^2,
\label{eq:R-vs-P}
\end{eqnarray}
where $R_0^2$ comes from the zero-point fluctuation.
\end{enumerate}

Our data confirm these relations with good precision.
We provide the data for $D=10$ since the results for $D=26$ are similar.

\begin{figure}[htbp]
\centering
\begin{subfigure}{.45\textwidth}
\rotatebox{0}{
\scalebox{0.4}{
\includegraphics{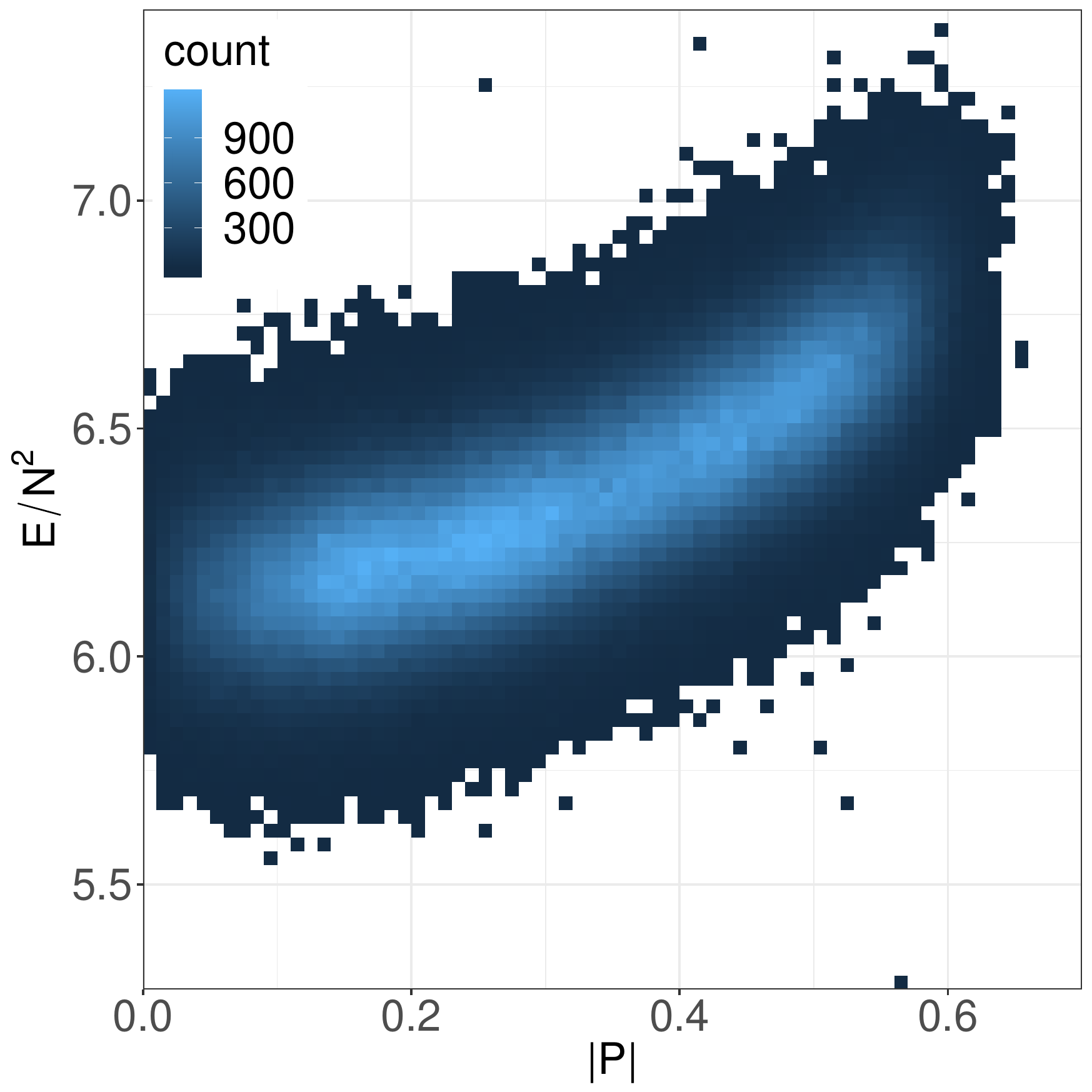}}}
\caption{\mbox{} }
\end{subfigure}
\begin{subfigure}{.45\textwidth}
\rotatebox{0}{
\scalebox{0.4}{
\includegraphics{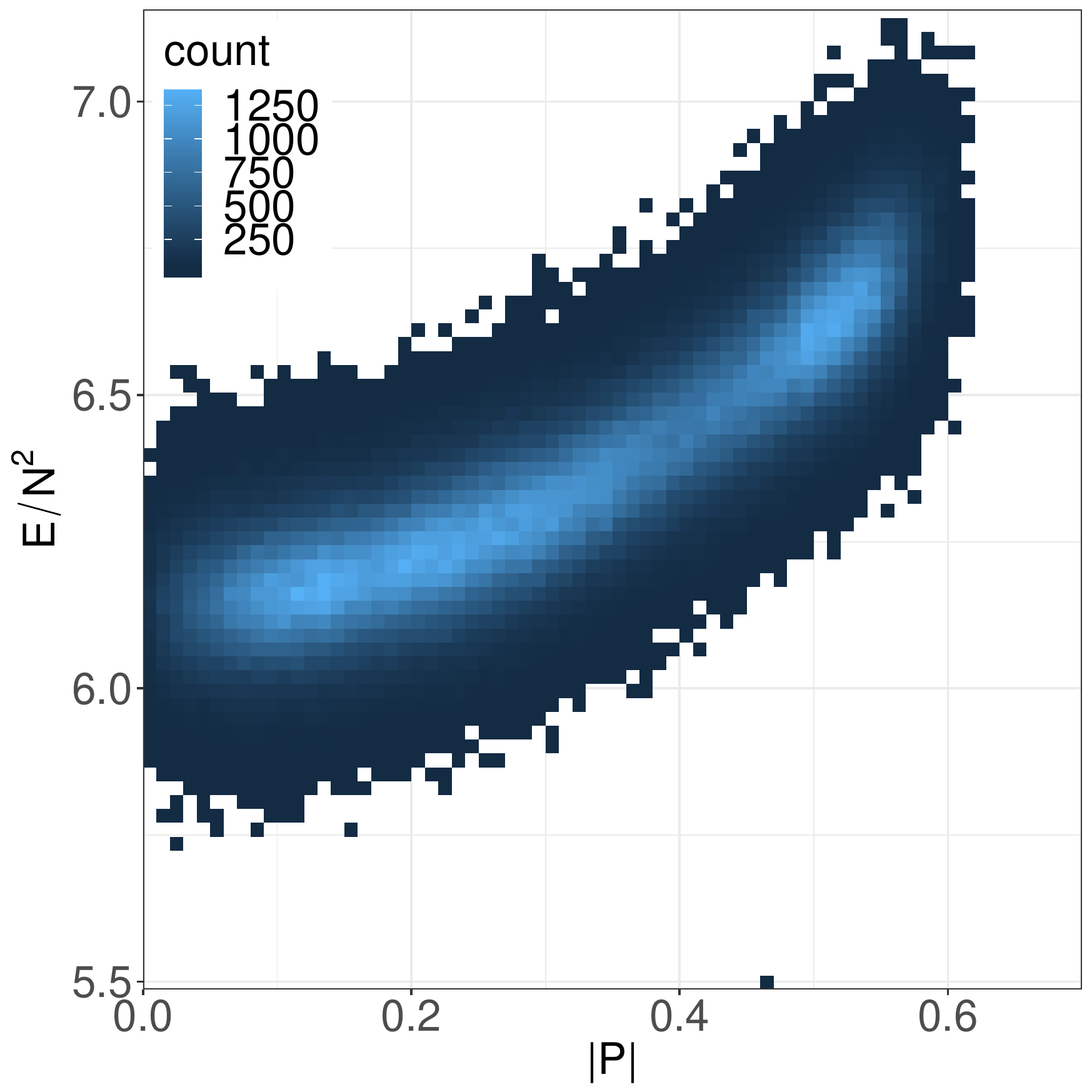}}}
\caption{\mbox{}}
\end{subfigure}
\begin{subfigure}{.45\textwidth}
\rotatebox{0}{
\scalebox{0.4}{
\includegraphics{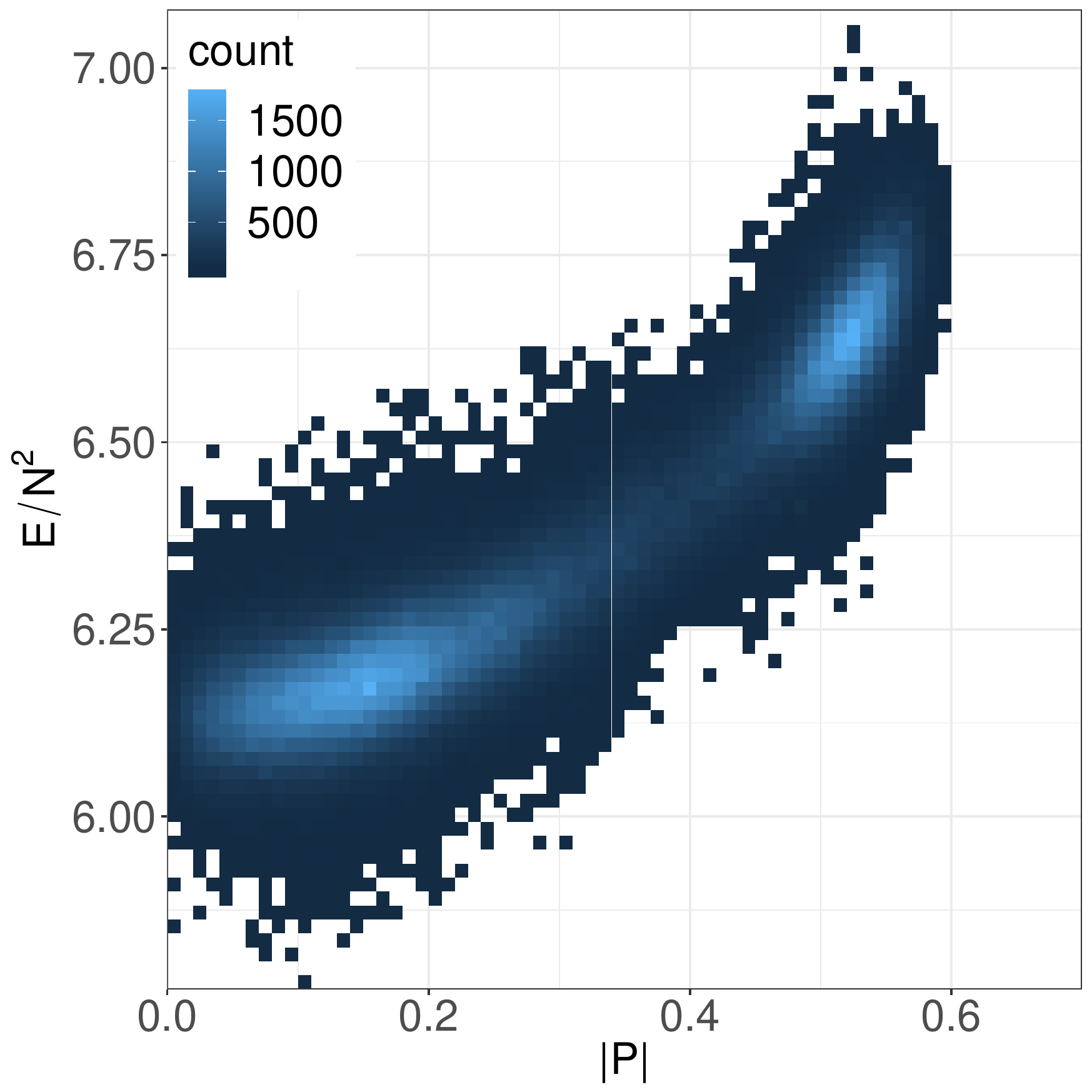}}}
\caption{\mbox{} }
\end{subfigure}
\begin{subfigure}{.45\textwidth}
\rotatebox{0}{
\scalebox{0.4}{
\includegraphics{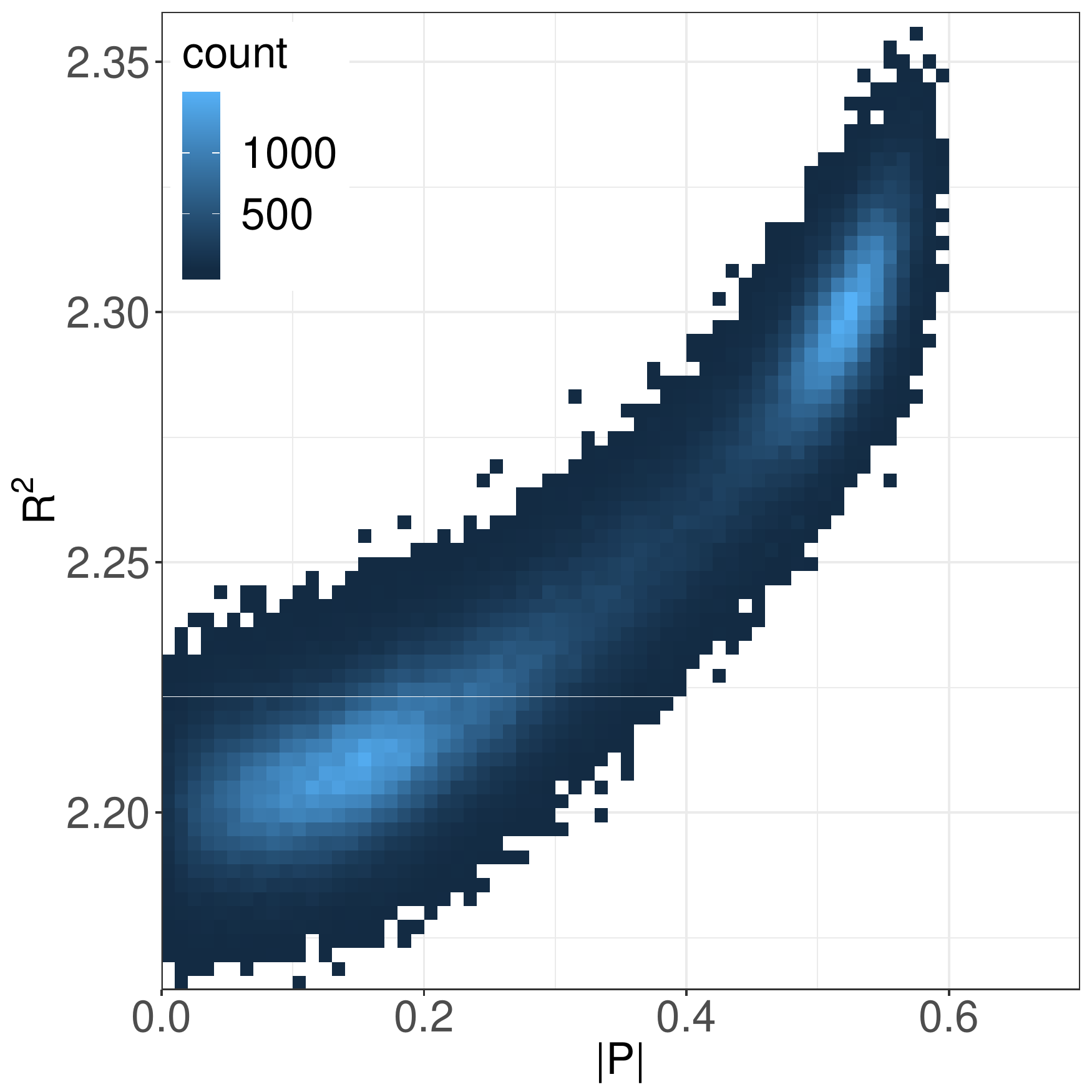}}}
\caption{\mbox{} }
\end{subfigure}
\caption{(a-c) Correlations of $E/N^2$ and $|P|$ for $L=24$, $N=32,48,64$ at $T=0.883, 0.885, 0.885$. d) Correlations of $R^2$ and $|P|$ for $L=24$, $N=64$ at $T=0.885$.}\label{fig:pd-D9-2d}
\end{figure}

As a first check, we have plotted the density distribution of the configurations in $(E,|P|)$-plane in figure~\ref{fig:pd-D9-2d}.
We have introduced bins in $(E,|P|)$-plane, counted the number of configurations in each bin, and repeated the same for $R^2$ as well.
As expected, the distribution becomes sharper as $N$ increases due to suppression of fluctuations in the planar limit.
This confirms, at least at a qualitative level, that  \eqref{eq:E-vs-P} and \eqref{eq:R-vs-P} are valid.

\begin{figure}[htbp]
\centering
\begin{subfigure}{.45\textwidth}
\rotatebox{0}{
\scalebox{0.4}{
\includegraphics{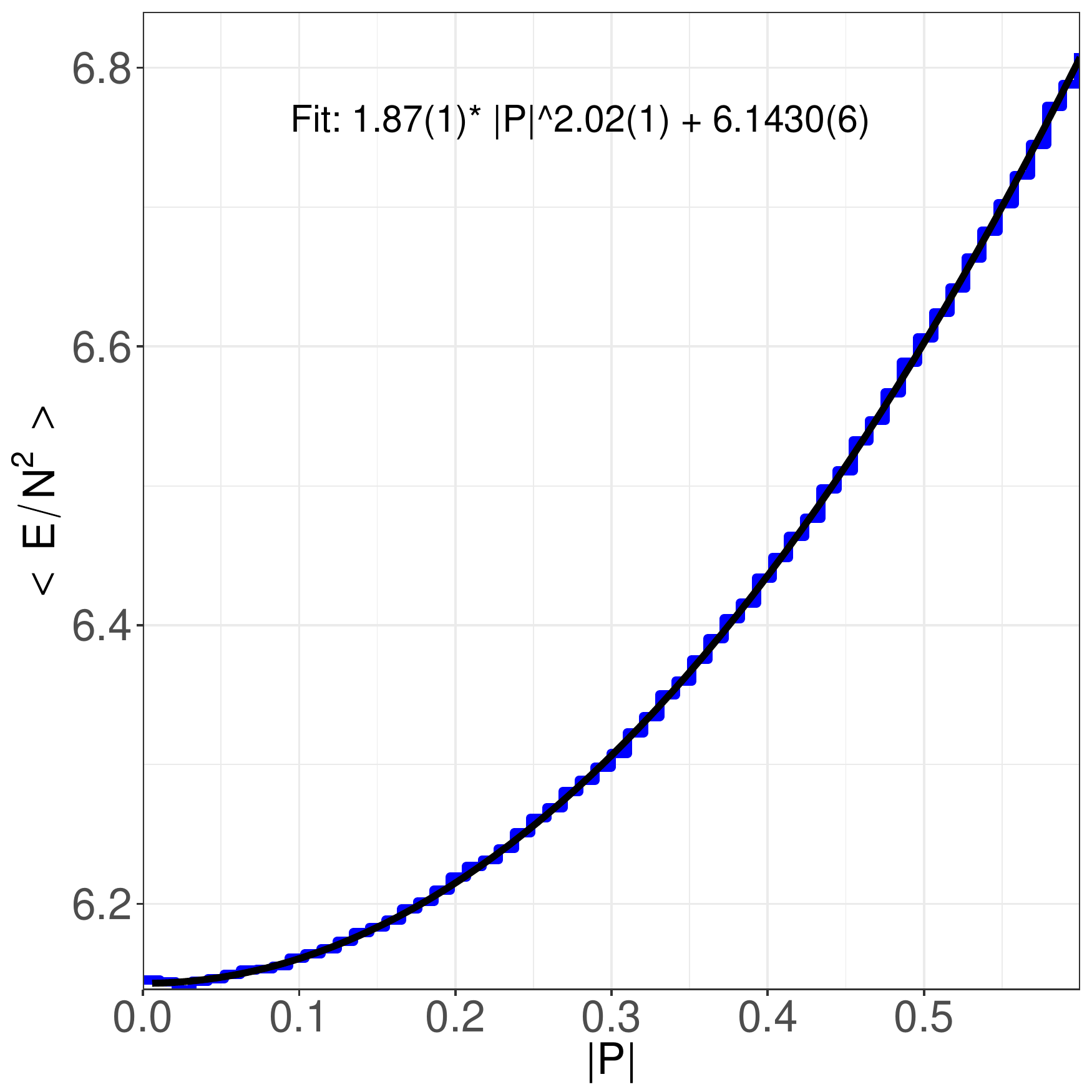}}}
\caption{\mbox{} }
\end{subfigure}
\begin{subfigure}{.45\textwidth}
\rotatebox{0}{
\scalebox{0.4}{
\includegraphics{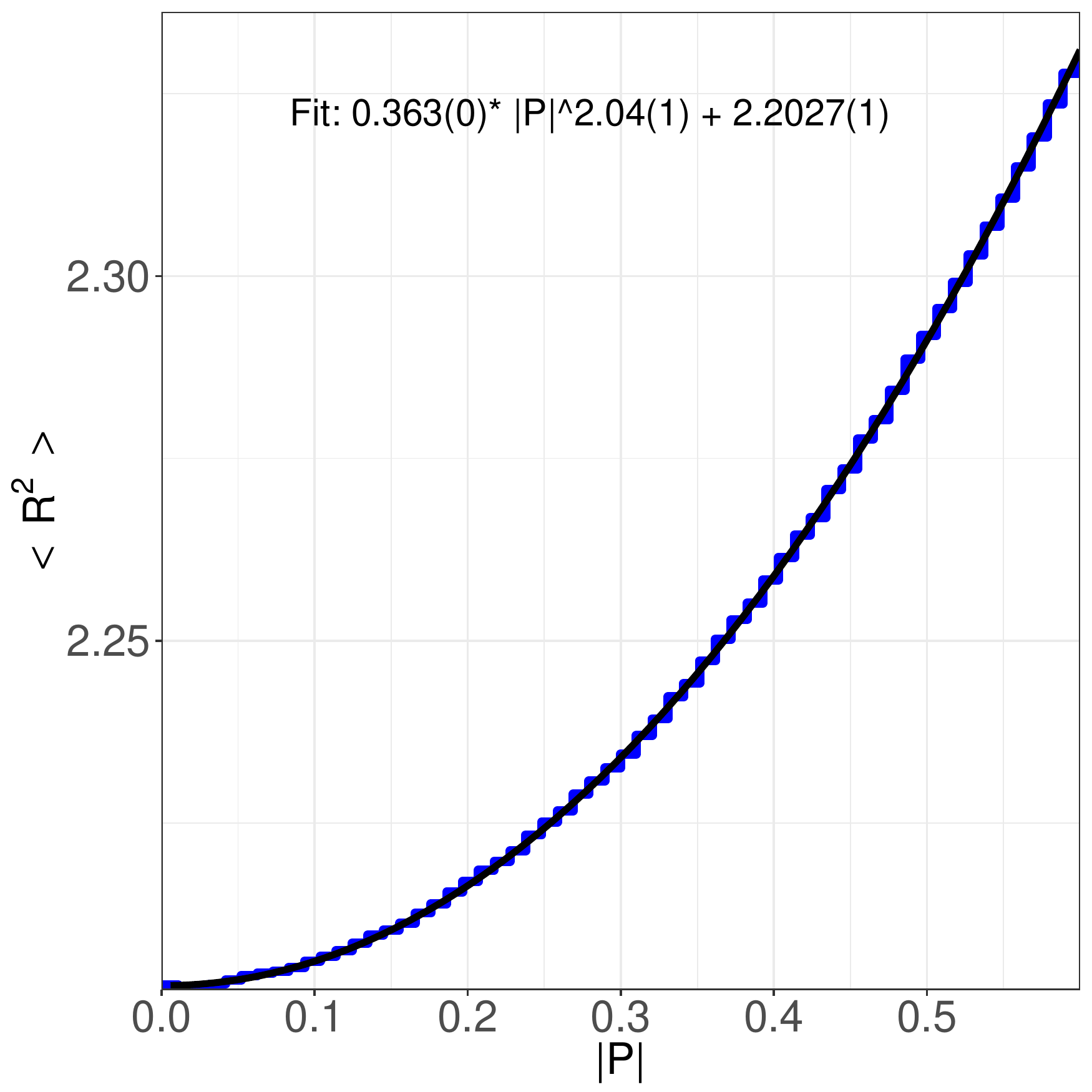}}}
\caption{\mbox{} }
\end{subfigure}
\begin{subfigure}{.45\textwidth}
\rotatebox{0}{
\scalebox{0.4}{
\includegraphics{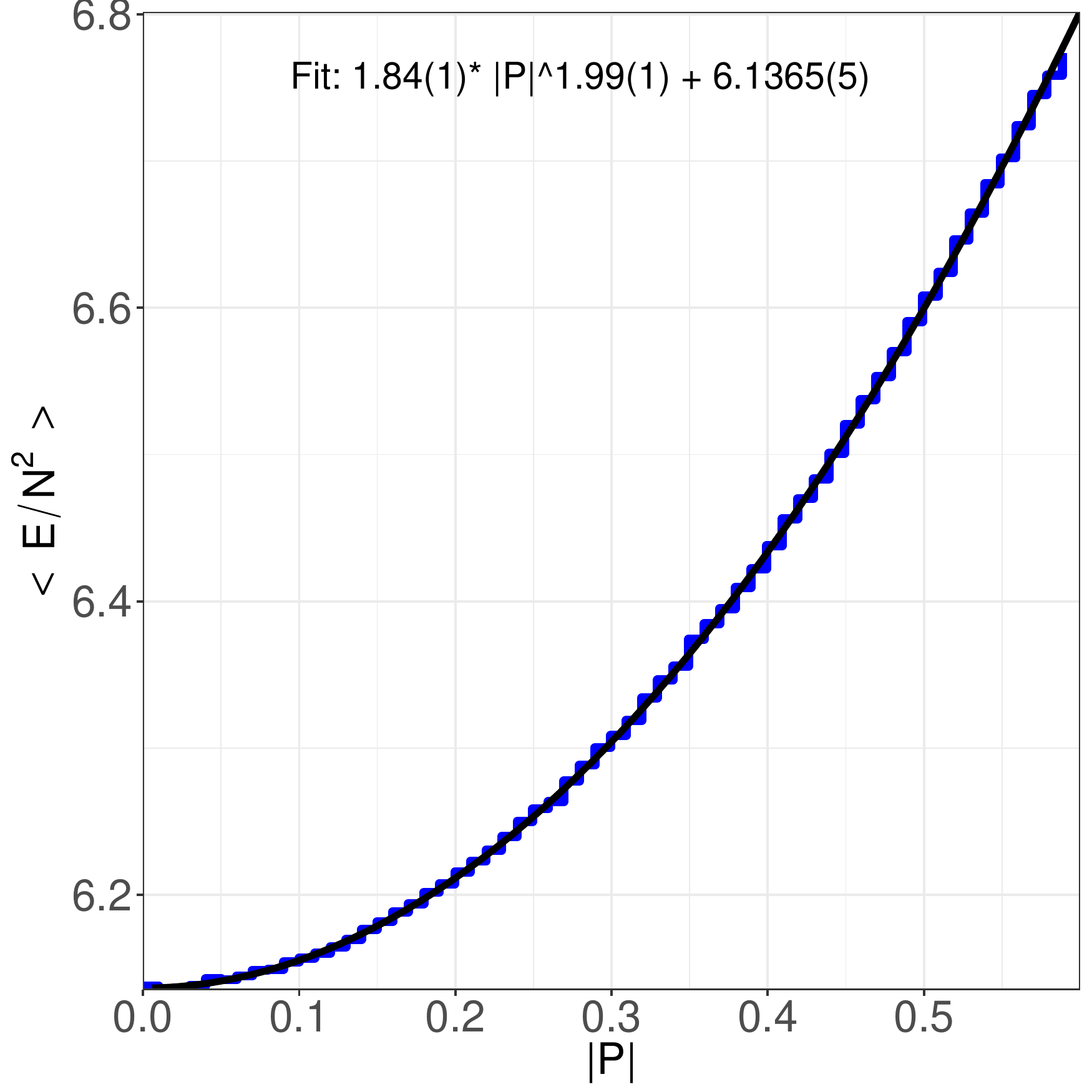}}}
\caption{\mbox{}}
\end{subfigure}
\begin{subfigure}{.45\textwidth}
\rotatebox{0}{
\scalebox{0.4}{
\includegraphics{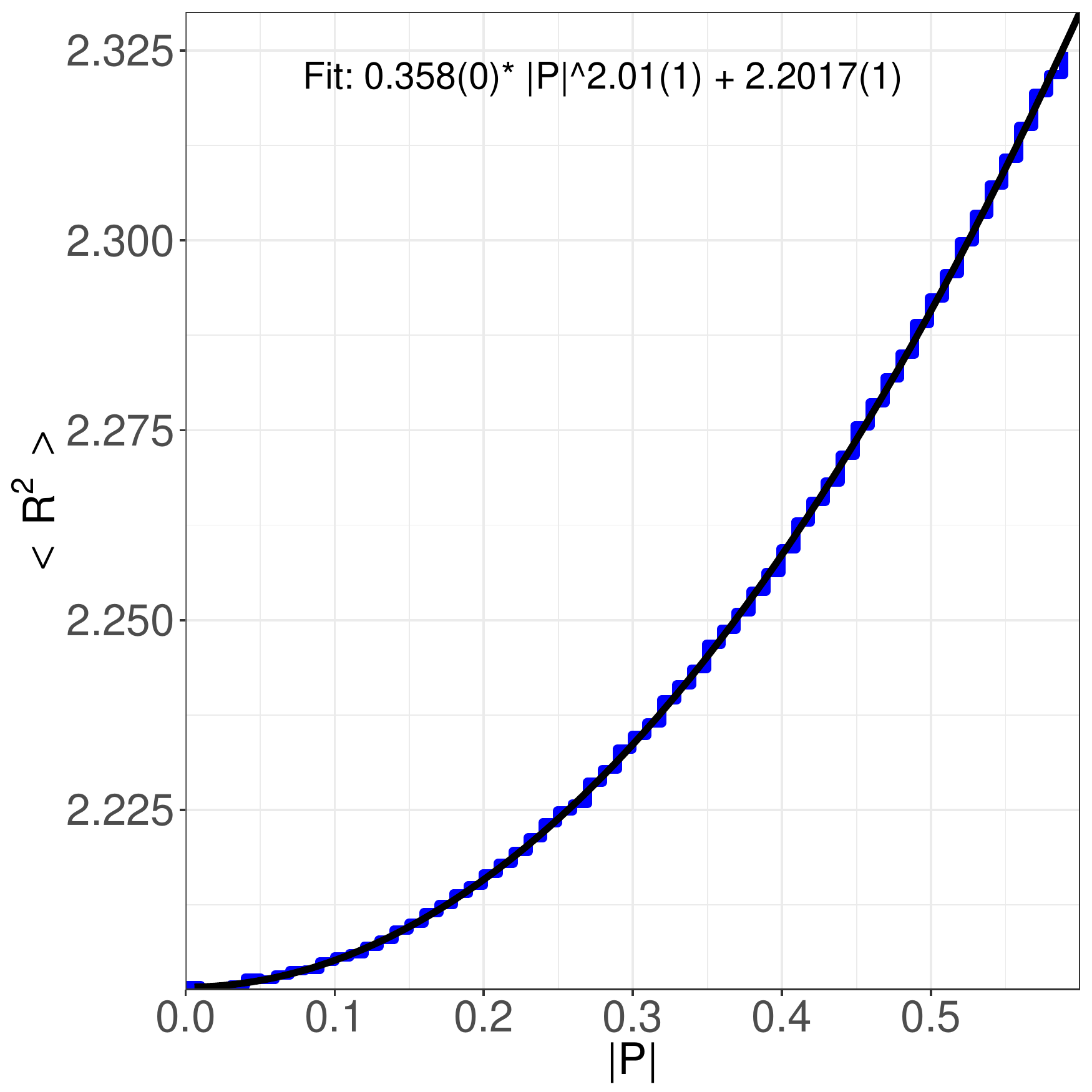}}}
\caption{\mbox{} }
\end{subfigure}
\caption{Several checks of the partial deconfinement prediction for $\vev{E}(|P|)$ and $\vev{R^2}(|P|)$. The blue curve shows binned values for $\vev{E/N^2}$ and $\vev{R^2}$. The fit to $a |P|^b + c$ is done in the range $0\leq |P| \leq 0.5$. a+b) $N=48$, $L=24$,  c+d) $N=64$, $L=24$; all $T=0.885$, $D=10$.}\label{fig:pd-D9}
\end{figure}

In order to confirm \eqref{eq:E-vs-P} and \eqref{eq:R-vs-P} quantitatively, we separate the configurations into bins according to their $|P|$ value.
In that way we obtain expectation values $\langle E\rangle (|P|,T)$ of the energy $E$ for fixed $|P|$ and in the same way $\langle R^2\rangle (|P|,T)$.
Figure~\ref{fig:pd-D9} shows the plot of $\frac{1}{N^2}\langle E \rangle  (|P|,T)$ as a function of $|P|$. We can see very good agreement with \eqref{eq:E-vs-P} and \eqref{eq:R-vs-P}.

\begin{figure}[htbp]
\centering
\begin{subfigure}{.45\textwidth}
\rotatebox{0}{
\scalebox{0.4}{
\includegraphics{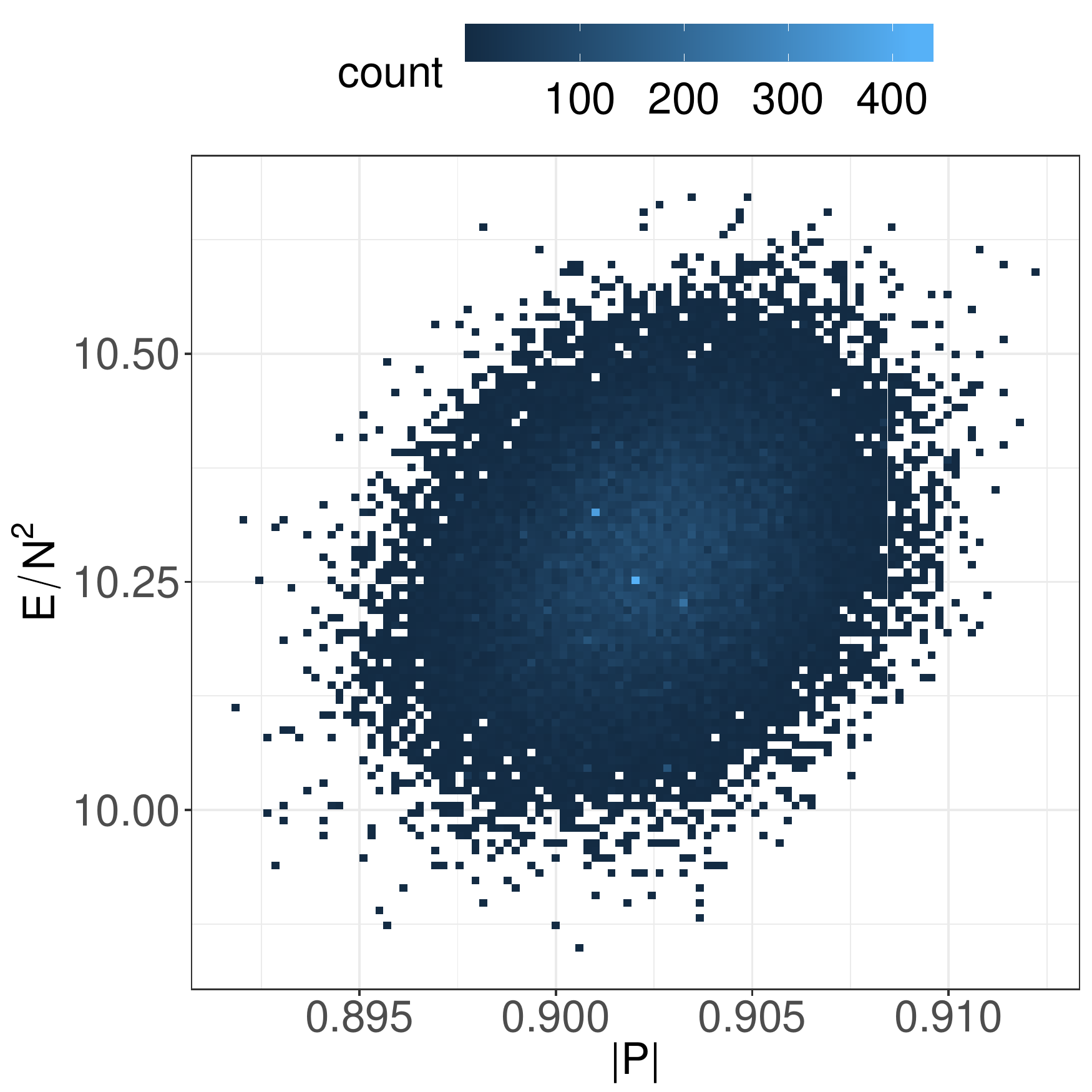}}}
\caption{\mbox{} }
\end{subfigure}
\begin{subfigure}{.45\textwidth}
\rotatebox{0}{
\scalebox{0.4}{
\includegraphics{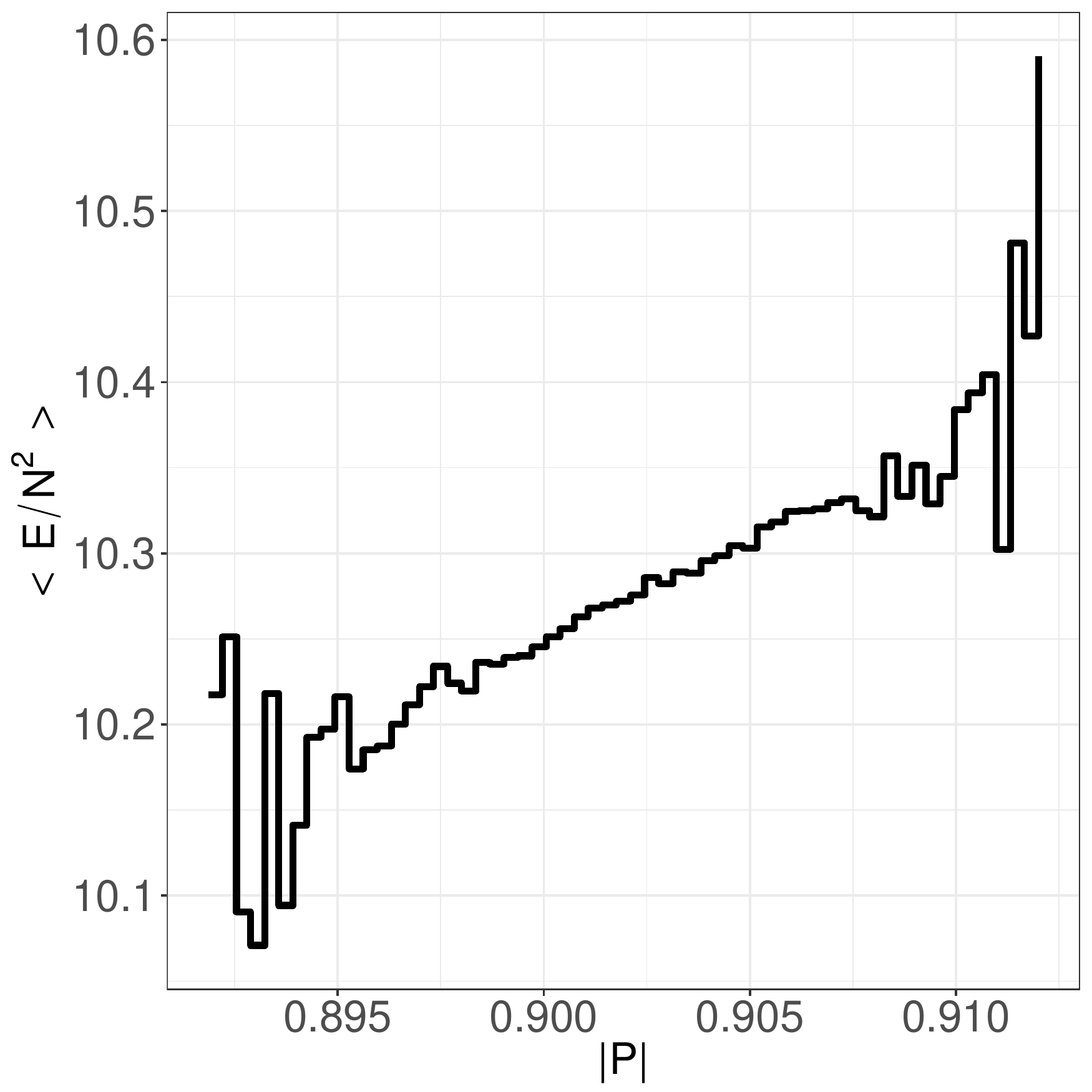}}}
\caption{\mbox{}}
\end{subfigure}
\caption{a) Distribution of $E$ and $|P|$ at $N=64, L=24, D=10, T=1.5$. b) $\vev{E}(|P|)$ for the same data set.}\label{fig:highT-correlation}
\end{figure}

In Ref.~\cite{Berkowitz:2016jlq}, the correlation between energy and Polyakov loop in the D0-brane matrix model has been studied in a parameter region with $|P|\gtrsim 0.7$, i.e.\ in the completely deconfined phase.
It was found that the energy and Polyakov loop are not correlated at all.
In contrast, we find that even for $T=1.5$ where $|P| \approx 0.9$, $\vev{E/N^2}$ and $\vev{R^2}$ are correlated, see figure~\ref{fig:highT-correlation}.
We also studied the low temperature phase at $T=0.5$, where there appears to be no correlation.

\section{Conclusion and future directions}\label{sec:discussions}

In this paper, we have studied the nature of deconfinement in the bosonic Yang-Mills matrix model \eqref{eq:MainAction}.
We have concluded that the transition is of first order for $D=10$.
By interpreting this model as the high-temperature limit of 2d maximal super Yang-Mills (SYM) it is natural to conclude that the phase diagram of 2d maximal SYM is like the left panel of figure~\ref{fig:2dSYM_phase_diagram}.
Via the gauge/gravity duality, we can rephrase this finding~\cite{Aharony:2004ig} to say that the $\alpha'$-corrections (which become more important at higher temperatures) do not alter the order of the phase transition between black hole and black string in canonical ensemble.

\begin{figure}[htbp]
\begin{center}
\scalebox{0.35}{
\includegraphics[trim = 0mm 20mm 0mm 40mm, clip]{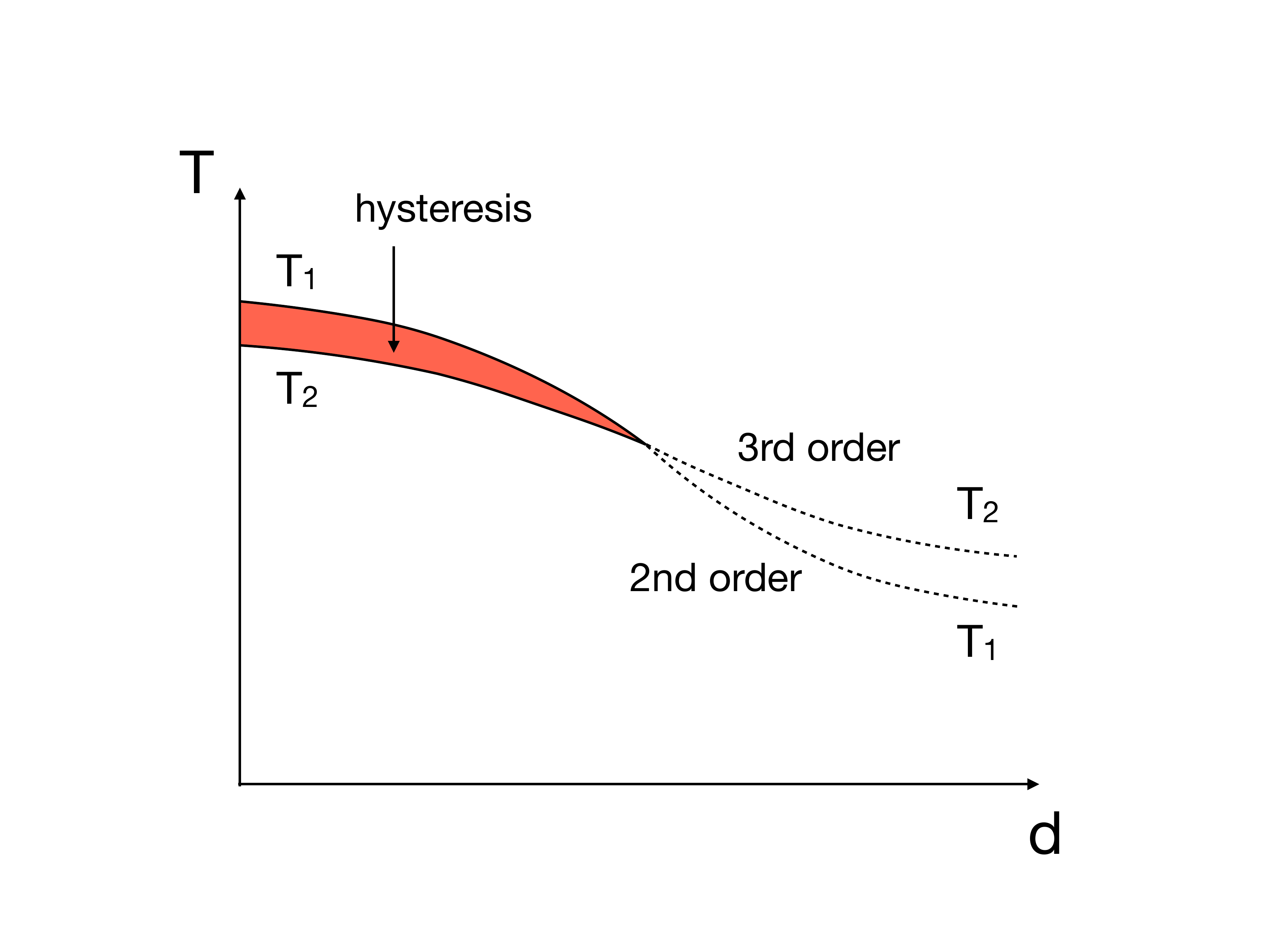}}
\scalebox{0.5}{
\includegraphics{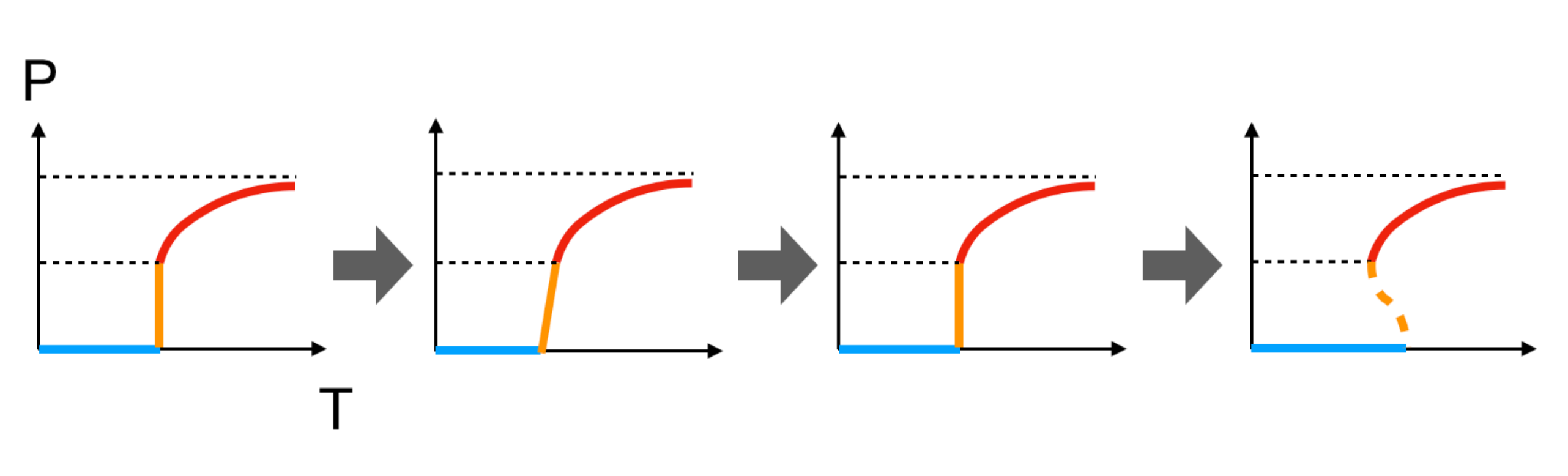}}
\end{center}
\caption{An educated guess about the $d$-dependence of the phase diagram, from $d=\infty$ to small $d$.
[Top] The dependence of the critical temperatures $T_1$ and $T_2$ on $d$.
[Bottom] The $P$-vs-$T$ plot for various $d$, from large (left) to small (right).
Blue, orange and red lines are confined, partially deconfined and completely deconfined phases, respectively.
At $d=\infty$, $T_1=T_2$, and at large but finite $d$, $T_1<T_2$ \cite{Mandal:2009vz}.
We have observed a first order transition, and hence $T_1>T_2$ for $D \leq26$.
}\label{fig:Pol-vs-T-educated-guess}
\end{figure}

We have also observed a first order transition for the $D=26$ case, which provides further insights about the validity of the large-$D$ expansion.
In comparison with the results of Ref.~\cite{Mandal:2009vz} and in agreement with \cite{Azuma:2014cfa}, we conjecture that there is a critical dimension $D_c>26$ with first order transition at all $D<D_c$, see figure~\ref{fig:Pol-vs-T-educated-guess}.
The large-$D$ picture with two separate transitions at $T_1$ and $T_2$ applies only for $D\geq D_c$.
The validity of this conjecture is not easy to confirm since the temperature separation $\Delta T$ decreases with increasing $D$ and higher accuracy is needed to determine the existence of separate transitions.

\begin{figure}[htbp]
\begin{center}
\rotatebox{0}{
\scalebox{0.4}{
\includegraphics[trim = 0mm 50mm 0mm 95mm, clip]{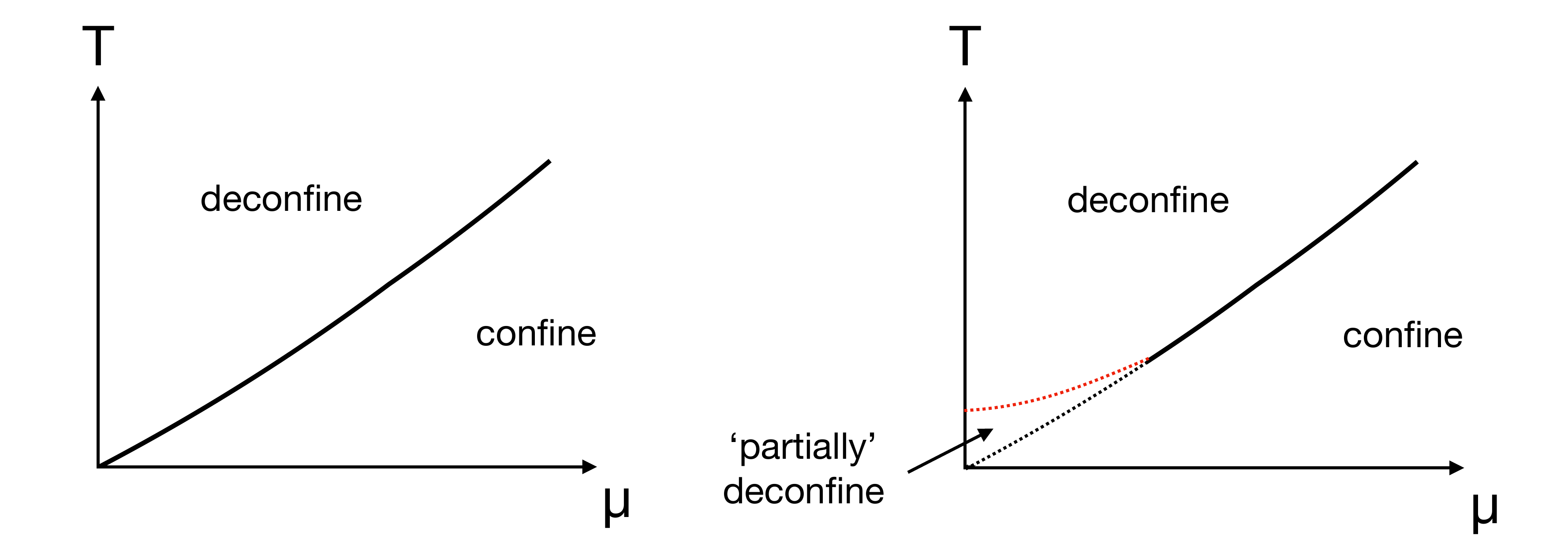}}}
\end{center}
\caption{Two kinds of conjectured phase diagrams of the BMN matrix model in the canonical ensemble.
The vertical axis is the temperature $T$.
The large-$\mu$ region admits perturbative calculation~\cite{Furuuchi:2003sy,Spradlin:2004sx}
and the transition is found to be of first order.
The small-$\mu$ region has been studied by using the dual gravity description~\cite{Costa:2014wya},
but the order of the transition has not been established.
See Refs.~\cite{Asano:2018nol,Catterall:2010gf} for lattice simulations along this direction.
}\label{fig:conjectured-phase-diagram-BMN}
\end{figure}

\begin{figure}[htbp]
\begin{center}
\rotatebox{0}{
\scalebox{0.4}{
\includegraphics[trim = 0mm 50mm 0mm 95mm, clip]{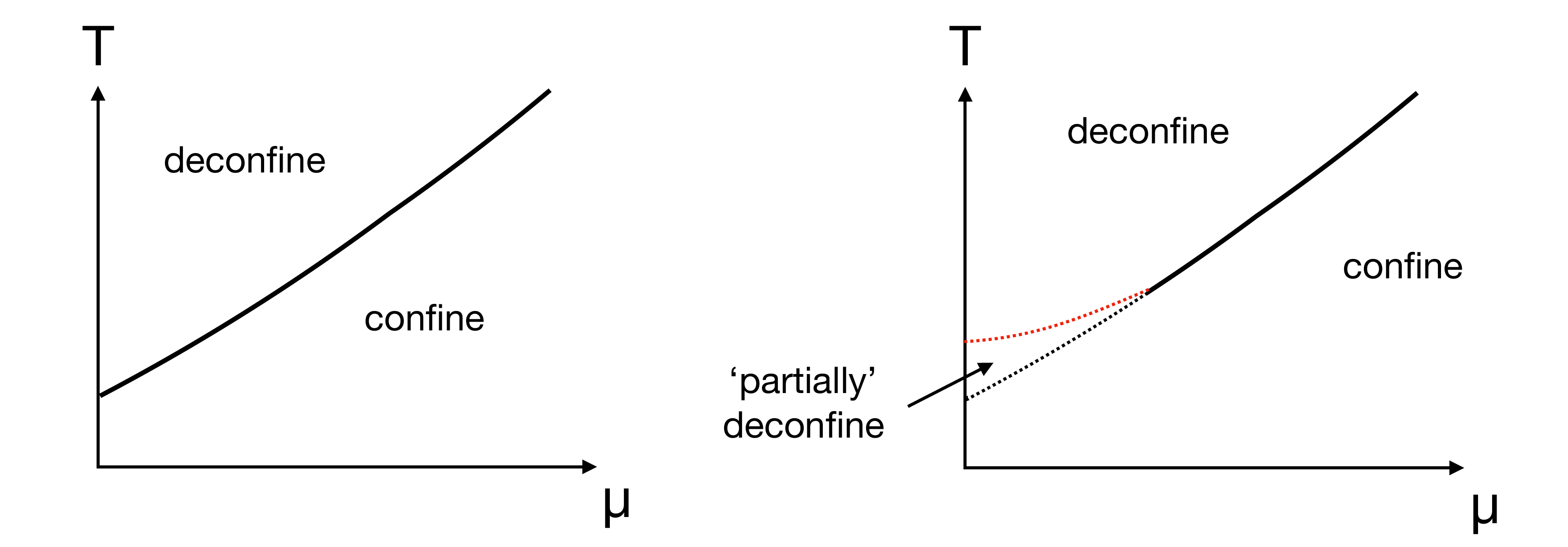}}}
\end{center}
\caption{Possible phase diagrams of the bosonic BMN matrix model in the canonical ensemble.
The vertical axis is the temperature $T$.
When the transition is of first order, an unstable phase corresponding to the Schwarzschild black hole should exist~\cite{Hanada:2018zxn}.
Our numerical simulations studied the $\mu=0$ axis, where we only found one transition, with no intermediate phase.
Hence the left panel is realized. (Although, strictly speaking, the transition may not be of first order at intermediate $\mu$.)
}\label{fig:bosonic-BMN-phase-diagram}
\end{figure}

There are various directions for future follow-up studies.
The Berenstein-Maldacena-Nastase (BMN) matrix model~\cite{Berenstein:2002jq}, which is a one-parameter deformation of the D0-brane matrix model~\cite{Banks:1996vh,deWit:1988wri,Witten:1995im} with the flux parameter $\mu$, would provide us with a numerically tractable setup with a weakly-coupled gravity dual~\cite{Costa:2014wya}.
There are two possible phase diagrams for the BMN matrix model, as shown in figure~\ref{fig:conjectured-phase-diagram-BMN}, and in particular the phase structure at $\mu=0$ is still unclear.
For an unambiguous determination of the phase structure, simulations at large enough $N$ and sufficiently low temperature are essential.
Despite the extensive numerical studies performed for the D0-brane matrix model, see Refs.~\cite{Anagnostopoulos:2007fw,Catterall:2008yz,Kabat:1999hp} and related studies, this has not been achieved so far, and the dual gravity interpretation~\cite{Banks:1996vh,deWit:1988wri,Itzhaki:1998dd} of the results might be affected.
The bosonic version of this model is an instructive exercise to test the simulation methods.
Before we had obtained our numerical data, we had two possibilities for its phase diagram in mind, which are illustrated in figure~\ref{fig:bosonic-BMN-phase-diagram}.
Our numerical simulation indicate that the picture on the left without a stable intermediate phase is realized at $\mu=0$.

The supersymmetric matrix models~\cite{Banks:1996vh,deWit:1988wri,Itzhaki:1998dd,Berenstein:2002jq} are dual to black zero-brane in type IIA string theory.
In order to understand how gauge theory degrees of freedom describe quantum gravity, it is important to determine which of the two possibilities shown in figure~\ref{fig:conjectured-phase-diagram-BMN} is realized in this case.
This requires an investigation of the very low temperature region, as previously studied in Ref.~\cite{Hanada:2013rga}, but also at sufficiently large $N$.
So far, the largest value of $N$ in the literature is $N=32$~\cite{Berkowitz:2016jlq}, and given the lesson from this paper, it might still be too small at low temperatures.

In this paper, we have not studied the details of the unstable partially deconfined phase, or equivalently the maximum of the free energy.
This phase, which is {\it not important} in the importance sampling for the canonical ensemble, actually contains very {\it important} information of the theory, because this is the phase connecting the completely deconfined phase and the confined phase in the microcanonical ensemble.
It should be possible to study the property of this phase by determining the location of the `dip' between the two peaks, and picking up the configurations from there.
A detailed numerical study of this phase would be useful for understanding black hole evaporation in the context of holography.
When the bosonic matrix model is interpreted as the high-temperature limit of 2d maximal SYM, a comparison with the dual gravity calculation at low temperature would be very interesting, because the low and high temperature regions resemble each other at least at the qualitative level as we have shown in this paper.

\acknowledgments
The authors would like to thank Oscar Dias, Pau Figueras, Goro Ishiki, Samuel Kov\'{a}\v{c}ik, Denjoe O'Connor, Andreas Rabenstein, Jorge Santos, and Hiromasa Watanabe for discussions.
They thank the ECT* for its hospitality during the workshop “Quantum Gravity meets Lattice QFT” where this work was initiated, and the Okinawa Institute of Science and Technology for the hospitality during the workshop "Quantum and Gravity in Okinawa" where this work was finalized.
G.~B.\ acknowledges support from the Deutsche
Forschungsgemeinschaft (DFG) Grant No.\ BE 5942/2-1.
N.~B.\ was supported by an International Junior Research Group grant of the Elite Network of Bavaria.
E.~R.\ was supported by a RIKEN Special Postdoctoral fellowship.
M.~H.\ thanks the members of Brown University and Dublin Institute for Advanced Study for the hospitality during his visits.
The numerical simulations were performed on ATHENE, the HPC cluster of the Regensburg University Compute Centre.
Computing support for this work came also from the Lawrence Livermore National Laboratory (LLNL) Institutional Computing Grand Challenge program.

\bibliographystyle{JHEP}
\bibliography{matrix-model}

\appendix
\section{Simulation details}

In the next pages we report some examples of Monte Carlo histories of the observables $|P|$, $E/N^2$ and $R^2$ around the critical region where their histograms show a double-peak structure.
The histories show clear tunneling effects between the two different phases in the critical temperature region for $N=64$, $L=24$, and both $D=10$  and $D=26$.
More details about the data, containing the full statistics accumulated to obtain and reproduce the results in this paper, will be released online.

\begin{figure}
	\centering
	\begin{subfigure}[b]{1.0\textwidth}
		\includegraphics[width=\textwidth]{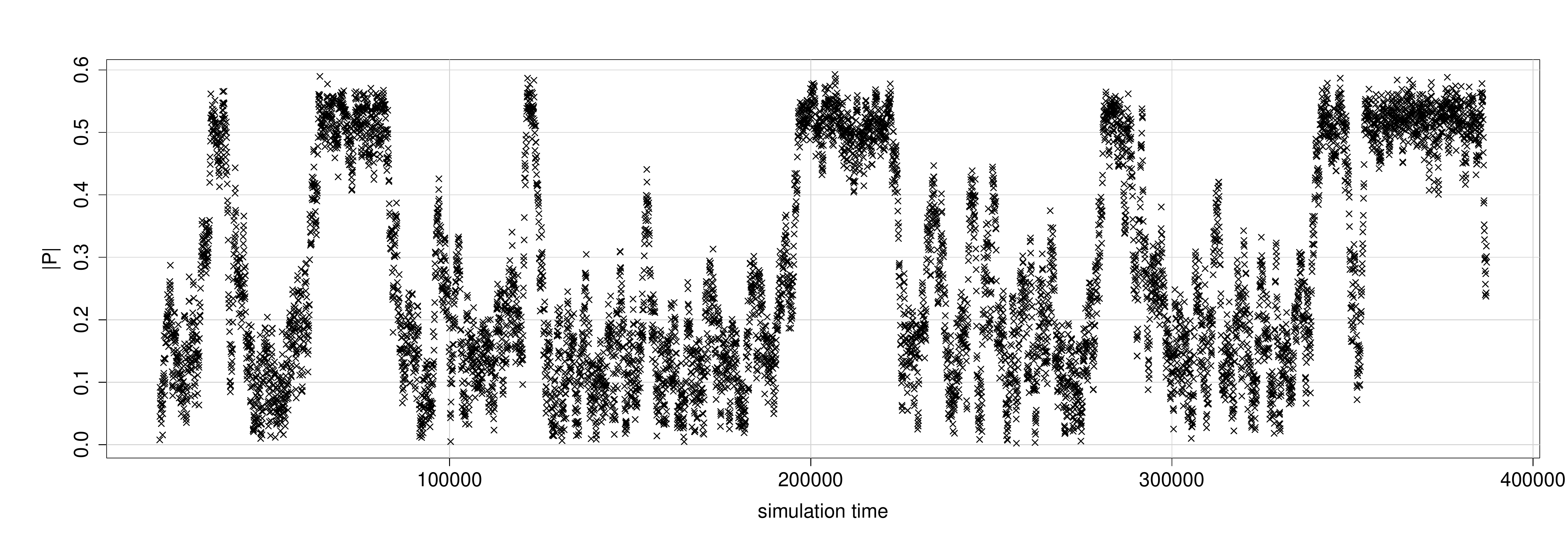}
	\end{subfigure}

	\begin{subfigure}[b]{1.0\textwidth}
		\includegraphics[width=\textwidth]{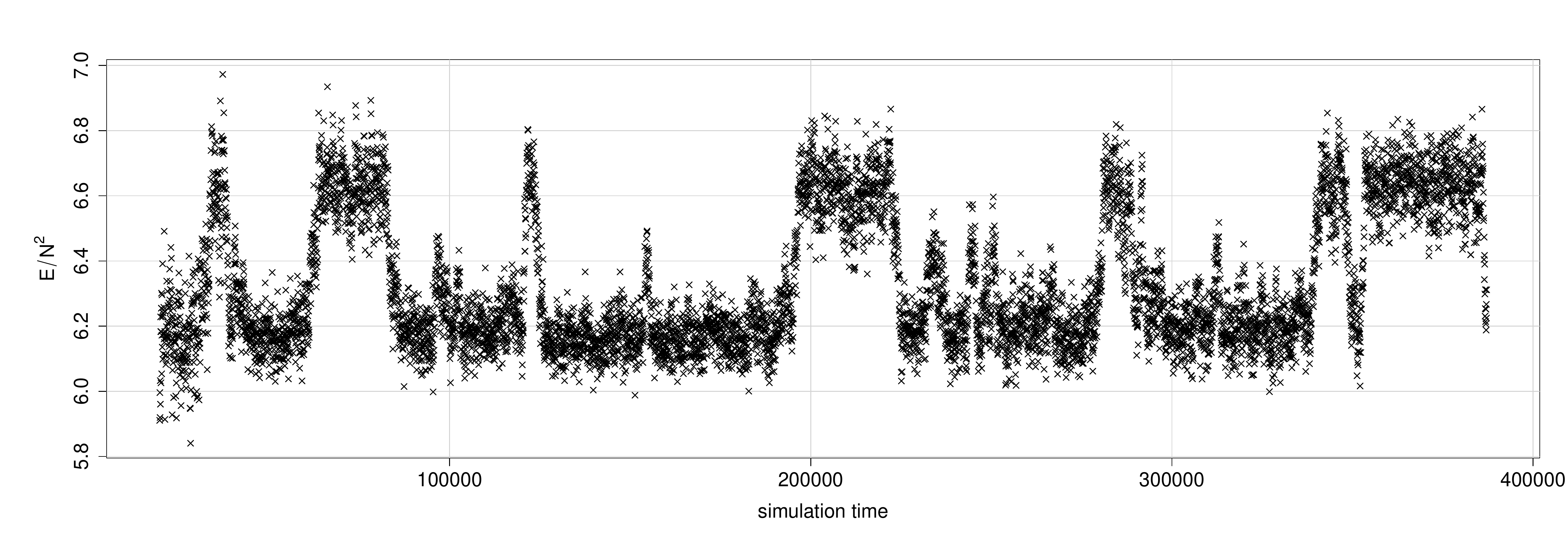}
	\end{subfigure}

	\begin{subfigure}[b]{1.0\textwidth}
		\includegraphics[width=\textwidth]{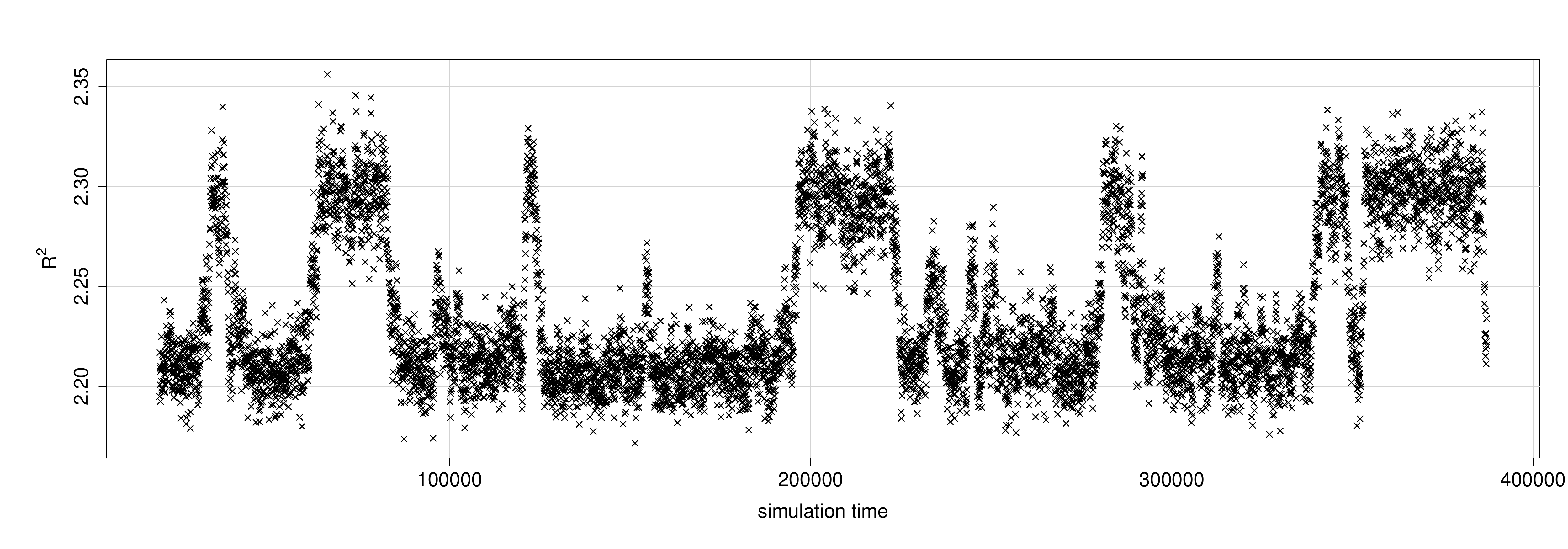}
	\end{subfigure}
	\caption{Simulation history for $N=64$, $L=24$, $D=10$, $T=0.885$. Only every 50th measurement is shown. Configurations are dropped at the beginning to remove thermalization effects.}\label{fig:HistoryD10}
\end{figure}

\begin{figure}
	\centering
	\begin{subfigure}[b]{1.0\textwidth}
		\includegraphics[width=\textwidth]{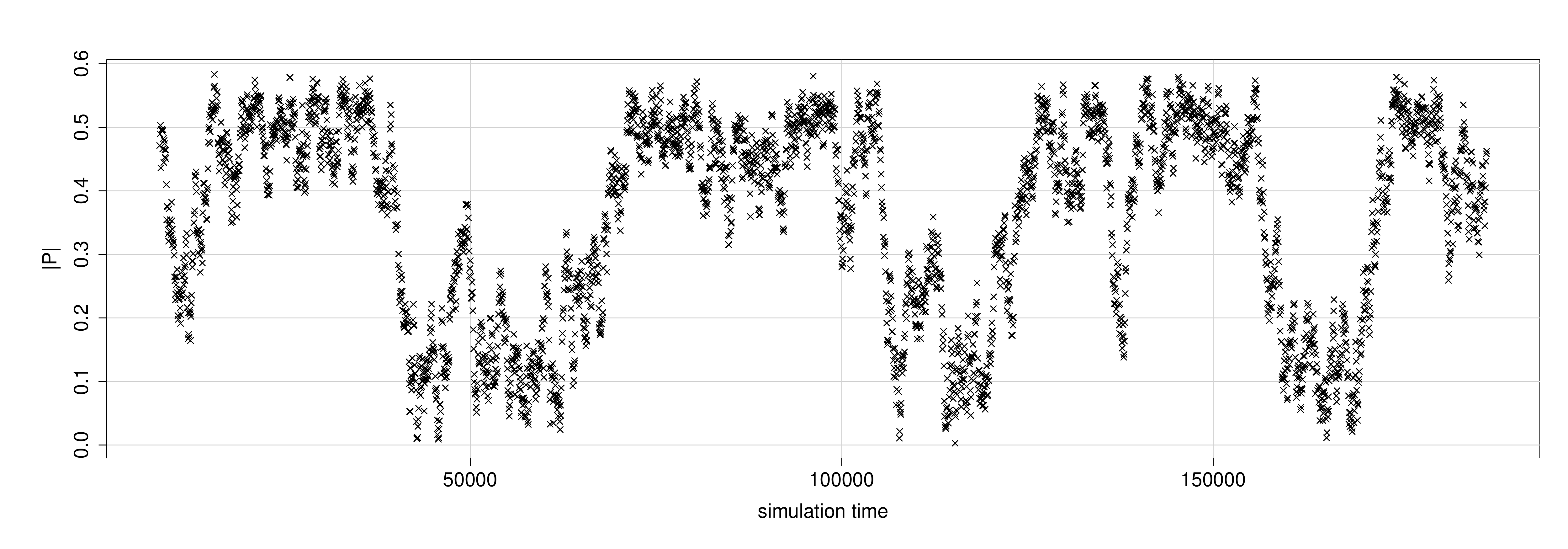}
	\end{subfigure}

	\begin{subfigure}[b]{1.0\textwidth}
		\includegraphics[width=\textwidth]{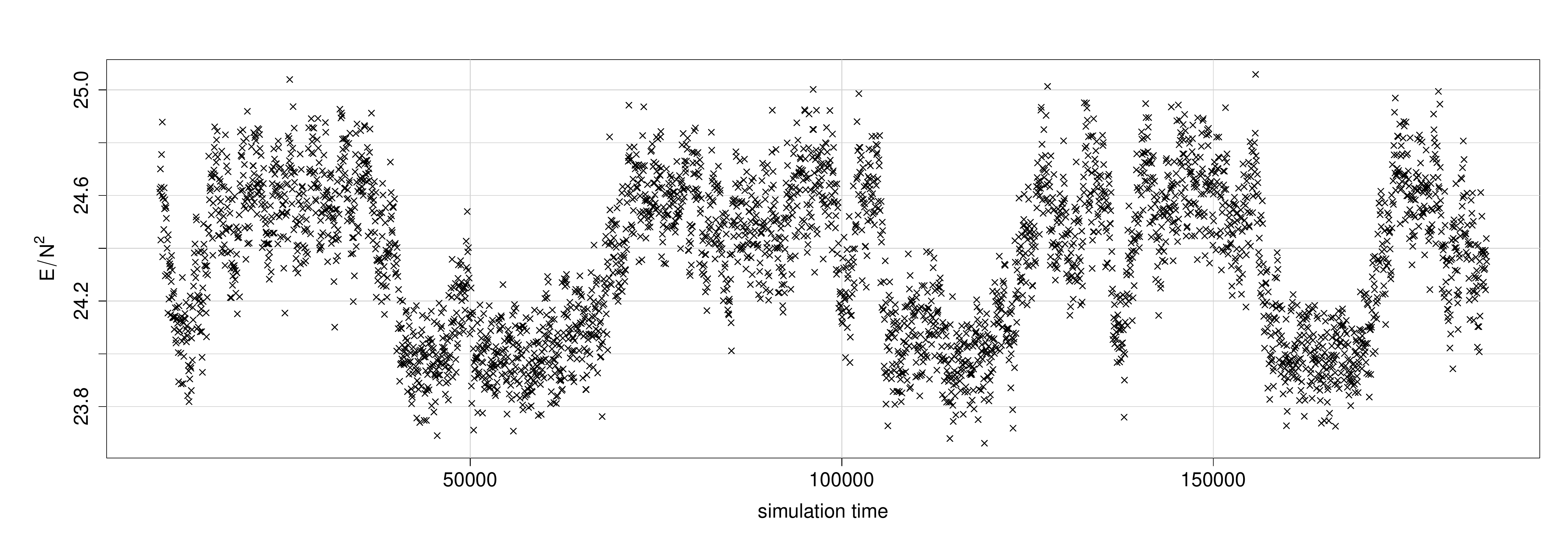}
	\end{subfigure}

	\begin{subfigure}[b]{1.0\textwidth}
		\includegraphics[width=\textwidth]{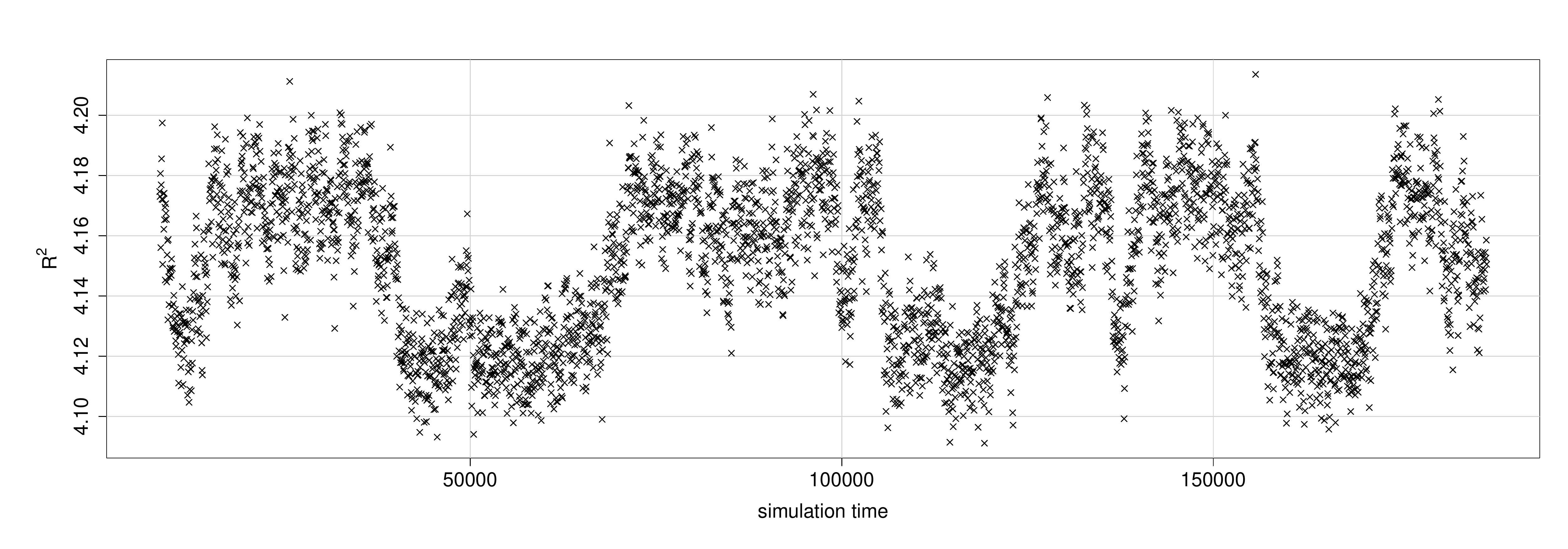}
	\end{subfigure}
	\caption{Simulation history for $N=64$, $L=24$, $D=26$, $T=0.8732$. Only every 50th measurement is shown. Configurations are dropped at the beginning to remove thermalization effects.}\label{fig:HistoryD26}
\end{figure}

\newpage

\section{Data for $N=32$}

\begin{figure}[h!]
	\centering
	\begin{subfigure}[b]{0.45\textwidth}
		\includegraphics[width=\textwidth]{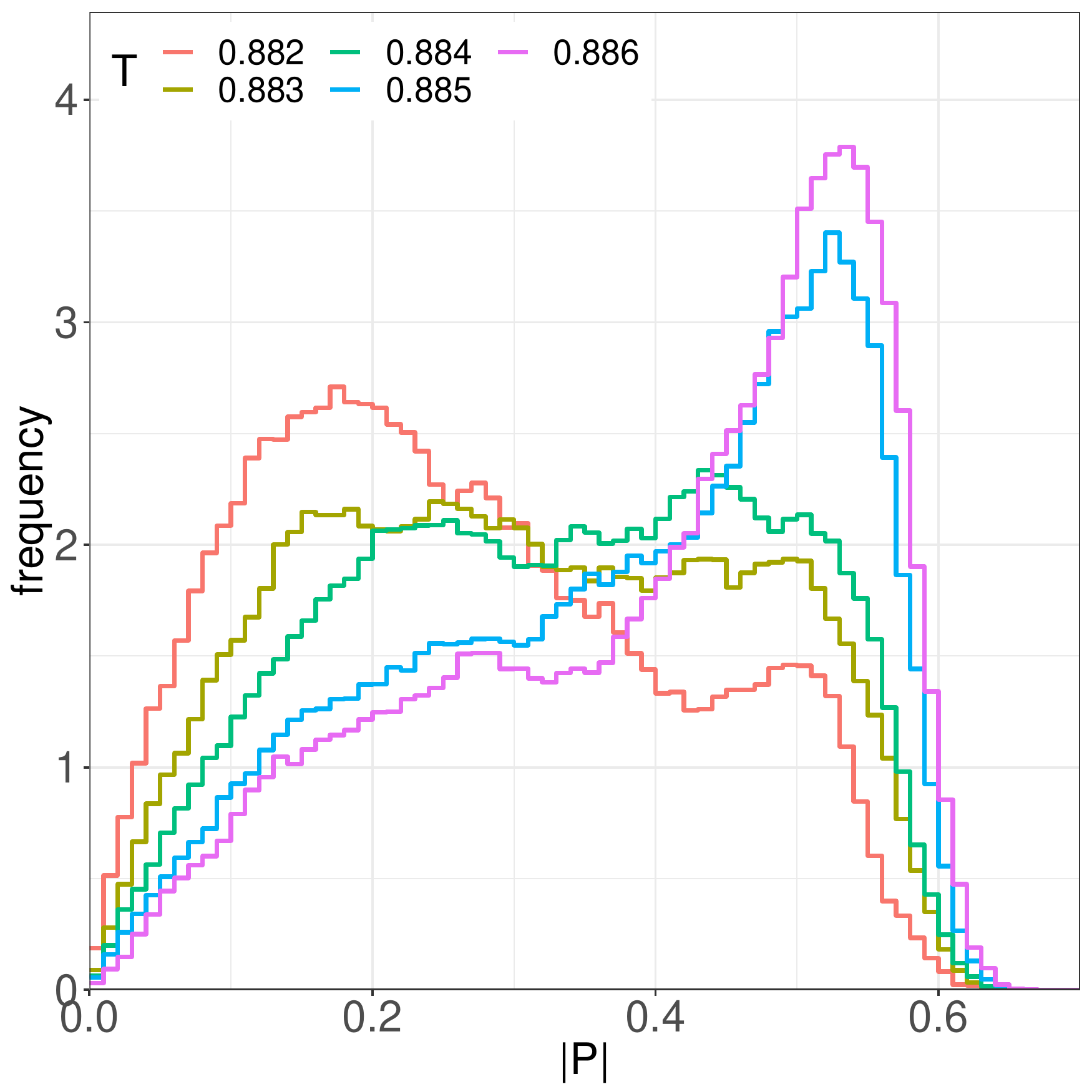}
	\end{subfigure}
	\begin{subfigure}[b]{0.45\textwidth}
		\includegraphics[width=\textwidth]{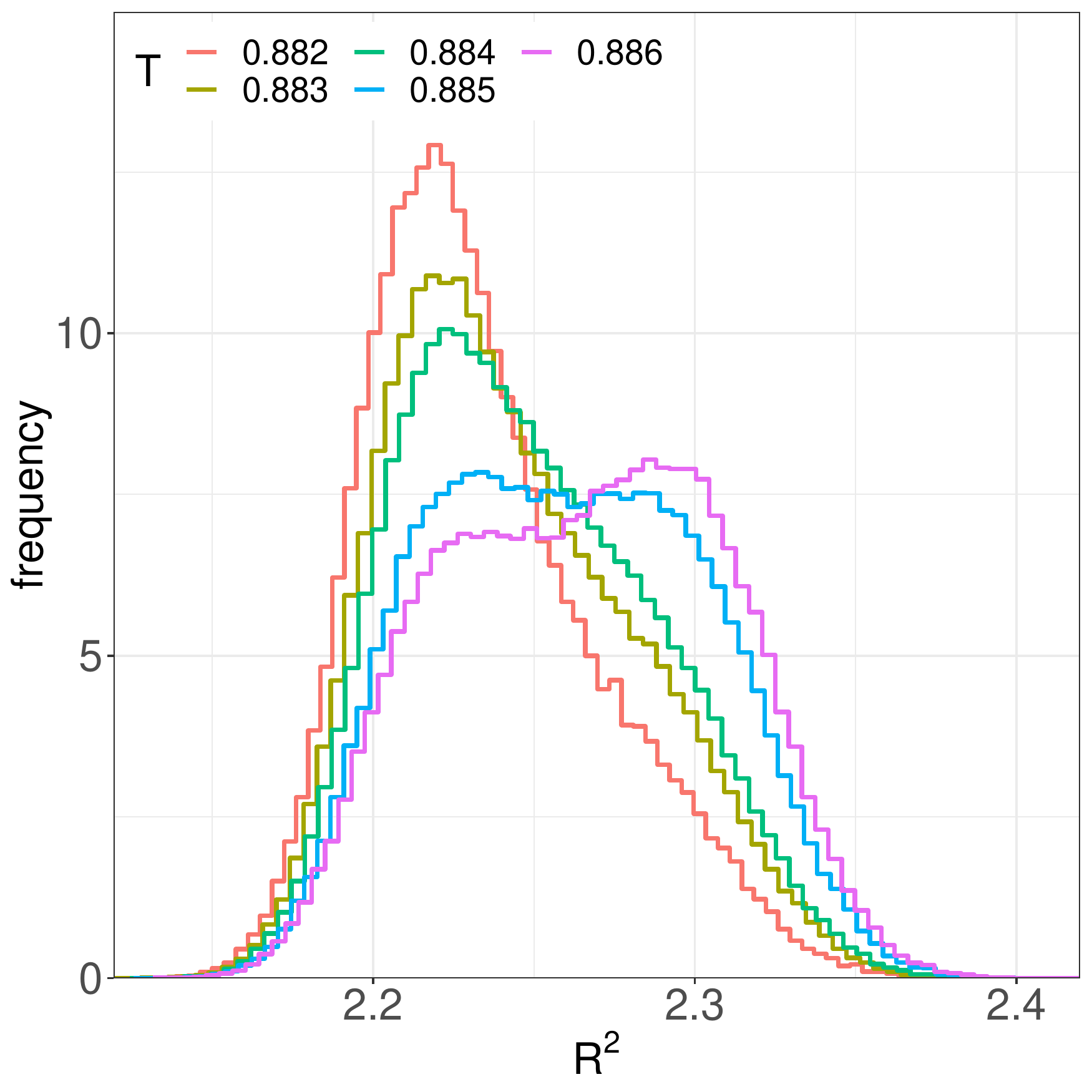}
	\end{subfigure}

	\caption{Histogram of the order parameter $|P|$ and the matrix size $R^2$ close to the transition temperature for the $D=10$ theory.
	Simulations are performed at $N=32$, $L=24$ and various temperatures.}\label{fig:N32}
\end{figure}

To compare with the results of Ref.~\cite{Azuma:2014cfa}, we provide histograms for $|P|$ and $R^2$ at $N=32$, $L=24$ in figure~\ref{fig:N32}.
Similarly to Ref.~\cite{Azuma:2014cfa}, the histogram of $|P|$ features a slight shoulder around $|P|= 0.25 - 0.3$ for the transition temperature $T=0.885$.
This may be taken as an indication for a first order transition.
However, as emphasized in the main text, a detailed analysis including larger $N$ and a continuum extrapolation is needed for a definite conclusion.

\end{document}